\documentclass[final,1p,times]{elsarticle}

\usepackage[T1]{fontenc} 

\usepackage{amsmath,amsthm,amsfonts}
\usepackage{amssymb}
\usepackage{graphicx}
\usepackage{csquotes}
\usepackage{float}

\newtheorem{lemma}{Lemma}

\newtheorem{corollary}{Corollary}
\usepackage{graphicx}
\usepackage{bcprules}

\usepackage{booktabs}
\usepackage{etoolbox}
\AtBeginEnvironment{tabular}{\footnotesize}

\usepackage{csquotes}

\usepackage{fancyvrb}
\usepackage{paralist}

\usepackage{todonotes}
\usepackage{multicol}
\usepackage{multirow}
\usepackage{subcaption}

\usepackage{setspace} 

\usepackage{hyperref}
\usepackage[noend,linesnumbered]{algorithm2e}

\usepackage{todonotes}

\theoremstyle{remark} 

\usepackage[inline]{enumitem}
\usepackage{stmaryrd}
\usepackage{url}
\usepackage{textcomp}

\usepackage{mathtools}

\usepackage{xcolor,colortbl}
\definecolor{Gray}{gray}{0.9}
\definecolor{LightCyan}{rgb}{0.88,1,1}

\usepackage{booktabs}

\usepackage{xcolor}

\colorlet{punct}{red!60!black}
\definecolor{background}{HTML}{EEEEEE}
\definecolor{delim}{RGB}{20,105,176}
\colorlet{numb}{magenta!60!black}

\usepackage{tikz}
\usepackage{ulem}
\newif\iflon
\iflon
\newcommand{\iflong}[1]{#1}
\newcommand{\ifshort}[1]{}
\else
\newcommand{\iflong}[1]{}
\newcommand{\ifshort}[1]{#1}
\fi

\newtheorem{definition}{Definition}

\newif\ifacm

\acmtrue    
\ifacm
\newtheorem{remark}{Remark}
\newtheorem{example}{Example}
\fi

\iflon
\usepackage[appendix=inline]{apxproof} 
\else
\usepackage[appendix=append]{apxproof}
\fi

\newtheoremrep{lemmax}[theorem]{Lemma}
\newtheoremrep{theox}[theorem]{Theorem}
\newtheoremrep{corx}[theorem]{Corollary}
\newtheoremrep{propx}[property]{Property}

  {\small \em \tt ccc\begin{table} aaa bbb}%
  {\end{table} ddd}

\newcommand{\json}{JSON}
\newcommand{\kw}[1]{\textbf{#1}}
\renewcommand{\kw}[1]{\ensuremath{\mathtt{#1}}}
\newcommand{\qkw}[1]{\ensuremath{\mathtt{\QQ{#1}\QQ}}}
\newcommand{\fkw}[1]{\ensuremath{\mathtt{\footnotesize\QQ{#1}\QQ}}}

\renewcommand{\kw}[1]{\akw{#1}}
\renewcommand{\qkw}[1]{\qakw{#1}}

\newcommand{\key}[1]{\ensuremath{\mathit{#1}}}

\newcommand{\akey}[1]{\ensuremath{\mathsf{#1}}}

\newcommand{\rkw}[1]{\ensuremath{\mathsf{#1}}}
\renewcommand{\rkw}[1]{\ensuremath{\mbox{\sf{\small #1}}}}
\newcommand{\akw}[1]{\ensuremath{\mbox{\tt{\small #1}}}}
\newcommand{\qakw}[1]{\QQ\akw{#1}\QQ}

\newcommand{\qnot}{\qkw{not}}
\newcommand{\xtrue}{\kw{true}}

\newcommand{\xfalse}{\kw{false}}

\newcommand{\xnull}{\kw{null}}
\newcommand{\qnull}{\qkw{null}}

\newcommand{\qone}{\qkw{oneOf}}

\newcommand{\qany}{\qkw{anyOf}}

\newcommand{\qall}{\qkw{allOf}}
\newcommand{\fall}{\fkw{allOf}}

\newcommand{\qreq}{\qkw{required}}

\newcommand{\qpattReq}{\qkw{patternRequired}}

\newcommand{\qtype}{\qkw{type}}

\newcommand{\qprops}{\qkw{properties}}

\newcommand{\qpropN}{\qkw{propertyNames}}
\newcommand{\fpropN}{\fkw{propertyNames}}

\newcommand{\qpattProps}{\qkw{patternProperties}}
\newcommand{\fpattProps}{\fkw{patternProperties}}

\newcommand{\qminP}{\qkw{minProperties}}

\newcommand{\qmaxP}{\qkw{maxProperties}}

\newcommand{\qaddProps}{\qkw{additionalProperties}}

\newcommand{\xunProps}{\kw{unevaluatedProperties}}
\newcommand{\qunProps}{\qkw{unevaluatedProperties}}

\newcommand{\qnotMof}{\qkw{notMultipleOf}}

\newcommand{\qpatt}{\qkw{pattern}}

\newcommand{\qnotPatt}{\qkw{notPattern}}

\newcommand{\quniqIts}{\qkw{uniqueItems}}
\newcommand{\xrepIts}{\kw{repeatedItems}}
\newcommand{\qrepIts}{\qkw{repeatedItems}}

\newcommand{\qcont}{\qkw{contains}}

\newcommand{\qcontAft}{\qkw{containsAfter}}

\newcommand{\qminC}{\qkw{minContains}}

\newcommand{\qmaxC}{\qkw{maxContains}}

\newcommand{\qminIt}{\qkw{minItems}}

\newcommand{\qmaxIt}{\qkw{maxItems}}

\newcommand{\qits}{\qkw{items}}
\newcommand{\fits}{\fkw{items}}

\newcommand{\qprefIts}{\qkw{prefixItems}}
\newcommand{\fprefIts}{\fkw{prefixItems}}

\newcommand{\qaddIts}{\qkw{additionalItems}}
\newcommand{\faddIts}{\fkw{additionalItems}}

\newcommand{\qunIts}{\qkw{unevaluatedItems}}

\newcommand{\xunStar}{\kw{unevaluated*}}
\newcommand{\qunStar}{\qkw{unevaluated*}}

\newcommand{\qdepS}{\qkw{dependentSchemas}}

\newcommand{\qconst}{\qkw{const}}

\newcommand{\qdref}{\qkw{\$ref}}
\newcommand{\fdref}{\fkw{\$ref}}

\newcommand{\qdrRef}{\qkw{\$recursiveRef}}

\newcommand{\qddefs}{\qkw{\$defs}}

\newcommand{\qobject}{\qkw{object}}

\newcommand{\qnumber}{\qkw{number}}

\newcommand{\qinteger}{\qkw{integer}}

\newcommand{\qstr}{\qkw{string}}

\newcommand{\qarray}{\qkw{array}}

\newcommand{\qboolean}{\qkw{boolean}}

\newcommand{\qdid}{\qkw{\$id}}

\newcommand{\qda}{\qkw{\$anchor}}

\newcommand{\qddRef}{\qkw{\$dynamicRef}}

\newcommand{\qdda}{\qkw{\$dynamicAnchor}}

\newcommand{\eqdef}{\ensuremath{\stackrel{\vartriangle}{=}}}

\newcommand{\gcomment}[1]{}
\newcommand{\oldversion}[1]{}
\newcommand{\hideforspace}[1]{}
\newcommand{\hide}[1]{}
\newcommand{\save}[1]{}
\newcommand{\code}[1]{}

\newcommand{\Iff}{\Leftrightarrow}
\newcommand{\Implies}{\Rightarrow}

\newcommand{\Or}{\vee}
\newcommand{\BigOr}{\bigvee}

\newcommand{\TypeOf}{\key{TypeOf}}
\newcommand{\comment}[1]{}

\renewcommand{\And}{\wedge}
\newcommand{\BigAnd}{\bigwedge}

\newcommand{\Not}{\neg}

\newcommand{\DNum}{\akey{DNum}}

\newcommand{\Str}{\akey{Str}}
\newcommand{\Nat}{\akey{Nat}}
\newcommand{\Int}{\akey{Int}}

\newcommand{\Bool}{\akey{Bool}}

\newcommand{\Min}{\akey{min}}

\newcommand{\keykey}[1]{\key{\underline{#1}}}

\newcommand{\Inf}{\infty}

\newcommand{\VerSix}{Draft-06}
\newcommand{\VerSeven}{Draft-07}
\newcommand{\VerEight}{Draft 2019-09}
\newcommand{\VerTwenty}{Draft 2020-12}

\newcommand{\DTwenty}{Draft 2020-12}
\newcommand{\JS}{JSON Sche\-ma}
\newcommand{\mJS}{Modern JSON Sche\-ma}
\newcommand{\MJS}{Modern JSON Sche\-ma}
\newcommand{\cJS}{Classical JSON Sche\-ma}

\newcommand{\custcom}[2]{\marginpar{\tiny #1: {#2}}}

\newcommand{\GG}[1]{\custcom{giorgio}{#1}}
\newcommand{\DC}[1]{\custcom{dario}{#1}}

\newcommand{\M}{\ |\ }

\newlength{\NL}
\newlength{\SaveNL}

\newcommand{\EmptySet}{\emptyset}

\newcommand{\NN}{\ensuremath{\ \hat{}\ }}
\newcommand{\QQ}{\textnormal{\textquotedbl}}
\newcommand{\Set}[1]{\{\,{#1}\,\}}

\newcommand{\SetOpen}{\{\!|}
\newcommand{\SetClose}{|\!\}}
\renewcommand{\Set}[1]{\SetOpen{#1}\SetClose}

\newcommand{\SetST}[2]{\SetOpen{#1}\,\mid\,{#2}\SetClose}
\newcommand{\SetIIn}[3]{\Set{#1}^{{#2}\in{#3}}}
\newcommand{\SetTo}[1]{\Set{1\ldots{#1}}}

\newcommand{\SetFromTo}[2]{\Set{{#1}\ldots{#2}}}
\newcommand{\List}[1]{[\!|\,{#1}\,|\!]}

\newcommand{\ListOpen}{[\mkern-4.8mu|}
\newcommand{\ListClose}{|\mkern-4.67mu]}
\renewcommand{\List}[1]{\ListOpen\,{#1}\,\ListClose}
\newcommand{\ListST}[2]{\ListOpen\,{#1}\,\mid\,{#2}\,\ListClose}

\newcommand{\str}{s}

\newcommand{\J}{{J}}
\newcommand{\keys}[1]{\K{Keys}(#1)}
\newcommand{\K}[1]{\mbox{\text{\it{#1}}}}

\newcommand{\semt}{\ensuremath{\mathit{JVal}}}

\newcommand{\rlan}[1]{L(#1)}

\newcommand{\PP}[1]{\!\!\NN\!{#1}}
\newcommand{\PPP}[1]{\!\NN\!{#1}\$}

\renewcommand{\SetOpen}{\{\!|\,}
\renewcommand{\SetClose}{\,|\!\}}
\renewcommand{\Set}[1]{\SetOpen{#1}\SetClose}
\renewcommand{\SetST}[2]{\SetOpen{#1}\,\mid\,{#2}\SetClose}

\newcommand{\Get}{\kw{get}}
\newcommand{\GetS}{\kw{gets}}
\newcommand{\GetK}{\kw{getk}}

\newcommand{\DGet}{\kw{dget}}

\newcommand{\Load}{\kw{load}}
\newcommand{\StackDash}[1]{\stackrel{#1}{\vdash}}
\renewcommand{\StackDash}[1]{\mathrel{\vdash\raisebox{.90ex}{\tt\kern-0.5em {\tiny {#1}\kern0.5em}}}}

\newcommand{\Judg}[4]{{#2}\StackDash{#1}{#3}\ ? \ {#4}}
\renewcommand{\Judg}[4]{{}\StackDash{#1}{#3}\ ? \ {#4}}
\newcommand{\JudgC}[3]{\Judg{#1}{C}{#2}{#3}}
\newcommand{\JudgCJ}[2]{\JudgC{#1}{J}{#2}}
\newcommand{\Ret}[1]{\,\,\stackrel{#1}{\rightarrow}\,\,}
\newcommand{\RetL}[1]{\,\,{\rightarrow}\,\,}
\renewcommand{\RetL}[1]{\,\,\stackrel{#1}{\rightarrow}\,\,}
\newcommand{\Kl}{\vec{K}}

\newcommand{\KJudg}[3]{\Judg{K}{#1}{#2}{#3}}
\newcommand{\KJudgC}[2]{\JudgC{K}{#1}{#2}}
\newcommand{\KJudgCJ}[1]{\JudgCJ{K}{#1}}

\newcommand{\KLJudgCJ}[1]{\JudgCJ{L}{#1}}

\newcommand{\SJudg}[3]{\Judg{S}{#1}{#2}{#3}}
\newcommand{\SJudgC}[2]{\JudgC{S}{#1}{#2}}
\newcommand{\SJudgCJ}[1]{\JudgCJ{S}{#1}}

\newcommand{\pk}{\ensuremath{\kappa}}
\newcommand{\ps}{\ensuremath{\sigma}}

\newcommand{\rl}{\vec{r}}

\newcommand{\anot}{\qakw{not}}
\newcommand{\atrue}{\akw{true}}
\newcommand{\afalse}{\akw{false}}
\newcommand{\anull}{\akw{null}}
\newcommand{\aone}{\qakw{oneOf}}
\newcommand{\aany}{\qakw{anyOf}}
\newcommand{\aall}{\qakw{allOf}}
\newcommand{\amin}{\qakw{minimum}}
\newcommand{\amax}{\qakw{maximum}}
\newcommand{\areq}{\qakw{required}}

\newcommand{\apattReq}{\qakw{patternRequired}}
\newcommand{\atype}{\qakw{type}}

\newcommand{\aprops}{\qakw{properties}}
\newcommand{\apropN}{\qakw{propertyNames}}
\newcommand{\apattProps}{\qakw{patternProperties}}
\newcommand{\aminP}{\qakw{minProperties}}
\newcommand{\amaxP}{\qakw{maxProperties}}

\newcommand{\aaddProps}{\qakw{additionalProperties}}

\newcommand{\aunProps}{\qakw{unevaluatedProperties}}
\newcommand{\aunIts}{\qakw{unevaluatedItems}}

\newcommand{\anotMof}{\qakw{notMultipleOf}}

\newcommand{\apatt}{\qakw{pattern}}
\newcommand{\anotPatt}{\qakw{notPattern}}
\newcommand{\auniqIt}{\qakw{uniqueItems}}

\newcommand{\acont}{\qakw{contains}}
\newcommand{\acontAft}{\qakw{containsAfter}}

\newcommand{\aminC}{\qakw{minContains}}
\newcommand{\amaxC}{\qakw{maxContains}}
\newcommand{\aminIt}{\qakw{minItems}}
\newcommand{\amaxIt}{\qakw{maxItems}}
\newcommand{\ait}{\qakw{items}}

\newcommand{\aits}{\qakw{items}}
\newcommand{\aprefIts}{\qakw{prefixItems}}

\newcommand{\aconst}{\qakw{const}}
\newcommand{\adref}{\qakw{\$ref}}

\newcommand{\addefs}{\qakw{\$defs}}

\newcommand{\ada}{\qakw{\$anchor}}
\newcommand{\addref}{\qakw{\$dynamicRef}}

\newcommand{\rnot}{\rkw{not}}
\newcommand{\rtrueS}{\rkw{trueSchema}}
\newcommand{\rfalseS}{\rkw{falseSchema}}

\newcommand{\rone}{\rkw{oneOf}}
\newcommand{\rany}{\rkw{anyOf}}
\newcommand{\rall}{\rkw{allOf}}

\newcommand{\rprops}{\rkw{properties}}

\newcommand{\rpattProps}{\rkw{patternProperties}}

\newcommand{\raddProps}{\rkw{additionalProperties}}

\newcommand{\runProps}{\rkw{unevaluatedProperties}}
\newcommand{\runIts}{\rkw{unevaluatedItems}}

\newcommand{\rcont}{\rkw{contains}}
\newcommand{\rcontAft}{\rkw{containsAfter}}

\newcommand{\rits}{\rkw{items}}
\newcommand{\rprefIts}{\rkw{prefixItems}}

\newcommand{\rdref}{\rkw{\$ref}}

\newcommand{\robject}{\rkw{object}}
\newcommand{\rnumber}{\rkw{number}}
\newcommand{\rstr}{\rkw{string}}
\newcommand{\rarray}{\rkw{array}}

\newcommand{\rddref}{\rkw{\$dynamicRef}}

\newcommand{\rother}{\rkw{other}}

\newcommand{\rklist}{\rkw{klist}}
\newcommand{\JObjOpen}{\{}
\newcommand{\JObjClose}{\}}
\newcommand{\JObj}[1]{\JObjOpen\,{#1}\,\JObjClose}
\newcommand{\JArrOpen}{[}
\newcommand{\JArrClose}{]}
\newcommand{\JArr}[1]{\JArrOpen\,{#1}\,\JArrClose}

\newcommand{\plus}{+}

\newcommand{\TTriv}{\rkw{Triv}}

\newcommand{\cat}{\!\cdot\!}
\newcommand{\lcat}{\cdot}
\newcommand{\catHash}{\cat\qkw{\#}\cat}

\newcommand{\RAsA}[1]{R\mbox{\ as\ }{A}}

\newcommand{\fstURI}{\kw{fstURI}}

\newcommand{\bshort}{\noindent\begin{minipage}{0.5\textwidth}}
\newcommand{\eshort}{ \end{minipage}}

\newcommand{\shortrulefirst}[3]{\hspace*{-2.3em}\bshort \infrule[{#1}]{#2}{#3} \eshort}\newcommand{\shortrule}[3]{\hfill\bshort \infrule[{#1}]{#2}{#3} \eshort}

\newcommand{\ES}{\EmptySet}

\newcommand{\setmax}{\key{max}}

\newcommand{\Mm}{\ \ }

\renewcommand{\Mm}{\ \M\ }

\renewcommand{\ps}{\pk}

\renewcommand{\Ret}[1]{\rightarrow}

\newcommand{\btrue}{\mathcal{T}}
\newcommand{\bfalse}{\mathcal{F}}

\newcommand{\exactly}[1]{\ensuremath{\,\hat{}\,{#1}\$}}

\newcommand{\MinEP}{\key{minEP}}
\newcommand{\MaxEP}{\key{maxEP}}
\newcommand{\MinEI}{\key{minEI}}
\newcommand{\MaxEI}{\key{maxEI}}
\newcommand{\MinEPK}[1]{\MinEP(\{{#1}\})}
\newcommand{\MaxEPK}[1]{\MaxEP(\{{#1}\})}
\newcommand{\MinEIK}[1]{\MinEI(\{{#1}\})}
\newcommand{\MaxEIK}[1]{\MaxEI(\{{#1}\})}
\newcommand{\ExEP}{\key{exEP}}
\newcommand{\ExEI}{\key{exEI}}

\newcommand{\PCs}{\mbox{\it{p-covers}}}
\newcommand{\PCd}{\mbox{\it{p-covered}}}
\newcommand{\ICs}{\mbox{\it{i-covers}}}
\newcommand{\ICd}{\mbox{\it{i-covered}}}
\newcommand{\Cs}{\key{covers}}
\newcommand{\Cd}{\key{covered}}

\newcommand{\ENF}{\key{ENF}}

\newcommand{\XDNF}{\key{DNF}^{X}}

\newcommand{\XNNOT}{\key{XNotNot}}
\newcommand{\XNNOTOf}[1]{\APPLY{\XNNOT}{#1}}
\newcommand{\XNOT}{\key{XNot}}
\newcommand{\XNOTOf}[1]{\APPLY{\XNOT}{#1}}

\newcommand{\APPLY}[2]{{#1}({#2})}
\newcommand{\LC}{\mathbin{+\!+}}
\newcommand{\LOr}{\key{Or}}
\newcommand{\LAnd}{\key{And}}
\newcommand{\LClose}{\key{Close}}
\newcommand{\qunReq}{\qkw{unevaluatedRequired}}

\newcommand{\Split}{\key{Split}}

\newcommand{\PUPkey}{\key{PushUnProps}}
\newcommand{\PUP}[2]{\PUPkey({#1},{#2})}
\newcommand{\PUPe}[2]{\key{PUPBranch}({#1},{#2})}
\newcommand{\PUI}[2]{\key{PushUnItems}({#1},{#2})}
\newcommand{\PUIe}[2]{\key{PUIBranch}({#1},{#2})}

\newcommand{\qallOf}[1]{\JObj{\qall : [ #1 ]}}
\newcommand{\qanyOf}[1]{\JObj{\qany : [ #1 ]}}
\newcommand{\qoneOf}[1]{\JObj{\qone : [ #1 ]}}
\newcommand{\qnotOf}[1]{\JObj{\qnot: {#1}}}

\newcommand{\Up}{\!\!\uparrow}
\newcommand{\Down}{\!\!\downarrow}
\newcommand{\ESC}{elementwise statically characterized}
\newcommand{\UElim}{\key{Elim}}

\newcommand{\SEP}[1]{\key{UnlessLastInS}({#1})}
\renewcommand{\SEP}[1]{\key{Sep}({#1})}
\newcommand{\Col}{\!\,:\!\,}

\newcommand{\Cal}[1]{\ensuremath{\mathcal{#1}}}
\newcommand{\ddefined}{named}
\newcommand{\TL}[2]{\langle{#1},{#2}\rangle}

\newcommand{\TLF}{\key{TS}_{\iK}}

\newcommand{\SOf}[1]{\Cal{S}_{(S,{#1})}}
\newcommand{\KOf}[1]{\Cal{K}_{(S,{#1})}}
\newcommand{\SKOf}[1]{\Cal{SK}_{(S,{#1})}}

\newcommand{\InSketch}[1]{#1}
\newcommand{\InAppendix}[1]{}
\newcommand{\pK}{\ensuremath{\mathit{pk}}}
\newcommand{\iK}{\ensuremath{\mathit{ik}}}
\renewcommand{\AA}{annotation-accepting}
\newcommand{\AAF}{{\AA} from}

 \usepackage{tikz-cd}
 \tikzcdset{every label/.append style = {font = \normalsize}}
 
\usepackage[skins]{tcolorbox}
\usepackage{listings}
\usepackage[framemethod=tikz]{mdframed}

\makeatletter
\newcommand\querysize{\@setfontsize\querysize\@vipt\@viipt}
\makeatother
\definecolor{mygray}{rgb}{0.643,0.643,0.643}
\lstdefinestyle{querynonumbers}{
    language=JSON,
    stepnumber=1,
    numbersep=1pt, 
    xleftmargin =-2pt,
    tabsize=4,
    showspaces=false,
    showstringspaces=false,
    basicstyle=\linespread{1}\fontfamily{lmtt}\selectfont\querysize,
    keywordstyle=\color{blue},
    stringstyle=\color{purple},
    upquote=true,
    breaklines=true,
    commentstyle=\color{CadetBlue},
}
\lstdefinestyle{oneliner}{
    language=JSON,
    aboveskip=0pt,
    belowskip=0pt,
    basicstyle=\linespread{1}\fontfamily{lmtt}\selectfont\scriptsize,
}
\lstdefinestyle{query}{
    language=JSON,
    numbers=left,
    stepnumber=1,
    numbersep=2pt, 
    numberstyle=\color{black!65},
    xleftmargin =-2pt,
    tabsize=4,
    showspaces=false,
    showstringspaces=false,
    basicstyle=\linespread{1}\fontfamily{lmtt}\selectfont\querysize,
    keywordstyle=\color{blue},
    stringstyle=\color{purple},
    upquote=true,
    breaklines=true,
    commentstyle=\color{CadetBlue}
}
\newtcolorbox{querybox}[2][]{%
    sidebyside align=top,
    enhanced,
    boxsep=0pt,
    arc=0pt,
    top=-3pt, bottom=-3pt,
    left=12pt, right=0pt,
    colback=background,
    colframe=gray!90,
    boxrule=0.5pt,
    leftrule=12pt,
    overlay unbroken and first ={%
    \node[rotate=90,
          minimum width=0.5cm,
          anchor=south,
          font=\scriptsize\rmfamily,
          yshift=-13.7pt,
          white]
    at (frame.west) {\#2};
    }
}

\lstdefinelanguage{json}{
    literate=
      {:}{{{\color{punct}{:}}}}{1}
      {,}{{{\color{punct}{,}}}}{1}
      {\{}{{{\color{delim}{\{}}}}{1}
      {\}}{{{\color{delim}{\}}}}}{1}
      {[}{{{\color{delim}{[}}}}{1}
      {]}{{{\color{delim}{]}}}}{1},
}

\journal{Theoretical Computer Science}

\begin{document}

\begin{frontmatter}

\title{Elimination of annotation dependencies in validation for Modern JSON Schema}

\author[1]{Lyes Attouche}\ead{lyes.attouche@dauphine.psl.eu}
\author[2]{Mohamed-Amine Baazizi}
\ead{baazizi@ia.lip6.fr}
\author[1]{Dario Colazzo} \ead{dario.colazzo@dauphine.fr}
\author[3]{Giorgio Ghelli} \ead{ghelli@di.unipi.it}
\author[4]{Stefan Klessinger} \ead{Stefan.Klessinger@uni-passau.de}
\author[5]{Carlo Sartiani\corref{cor1}} \ead{carlo.sartiani@unibas.it}
\author[4]{Stefanie Scherzinger} \ead{stefanie.scherzinger@uni-passau.de}
\cortext[cor1]{Corresponding author}

\affiliation[1]{organization={Universit\'e Paris-Dauphine, PSL Research University}, 
                    addressline={Place du Mar\'echal de Lattre de Tassigny},
                    city={Paris},
                    postcode={75775},
                    country={France}}
\affiliation[2]{organization={Sorbonne Universit\'e, LIP6 UMR 7606}, 
                    addressline={4 place Jussieu},
                    postcode={75252}, 
                    city={Paris}, 
                    country={France}}
\affiliation[3]{organization={Dipartimento di Informatica, Universit\`a di Pisa},
                    addressline={Largo Bruno Pontecorvo, 3}, 
                    city={Pisa}, 
                    postcode={56127},
                    country={Italy}}
\affiliation[4]{organization={Universit\"at Passau},
                    addressline={Innstr. 43}, 
                    city={Passau}, 
                    postcode={94032},
                    country={Germany}}                     
\affiliation[5]{organization={DiING, Universit\`a della Basilicata},
                    addressline={Via dell'Ateneo Lucano, 10}, 
                    city={Potenza}, 
                    postcode={85100},
                    country={Italy}}

\begin{abstract}
{\JS} is a logical language used to define the structure of JSON values.
{\JS} syntax is based on nested \emph{schema} objects.
In all versions of {\JS} until {\VerSeven}, collectively known as {\cJS},
the semantics of a \emph{schema} was entirely described by the set of 
JSON values that it  validates. This semantics was the basis for a thorough theoretical study and for the development
of   tools to decide satisfiability and equivalence of schemas.
Unfortunately, {\cJS}  suffered a severe limitation in its ability to express extensions of object schemas,
which caused the introduction, with {\VerEight}, of two disruptive 
features: annotation dependency and dynamic references.

Annotation dependency means that the validation of an instance~$J$ by a schema~$S$ produces a
yes/no answer, as in {\cJS}, and also an ``annotation'', which is a set of \emph{evaluated fields/items}, 
and this annotation is passed to the annotation-dependent keywords, {\qunProps} and {\qunIts}, whose behavior
depends on it. Dynamic references is a distinct mechanism that we do not describe here.
This drastic change in the semantics
is the reason why the versions of {\JS} from {\VerEight} onwards are collectively called 
{\mJS}.

These new features undermine the previously developed semantic theory, and
the algorithms used to decide satisfiability and inclusion for {\cJS} are not easy to extend
to the new setting.
One possible solution is ``elimination'' --- rewriting a schema written in {\mJS} into an equivalent 
schema in {\cJS}.

%
In a recent paper, we provided an algorithm to eliminate dynamic references, but the elimination
of annotation dependencies remained open.

In this paper we prove that the elimination of annotation dependent keywords cannot, in general, avoid an exponential increase 
of the schema dimension. We then provide an algorithm to eliminate these keywords that, despite the theoretical  lower bound, behaves quite well in practice, as we verify with an extensive set of experiments.
\end{abstract}


\end{frontmatter}

\section{Introduction}\label{sec:intro}

{\JS} is a logical language used to define the structure of JSON values, to allow different pieces of software to cooperate.
{\JS} syntax is based on nested \emph{schema} objects.
In all versions of {\JS} until {\VerSeven}, which are collectively known as {\cJS},
the semantics of a \emph{schema} was entirely described by the set of 
JSON values that it validates. This semantics was the basis for a thorough theoretical study and for the development
of some semantic tools for {\JS}, notably tools to decide satisfiability and equivalence of schemas.
Unfortunately, {\cJS} suffered a severe limitation in its ability to express extensions of object schemas.
After a lengthy discussion, this led to the introduction of two disruptive features in version
 {\VerEight}: annotation dependency and dynamic references.

Annotation dependency means that the validation of an instance $J$ by a schema $S$ produces a
yes/no answer, as in {\cJS}, and also an ``annotation'', which is a set of \emph{evaluated fields/items}. 
This annotation is passed to the annotation-dependent keywords {\qunProps} and {\qunIts}, whose behavior
depends on it. Dynamic references are a distinct mechanism that we do not describe here.
This drastic change in the semantics
is the reason why the versions of {\JS} from {\VerEight} onward are collectively called
{\mJS}.

These new features undermine the previously developed semantic theory, and
the algorithms used to decide satisfiability and inclusion for {\cJS} are not easy to extend
to the new setting.
One possible solution is ``elimination'' --- rewriting a schema written in {\mJS} into an equivalent
schema in {\cJS}.

\subsection{{\JS}, schema factorization, and annotation dependency}

The definition of object schemas by extending other schemas is a crucial ability, used in many application
fields. Unfortunately, in {\cJS}, this is quite difficult when the resulting object schema is required to be \emph{closed},
where \emph{closed} means that no other properties are allowed apart from those explicitly described.

We use here some examples to introduce {\JS}, to exemplify the problem, and to show its solution in {\mJS}
(the classification of {\JS} Drafts as Classical vs.\ Modern has been introduced by Henry Andrews in \cite{modern}).

The following schema defines the shape of the parameters of a protocol in which every exchanged message has \qkw{from} 
and \qkw{class} 
properties, has either a \qkw{payload} or an \qkw{errorCode} property, and has no other property.

\begin{Verbatim}[fontsize=\small,frame=lines]
{ "anyOf": [ { "properties": { "from": { "$ref": ".../address" }, 
                               "class": { "const": "info" },
                               "payload": { "type": "string" } },
               "required": [ "from", "class", "payload" ],
               "additionalProperties": false },
             { "properties": { "from": { "$ref": ".../address" },
                               "class": { "const": "error"},
                               "errorCode": { "type": "integer" } },
               "required": [ "from", "class", "errorCode" ],
               "additionalProperties": false }
  ]
}
\end{Verbatim}

The boolean operator ${\qany : [S_{1}, S_{2}]}$ specifies that a {\json} value must satisfy either schema $S_{1}$ or schema~$S_{2}$, or both. The keyword ${\qprops: \JObj{\qkw{a}: S}}$   specifies that, \emph{if} the \qkw{a} property is present, then its value must satisfy
$S$, while $\qreq$ forces a property to be present; $\qaddProps : S$ specifies that every property that does not match any name introduced by an ``adjacent'' $\qprops$ keyword must have a value that satisfies $S$; when $S=\xfalse$, 
then no such property can be present, since no value satisfies $\xfalse$;
two keywords are ``adjacent'' when they are top-level keywords in the same object. In particular, in the above example, the first $\qaddProps : \xfalse$ forbids any name different from \qkw{from}, \qkw{class},  and \qkw{payload}, while the second forbids any name different from \qkw{from}, \qkw{class} and \qkw{errorCode}.
The ${\qdref:u}$ keyword allows any other schema to be referred using a URI $u$.

We would like to specify these two formats in a different way, by specifying that there is a basic kernel that can be extended in two different ways.
This ability is not
very relevant in this toy example, but it is crucial in practice.
Hence, we may try and rewrite the above schema as follows.

\begin{Verbatim}[fontsize=\small,frame=lines]
{ "properties": { "from": { "$ref": ".../address" } },
  "required": [ "from" ],
  "anyOf": [
    { "properties": { "class": { "const": "info" }, 
                      "payload": { "type": "string" } },
      "required": [ "class", "payload" ] },
    { "properties": { "class": { "const": "error" }, 
                      "errorCode": { "type": "integer" } },
      "required": [ "class", "errorCode" ] } ],
  "additionalProperties": false
}
\end{Verbatim}

This schema enforces the same constraints on \qkw{from}, \qkw{class}, \qkw{errorCode}, and \qkw{payload} as the previous schema. However, {\qaddProps} is applied to any property that is not defined by an \emph{adjacent} 
{\qprops} keyword;  therefore, in this schema, it would consider \qkw{class}, \qkw{errorCode}, and \qkw{payload} as ``additional'' since the {\qprops} that introduces them is not
adjacent to {\qaddProps}, hence {\qaddProps: \afalse} would be applied, and fail, to any  \qkw{class}, \qkw{errorCode}, or \qkw{payload} property.
If we moved {\qaddProps: \afalse} inside the branches of the {\qany}, as in the previous version, then {\qaddProps} would be adjacent to the internal
{\qprops} keywords, hence it would consider a \qkw{from} property as ``additional'', and validation of any object with a \qkw{from} property
would fail.

The problem stems from the fact that the definition of {\qaddProps} depends on the syntactic feature of ``being adjacent'' to {\qprops}, hence is not robust
with respect to schema factorization.
To solve this problem, {\JS} added, with version {\VerEight}, a new keyword called $\qunProps$, and changed the validation semantics.
Now, validation of an instance $J$ against a schema $S$ returns both a boolean result (\emph{valid} or \emph{not valid}) 
and a set of \emph{evaluated properties} --- the properties of the JSON instance under validation
that have been evaluated by a $\qprops$ keyword invoked directly or indirectly
by the operator. In the rewritten schema above, the first $\qprops$ keyword directly evaluates $\qkw{from}$, the $\qany$ keyword indirectly evaluates
either \qkw{class} and \qkw{payload} or \qkw{class} and \qkw{errorCode}, and, if we substitute $\qaddProps: \xfalse$ with $\qunProps: \xfalse$, this new
keyword would regard the properties evaluated by the adjacent $\qprops$ and $\qany$ as ``evaluated''. Hence, their presence in the object
would not cause any problem, while any other property would be regarded as ``unevaluated'', and in this case the keyword would fail
(because of the $\xfalse$ parameter), as desired. As a consequence, this new version of the schema, thanks to
{\qunProps}, would behave as the original one.

In summary, $\qunProps$ allows the factorization of the definition of ``closed'' object types, that is, types that admit no other fields apart from those that
are explicitly specified, and it
relies on the fact that, in {\mJS}, validation
returns both a boolean and an \emph{annotation}, specifying which object properties have been evaluated.
{\MJS} also defines the operator $\qunIts$, which behaves in a similar way for array schemas.

\subsection{Motivation of this research}

The evaluation model of {\JS} validation before {\VerEight} (\cJS) has been studied thoroughly~\cite{DBLP:conf/www/PezoaRSUV16,DBLP:conf/pods/BourhisRSV17}, and is quite simple:
the behavior of a schema is defined by the set of instances that it validates.
With
{\VerEight} (\mJS) the evaluation model changes in two ways:
\begin{itemize}
\item the addition of the $\qunProps$ and $\qunIts$ keywords requires that validation returns both a boolean (as before) and
    an ``annotation'';
\item the addition of the $\qdrRef$ operator, 
   generalized by the $\qddRef$ operator in {\VerTwenty},
   collectively called \emph{dynamic references}, causes validation to depend on the \emph{dynamic context}, a complex notion that 
   may be defined as ``the list of references 
    that have been followed by the operators that caused the invocation of the current operator'' \cite{Draft12}. 
\end{itemize}

Validation of {\cJS} is known to be  P-complete \cite{DBLP:conf/www/PezoaRSUV16}.
 In \cite{DBLP:journals/pacmpl/AttoucheBCGSS24}, it has been proved that the addition of 
annotation-dependent keywords alone leaves validation inside the P class, while the addition of dynamic references alone
makes validation PSPACE-complete, as does the combination of the two. In the same paper, an algorithm has been presented to eliminate $\qddRef$, that is, to rewrite
all instances of $\qddRef$ into instances of $\qdref:u$ keyword, which simply retrieves the schema referred by $u$ with no dependency
on the dynamic context. This rewriting requires, in general, an exponential increase in the size of the rewritten schema.
In that paper, no result was given on the elimination of {\qunProps}; however, the fact that validation with {\qunProps} is
still in P left the possibility that {\qunStar} elimination could be performed with a polynomial size increase.



In this paper, we study how to eliminate {\qunStar} keywords by rewriting them in terms of annotation-independent keywords.
Reasons for doing this, apart from scientific curiosity, are the following:
\begin{itemize}
\item Reuse of algorithms and tools: there is a vast ecosystem of tools for {\cJS}; some of them are quite easy to re-engineer for {\mJS} (such as editors or validators),
  but for tools that decide inclusion, satisfiability, or equivalence, such as those described in \cite{DBLP:conf/issta/HabibSHP21}
   and \cite{DBLP:journals/pvldb/AttoucheBCGSS22}, no technique is known to extend them to {\mJS},
   since they depend on the ability to eliminate
   conjunction by merging instances of {\qprops}, {\qpattProps} and {\qaddProps} that are argument of a logical conjunction into just one
   instance of {\qpattProps} (\emph{and-merge} operation). No algorithm is known to do the same with {\qunProps};
   a translation from {\mJS} to {\cJS} would solve this problem;
\item Aid for comprehension: the {\qunProps} operator is regarded as complex by the same community who designed it, as testified
   by many public discussions \cite{discussion76465431}; a translation to {\cJS} simpler operators may be useful to understand the operator and
   its properties, both in general, and in some specific 
   examples;
\item Schema comparison: a translation would make it easier to compare schema versions written in {\mJS} with previous versions
  written in {\cJS}, and, specifically, to verify whether the two are equivalent or not.
\end{itemize}

The first point is crucial. While quite technical, it forms the central motivation of this work: all known algorithms for analyzing satisfiability or inclusion in {\JS} require, at some point, merging a conjunction of many instances of {\qprops}, {\qpattProps}, and {\qaddProps} into one. The same must be done with the corresponding array operators. However, the fact that {\qunProps} and {\qunIts} depend on annotations makes it impossible to extend this crucial step to the new operators. No technique, apart from elimination, has been proposed to address this challenge.



\subsection{Contributions}

In the following, we use the term \emph{Static {\mJS}} to indicate {\mJS} without dynamic references;
in this paper we focus on {Static {\mJS}} since the problem of eliminating dynamic references has already been solved in
\cite{DBLP:journals/pacmpl/AttoucheBCGSS24}.

In this paper, we provide the following contributions:
\begin{enumerate}
\item Exponential lower bound: we show that there exist schemas in Static {\mJS} that contain $\qunProps$ such that any equivalent schema
    expressed in {\cJS} has an exponentially bigger size, and we show that one occurrence of  $\qunProps$ is sufficient for that;
    we show the same result for the $\qunIts$ operator;
\item Elimination algorithm: we provide an algorithm to rewrite every 
    $\qunIts$ and $\qunProps$  in terms of annotation-independent keywords; the algorithm is based on a notion of Evaluation Normal Form (ENF) 
    that is studied here for the first time, and which allows us to avoid
    the exponential blow-up in most practical cases;

\item We crawled more than 300 schemas from GitHub that use {\qunProps} and {\qunIts}, and we applied our algorithm
  on these schemas; our results confirm that, despite its exponential complexity, the algorithm based on the ENF is fast
  on real-world schemas, and it produces a small size expansion.
\end{enumerate}

\hide{
In \cite{DBLP:journals/pacmpl/AttoucheBCGSS24} it has been proved that the elimination of dynamic references may induce an exponential increase of the input schema.
Here we show that an elimination of   $\qunProps$ and  $\qunIts$ may induce a second exponential increase, which may result in
a total increase that is in $O(2^{2^{N}})$.
While the exponential increase cannot be avoided for the separate operations, we still do not know whether the combined elimination
really requires a double exponential, or whether it is possible to combine elimination of dynamic references and elimination
of annotation-dependent operators into a single operation that can be accomplished with just
one exponential increase.
}

For all the properties that we present in the paper, a full proof is presented in the Appendix.

\section{Related work}\label{sec:relwork}

JSON Schema and its decision problems have been studied in several papers. In \cite{DBLP:conf/www/PezoaRSUV16} Pezoa et al.\ formalized the semantics of JSON Schema Version 4 \cite{Draft04}, proved that validation is P-complete and that satisfiability is EXPTIME-hard. These results have been refined and extended by Bourhis et al.\ in \cite{DBLP:conf/pods/BourhisRSV17}; in that paper,
the authors mapped {\cJS} onto a modal logic called JSL, and proved that satisfiability is EXPTIME-complete for {\cJS}  without $\quniqIts$ and in 2EXPTIME when $\quniqIts$ is added to the language.  Inclusion for {\cJS} can be reduced to satisfiability, since
$S_{1} \subseteq S_{2}$ if and only if $\qallOf{S_{1} ,\qnotOf{S_{2}}}$ is unsatisfiable;
hence these results also apply to inclusion and equivalence.

Inclusion for {\cJS} has been studied by Habib et al.\ in \cite{DBLP:conf/issta/HabibSHP21}, where an incomplete, rule-based inclusion algorithm has been introduced. In \cite{DBLP:journals/pvldb/AttoucheBCGSS22}, Attouche et al.\ tackled the same problem from a different perspective, by providing the first sound and complete witness generation algorithm for {\cJS} without $\quniqIts$; the algorithm generates a value $J$ satisfying a schema $S$ if and only if $S$ is satisfiable.

All these papers study {\cJS}. To the best of our knowledge, the only paper studying {\mJS} is \cite{DBLP:journals/pacmpl/AttoucheBCGSS24} by Attouche et al., where the authors formalized the new version of the language, and proved that validation becomes PSPACE-complete when dynamic references come into play; they also proved that $\qunProps$ and  $\qunIts$ leave validation in P, and they provided an algorithm to eliminate dynamic references, at the price of an exponential space growth, in the worst case.

\section{Syntax and semantics of {\mJS} and {\cJS}}

We refer here to {\mJS} as defined in {\DTwenty} \cite{specs2020}.
This language is formalized in \cite{DBLP:journals/pacmpl/AttoucheBCGSS24}; here we define a subset of the language
that is rich enough for our aims.

\subsection{Syntax of {\mJS} and {\cJS}}

{\json} instances $\J$ are either base values or nested arrays and objects; the order of object properties (or \emph{fields}) is irrelevant; the names of
two different fields in an object must be different, as formalized by the following grammar, where $\DNum$ is the set of all decimal
numbers (numbers that admit a finite representation in decimal notation).
\[
\begin{array}{lllllllllll}
\multicolumn{2}{l}{ s\in\Str, d\in\DNum, n\in \Nat, l_{i} \in \Str } \\[\NL]
\J ::=  & \anull \mid \atrue \mid \afalse \mid d \mid \str         \mid  \JArr{\J_1, \ldots, \J_n}  \mid   \JObj{l_1:\J_1,\ldots,l_n:\J_n } &  i\neq j \Rightarrow l_i\neq l_j
\end{array}
\]

A {\JS} schema is expressed in {\json} notation.
The keywords in a schema appear in any order, but they are evaluated in an order that
respects the dependencies among the keywords.
Following the approach of \cite{DBLP:journals/pacmpl/AttoucheBCGSS24}, we formalize this behavior by assuming that,
before validation, each schema is
reordered to respect the grammar in Figure \ref{fig:grammar}.

The grammar specifies that a schema $S$ is either a boolean schema that matches any {\json} instance ($\xtrue$) or no
instance at all ($\xfalse$), or it begins with a 
possibly empty sequence of \emph{Independent Keywords} 
(IK), followed by
a possibly empty sequence of \emph{Statically Dependent} keywords (SDK), followed by
a possibly empty sequence of \emph{Annotation Dependent} keywords (ADK).
%
Specifically, the two keywords in \key{SDK}, {\qaddProps} and {\qits},
depend on the syntactic content of three keywords in
\key{IK} ($\qprops$, $\qpattProps$, $\qprefIts$),
and the two keywords in \key{ADK} depend on the annotations returned by the keywords in \key{SDK}
and in  \key{IK}.
Observe that in {\JS} jargon, the term \emph{keyword} indicates the entire field,
that is, the pair \emph{keywordName: keywordValue}, such as ${\qtype: \key{Tp}}$,
not just the ``keyword-name''; this can be a bit confusing.
In the rest of the paper, we use just~$K$ in order to range on keywords, whenever the distinction
$\key{IK}$-$\key{SDK}$-$\key{ADK}$ is irrelevant.

In the grammar, square brackets ``['' and ``]'', and curly brackets ``\{'' and ``\}'',
are terminal symbols, while we use
$(X\ \SEP{,})^*$ to indicate a repetition of zero or more instance of the terminal $X$ each followed
by the separator  ``,'', and we use $( E )^?$ to indicate optional $E$; we only use them in the
last production of $S$ to indicate the fact that all elements are separated by commas.

{\mJS} introduced the new operators we described and also a minor syntactic reorganization of the annotation-independent array operators, which were called $\qits$ and $\qaddIts$ in {\cJS} and became
$\qprefIts$ and $\qits$ in {\mJS}. This can be quite confusing; therefore, in this paper, we will always use the {\mJS} syntax for these operators, and we will formalize {\cJS} as ``Static {\mJS} without
the \key{ADK} production''. Translating from this language to an actual dialect of {\cJS} such as
{\VerSix} just requires
a trivial translation of the annotation-independent array operators.


\begin{figure}[h!]
$$
\begin{array}{lllllll}
\multicolumn{3}{l}{
d\in \DNum, i \in \Nat, n \in \Nat, k \in \Str, b \in \Bool, 
u \in \key{URIs}, 
p \in \key{POSIX\ patterns}, J \in \semt}\\[\NL]
\key{Tp} &::=& \qobject \Mm \qnumber \Mm \qinteger \Mm \qstr  \Mm \qarray \Mm \qboolean \Mm \qnull  \\[\NL]
\key{S} &::=
& \xtrue \ \M\ \xfalse
      \M\ \JObjOpen\  ((\key{IK}\ \SEP{,})^*\ ,)^?\ 
                              ((\key{SDK}\ \SEP{,})^*\ ,)^?\ 
                               (\key{ADK}\ \SEP{,})^*
             \ \JObjClose 
            \\[\NL]
\key{IK} & ::= &
 \amin: d    \Mm
 \amax: d    \Mm
 \apatt: p    \Mm
 \aconst: J   \Mm
  \atype: \key{Tp}  \\ && \Mm
  \aany: \JArr{S_1,\ldots,S_n}   \Mm
  \aall: \JArr{S_1,\ldots,S_n}   \Mm
  \aone: \JArr{S_1,\ldots,S_n}   
   \\ && 
 \Mm
  \anot: S  
 \\ && 
 \Mm
 \apattProps: \JObj{ p_1 : S_1,\ldots,p_n : S_n }   
  \\ && 
 \Mm
 \aprops: \JObj{ k_1 : S_1,\ldots,k_n : S_n }   
 \\ && 
  \Mm \areq: \JArr{k_1,\ldots,k_n} 
  \\ &&
  \Mm \aminP : i \Mm \amaxP : i \Mm \apropN : S
   \\ & & \Mm
 \aprefIts: \JArr{S_1,\ldots,S_n}  \Mm \acont: S   \\ &&
 \Mm \aminIt : i \Mm \amaxIt : i \Mm \auniqIt : b \\ &&
  \Mm \adref: u  
 \Mm \addefs :  \JObj{k_1: S_1,\ldots,k_n:S_n}  
 \Mm \ada: \emph{plain-name}    
 \\ && 
 \Mm
 k: J  \text{\ \ (with $k$ not previously cited)} 
 \\[\NL]
$\key{SDK}\!\!$ & ::= &   
 \aaddProps: S  \Mm
 \ait: S \\ 
$\key{ADK}\!\!$  & ::= & 
 \aunProps: S    \Mm
 \aunIts: S 
\end{array}
$$
\caption{Grammar of a core subset of Static {\mJS}, Draft 2020-12.}
\label{fig:grammar}
\end{figure}

In this grammar, $\semt$ is the set of
all JSON values, and \emph{plain-name} denotes any alphanumeric string starting with a letter.
A valid schema must also satisfy two more constraints: (1)~every URI that is the argument of
$\qdref$ 
must reference a schema, and  (2)~any two adjacent keywords must have different names,
where \emph{adjacent} indicates two fields in the same object.

\subsection{Validation judgments}\label{sec:judg}

\renewcommand{\Ret}[2]{\rightarrow {#2}}
\renewcommand{\ps}{\pk}

\paragraph{Formalizing the Boolean keywords} 
The behavior of validation is specified by a judgment
$\SJudgC{J}{S}\Ret{r}{(r,\ps)}$ that specifies that the validation of an instance $J$ by a schema $S$  returns a boolean $r$, chosen between $\btrue$ and $\bfalse$, 
and an annotation $\ps$, which indicates which of the fields, or items, of~$J$ have been evaluated,
while
the judgment $\KJudgC{J}{IK}\Ret{r}{(r,\ps)}$ specifies the validation behavior of an independent keyword $IK$.
The small letter on top of $\StackDash{}$ is not a variable but a symbol used to differentiate among
schema judgments $\StackDash{S}$, keyword judgments $\StackDash{K}$, and list-of-keywords judgments $\StackDash{L}$,
introduced later on.
We will use $\SJudgC{J}{S}$ to say that $S$ validates $J$, that is, that $\SJudgC{J}{S}\Ret{}{(\btrue,\ps)}$ for some
$\ps$.

The $\StackDash{S}$ and $\StackDash{K}$ judgments are specified by mutual induction by formal rules such as the following, which
specifies that the application of a conjunction keyword $\aall:[S_1,...,S_n]$ to any instance $J$ yields the
conjunction of all the results $r_i$ obtained by applying each $S_i$ schema to $J$.
It also specifies that the set of fields/items evaluated by the $\qall$ keyword is the union of all the fields/items $\ps_i$
that are evaluated by the $S_i$ arguments of the keyword.

\infrule[\rall]
{
\forall i\in \SetTo{n}.\ \SJudgC{J}{S_i}\Ret{r_i}{(r_i,\ps_i)}
}
{\KJudgCJ{\aall:[S_1,...,S_n]}
 \Ret{r}
 (\BigAnd_{i\in\SetTo{n}}{r_i},\bigcup_{i\in\SetTo{n}}\ps_i)
}

The disjunction rule is almost identical,
but has $\Or$ instead of $\And$; observe that this rule prevents any short-circuit evaluation: Even if the first
argument $S_1$ was satisfied by $J$, all the other branches must still be evaluated, since one must collect the
annotations that they produce.

\infrule[\rany]
{
\forall i\in \SetTo{n}.\ \ \SJudgC{J}{S_i}\Ret{r_i}{(r_i,\ps_i)} 
}
{\KJudg{C}{J}{\aany:[S_1,...,S_n]}
 \Ret{r}
 (\BigOr_{i\in\SetTo{n}}{r_i},\bigcup_{i\in\SetTo{n}}\ps_i)
}

Finally, rule $(\rone)$ (see Figure \ref{fig:valrules}) specifies that, to satisfy $\qone:[S_1,\ldots,S_n]$,
$J$ must satisfy one, and only one, of  $S_1,\ldots,S_n$.
Rule $(\rnot)$ (Figure \ref{fig:valrules}) specifies that to satisfy $\qnot:S$,
$J$ must not satisfy $S$.

\paragraph{Terminal keywords}
The behavior of keywords that do not refer to any other schemas, which we call \emph{terminal keywords} and which are listed in Table \ref{tab:indk}, 
is completely defined by a type and a condition, since they do not return any annotation.

For example,
$\qreq\!:\![k_1,\ldots,k_n]$ is defined by the type $\TypeOf(\qreq)=\qobject$, by the condition
$ \rkw{cond}(J,\qreq\!:\![k_1,\ldots,k_n])=\forall i\!\in\!1..n.\ k_i\in \kw{names}(J)$, 
and by the two following rules:

\begin{multicols}{2}
\infrule[\key{kw}\TTriv]
{\qquad\ \ \ 
\TypeOf(J) \neq  \TypeOf(\key{kw})
\qquad\ \ \ 
}
{\KJudgCJ{(\key{kw}\!:\!{J'})}
 \Ret{\btrue}
 (\btrue,\ES)
}

\columnbreak

\infrule[\key{kw}]
{
\TypeOf(J) =  \TypeOf(\key{kw})
\quad
 r = \rkw{cond}(J,\key{kw}\!:\!{J'})
}
{\KJudgCJ{(\key{kw}\!:\!{J'})}
 \Ret{r}
 (r,\ES)
} 
\end{multicols}
 
The generic rules, once instantiated as indicated in Table \ref{tab:indk}, become as follows.
The first rule
says that the keyword is satisfied by any instance that is not an object; the second rule says that, when $J$ is an object,
then $J$ satisfies $\areq : \JArr{k_1,\ldots,k_n}$ iff every $k_i$ belongs to $\kw{names}(J)$, where we use
$\kw{names}(J)$ to indicate the set of the field names of $J$.

\begin{multicols}{2}
\small
\infrule[\rkw{required}\TTriv]
{
\TypeOf(J) \neq  \robject
}
{\KJudgCJ{\areq : \JArr{k_1,\ldots,k_n}}
 \Ret{\btrue}
 (\btrue,\ES)
}

\columnbreak

\infrule[\rkw{required}]
{
\TypeOf(J) =  \robject
\quad\ 
 r = \forall i\!\in\!1..n.\ k_i\in \kw{names}(J))
}
{\KJudgCJ{\areq : \JArr{k_1,\ldots,k_n}}
 \Ret{r}
 (r,\ES)
} 
\end{multicols}

Table \ref{tab:indk} provides a full specification of all terminal keywords that we consider in this paper.
In the first two lines, the notation $\TypeOf(\atype: \rkw{Tp}) = \mbox{\emph{no\ type}}$ indicates that these keywords do not have the $\key{kw}\TTriv$ rule, since they are not specific to a single type. In the $\qpatt$ line we use the notation $\rlan{p}$ to indicate the set of strings that match $p$;  in the \qkw{min*} and \qkw{max*} rules, we use the symbol $|J|$ to indicate the number of 
fields of an object $J$ and the number of elements of an array $J$.\footnote{Actually, $\fpropN:S_n$ is not really a ``terminal'' keyword,
since $S_n$ is a schema parameter. As this keyword neither generates nor passes any annotation, it is still fully defined by a type and a condition,
hence we can provide its definition in this table.}

\begin{table}
\caption{Terminal keywords.}
\label{tab:indk}

\begin{tabular}{| l | l | l |}
\hline 
\textbf{assertion \key{kw{\Col}J'}} & \textbf{\TypeOf(\key{kw})} & \textbf{\rkw{cond}($J,\key{kw}{\Col}J'$)} \\ \hline \hline
$\aconst: J_c $  & \emph{no type} &  $J = J_c$ \\ \hline
$\atype: \rkw{Tp}$ & \emph{no type} &   $\TypeOf(J)= \rkw{Tp}$ \\ \hline
\amin: d   &  \rnumber &  $J \geq$ d \\ \hline
\amax: d   &  \rnumber &  $J \leq$ d \\ \hline
\apatt: p  &  \rstr &  $J \in \rlan{p}$ \\ \hline
$\apropN : S$  & \robject &    $\forall k\in \kw{names}(J).\ \SJudgC{k}{S}$ \\ \hline
\aminP: i  & \robject &  $|J| \geq i$  \\ \hline
\amaxP: i  & \robject &  $|J| \leq i$   \\ \hline
$\areq: \JArr{k_1,\ldots,k_n}$  & \robject &  $\forall i.\ k_i\in \kw{names}(J)$   \\ \hline
\auniqIt: \xtrue  & \rarray &  $J=[J_1,\ldots,J_n]\mbox{\ with\ } n\geq 0 $  
$\And\ \forall i,j.\ 1 \leq i\neq j \leq n \Implies J_i\neq J_j  $  \\ 
\hline
\auniqIt: \xfalse  & \rarray & True  \\ 
\hline
\aminIt: i  & \rarray &  $|J| \geq i$    \\ \hline
\amaxIt: i  & \rarray &  $|J| \leq i$  \\ \hline
\end{tabular}

\end{table}

\paragraph{Non-terminal independent structural keywords} Boolean operators collect and transmit the $\ps_i$ annotations of their arguments, 
but annotations are originated by the object and array keywords.

For example, this is the rule for the $\apattProps$ keyword, which is best explained by an example.
Consider an object $J=\JObj{\qkw{age}: 1, \qkw{phone}: \anull}$ and a keyword 
$$K=\qpattProps:\JObj{\qkw{a}: S_{a}, \qkw{g}: S_{g}, \qkw{x|y}: S_{xy}}.$$
To compute $\KJudgC{J}{K}$, one first collects in $\Pi$ all pairs ($J$-field, $K$-field) where the $J$-field name matches
the $K$-field pattern:\footnote{The pattern 
$\fkw{a}$ matches any string that contains $\fkw{a}$, hence it matches $\fkw{age}$.}
$$\Pi=\Set{(\qkw{age}:1,\qkw{a}:S_{a}),(\qkw{age}:1,\qkw{g}:S_{g})}.$$
(In this paper we use $\Set{\ldots}$ to indicate sets since we use $\JObj{\ldots}$ for JSON objects.)
For each matching pair $(k_i{\Col}J_i,p_{j}{\Col}S_j)$, we check that $J_i$ is validated by $S_j$;
observe that $J$ can contain fields that are not matched by $K$ (such as $\qkw{phone}{\Col} \ \anull$), and $K$
can contain fields that do not match~$J$ (such as $\qkw{x|y}{\Col}S_{xy}$); such fields are just ignored.
The object fields \emph{evaluated} by $K$
are all and only those that appear in $\Pi$, as specified in the returned pair ${(\dots,\SetIIn{k_i}{(k_i{\Col}J_i,p_{j}{\Col}S_j)}{\Pi})}$.

\infrule[\rpattProps]
{
J = \JObj{k_1{\Col}J_1,\ldots,k_n{\Col}J_n} \andalso
\Pi = \SetST{(k_i{\Col}J_i,p_{j}{\Col}S_j)}{k_{i} \in \rlan{p_{j}}} \\[\NL]
\forall \pi=(k_i{\Col}J_i,p_{j}{\Col}S_j) \in \Pi.\ 
\SJudgC{J_{i}}{S_{j}}\Ret{r_{\pi}}{(r_{\pi},\ps_{\pi})}  
}
{
\KJudgCJ{\apattProps: \JObj{ p_1 {\Col} S_1,\ldots,p_m {\Col} S_m }} 
 \Ret{r}
 {(\BigAnd_{\pi\in\Pi}{r_\pi},\SetIIn{k_i}{(k_i{\Col}J_i,p_{j}{\Col}S_j)}{\Pi})}
}

Observe that the annotations $\ps_{\pi}$ returned by the judgment in the premise are discarded.
The reason is that each rule only returns fields/items that belong to the current instance $J$,
such as the $k_{i}$ fields that are returned by this rule, while the judgment in the premise
does not apply $S_{j}$ to the current instance $J$ but to the value of a field $k_i\Col J_{i}$ of $J$, so that $\ps_{\pi}$ 
contains fields that are not fields of $J$, but are fields of $J_i$.
This was not the case for the boolean operators, which reapply their arguments $S_i$ to the same instance~$J$.
In the {\JS} jargon, the keywords that analyze the content of the instance are called \emph{structural} keywords, while those
that  
just reapply their schema arguments to the same $J$
are called \emph{in-place} applicators. Structural keywords may originate annotations, but they do not transmit annotations
generated by others, while
in-place keywords transmit annotations but do not originate them.



The above rule can only be applied to an instance $J$ that is an object; in the other cases, the following ``trivial'' rule applies.
All keywords that refer to one specific type have their own version of this rule: they are trivially \emph{satisfied} by any instance that
belongs to any other type.

\infrule[\rpattProps\TTriv]
{
\TypeOf(J) \neq \qobject
}
{\KJudgCJ{\apattProps:J'}
 \Ret{\btrue}
 {(\btrue,\ES)}
}

It is worth repeating that this operator does not require the presence of any fields; it only constrains
the fields that are present to satisfy the associated schemas.

\hide{
The $\qpattProps$ keyword is universally quantified over the object fields: 
 it does not require the presence of any specific field, and
it is always satisfied by the empty record (we adopt the usual convention that $\And(\Set{})$ is true); for this reason, we say that
it is a \emph{constraint} keyword, similarly to the $\qmaxP:m$ keyword, that imposes an upper bound on object fields.
On the contrary, the $\qreq$ keyword, that we have exemplified in the Introduction and formalized during the discussion about terminal keywords, is existentially quantified
over the object fields. While $\qpattProps$ may be violated by adding a ``wrong'' field,  $\qreq$ may be violated by the lack of a field;
for this reason, we say that it is a \emph{requirement} keyword, similarly to the $\qminP:m$ keyword, that imposes a lower bound on object fields.
Constraints and requirements exhibit some forms of De Morgan duality, which are explored in depth in \cite{DBLP:journals/tcs/BaaziziCGSS23}.
}

We give now a rapid overview of the other non-terminal structural rules for the independent keywords
(those of the $IK$ production in the grammar);
for a complete explanation, we refer to~\cite{DBLP:journals/pacmpl/AttoucheBCGSS24}.
The rule $(\rprops)$ is almost identical to rule $(\rpattProps)$. The only difference is that we do not match an instance field $k_i:J_i$ in $J$ with the many, or zero, keyword fields $p_j:S_j$ such that $k_i\in\rlan{p_j}$, but with
the one, or zero, keyword field $k_j:S_j$ such that $k_i=k_j$.
Rule $(\rprefIts)$ specifies that $J$ satisfies $\qprefIts:[S_1,\ldots,S_m]$ if every element $J_i$ in a position $i\leq m$ satisfies $S_i$. Similarly to $\qprops$, it does not require the presence
of any item --- it is always satisfied by an empty array --- but it constrains the items that are present to satisfy
the corresponding schemas.
Rule $(\rcont)$ specifies that $J$ satisfies $\qcont:S$ if it contains at least one element that satisfies
$S$; the positions of all and only the items of $J$ that satisfy $S$ are returned in $\pk$ as ``evaluated''.
\hide{
this rule is specified by an existential quantification over the items of $J$, hence we say that it is a \emph{requirement}:
it \emph{requires} the presence of elements in the array, it always fails with an empty array.}

\paragraph{Dependent keywords}
The keyword judgment $\KJudgCJ{IK}\Ret{r}{(r,\pk)}$ specifies the behavior of a single independent keyword.
The behavior of a sequence of keywords $\List{K_1,\ldots,K_n}$, where each $K_{j}$ may depend on the
keywords $K_{i}$ with $i< j$, 
is specified by a new judgment $\KLJudgCJ{\List{K_1,\ldots,K_n}}\Ret{\rl}{(r,\pk)}$,
which, for the independent keywords, takes the conjunction of the validation results and the union of the evaluated fields;
the judgment is inductive; hence we also have a rule for the empty case.
Here, $\List{\ldots}$ indicates a list, $\List{}$ is an empty list,
$\Kl$ indicates a list of keywords, and $\Kl+K$ is $\Kl$ extended with $K$.

\begin{multicols}{2}
\infrule[\rklist-IK]
{
K\in\key{IK} \andalso
\KLJudgCJ{\Kl}\Ret{\rl}{(r_l,\pk_l)} \\[\NL]
\KJudg{C}{J}{K}\Ret{r}{(r,\pk)} 
}
{\KLJudgCJ{(\Kl\plus K)}
\Ret{\rl\plus r}{(r_l \And r,\pk_l \cup \pk)}
}

\columnbreak
\infax[\rklist-0]
{\ \ \\
\ \ \\
\KLJudgCJ{\List{}}   \Ret{\List{}}{(\btrue,\ES)}
\ \ \\
\ \ \\
\ \ \\
}

\end{multicols}

This keyword-list judgment allows us to specify the behavior of the annotation-dependent keywords, which are
$\qunProps$ and $\qunIts$ (production $ADK$ in the grammar).

The rule for $\aunProps$ is the following.

\infrule[\runProps]
{
J=\JObj{k_1 : J_1,\ldots,k_n:J_n}  \andalso
\KLJudgCJ{\Kl}\Ret{\rl}{(r,\pk)} \andalso
\Pi = \SetST{(k_i{\Col}J_i)}{1\leq i \leq n \And\ k_i \not\in \pk} \\[\NL]
\forall \pi=(k_i{\Col}J_i) \in \Pi.\ 
\SJudgC{J_{i}}{S}\Ret{r_{\pi}}{(r_{\pi},\ps_{\pi})}
}
{
\KJudgCJ{(\Kl\plus\aunProps: S)}
 \Ret{}
 {(r\And \BigAnd_{\pi\in\Pi}{r_\pi},\Set{k_{1}\ldots,k_{n}})}
}

The rule first computes the annotation $\pk$ produced by the application of all keywords in
$\Kl$ to~$J$, and the set $\Pi$ that contains all fields of $J$ that are not listed in
$\pk$. Then, it checks that these unevaluated fields satisfy $S$.
The result combines the boolean $r$ returned by the keywords~$\Kl$ with the conjunction of the results of the tests performed by 
the {\qunProps} keyword. 
The judgment returns, as evaluated fields, the complete list of all object fields ($\Ret{}{(\ldots,\Set{k_{1}\ldots,k_{n}})}$),
because the combination
$(\Kl\plus\aunProps: S)$ evaluates all fields of $J$.

As we explained in the Introduction,
the keyword $\qunProps$ has been added in {\mJS} to generalize the behavior of $\qaddProps$.
The keyword $\qaddProps$, which has been retained in {\mJS}, only excludes those fields that are evaluated by an adjacent
object keyword $\qpattProps$ or $\qprops$ --- it ignores what is evaluated by adjacent in-place applicators such as the boolean
applicator or the reference applicators. It is thus less powerful than $\qunProps$, but it can be defined and implemented in a way
that does not depend on annotation passing but relies only on a static analysis of the patterns that are listed in the adjacent object keywords.
This is formalized by the rule below, where the set $\Pi$ does not depend on the
annotations $\pk$ produced by $\Kl$, but on the patterns and keywords that can be statically extracted from the object keywords in $\Kl$, as formalized by the function $\akw{propsOf}(\Kl)$, defined below the rule.

\infrule[\raddProps]
{
J=\JObj{k_1 : J_1,\ldots,k_n:J_n}  \andalso\!\!\!
\KLJudgCJ{\Kl}\Ret{}{(r,\pk)} \andalso\!\!\!
\Pi = \SetST{(k_i{\Col}J_i)}{1\!\leq\! i \!\leq\! n \And\ k_i \not\in \rlan{\akw{propsOf}(\Kl)}} \\[\NL]
\forall \pi=(k_i{\Col}J_i) \in \Pi.\ 
\SJudgC{J_{i}}{S}\Ret{r_{\pi}}{(r_{\pi},\ps_{\pi})}
}
{
\KJudgCJ{(\Kl\plus\aaddProps: S)}
 \Ret{}
 {(r\And \BigAnd_{\pi\in\Pi}{r_\pi},\Set{k_{1}\ldots,k_{n}})}
}

$$
\begin{array}{llll}
\akw{propsOf}(\qprops: \JObj{ k_1 : S_1,\ldots,k_m : S_m }) &=& \keykey{k_1} \cat \qkw{|}\cat\ldots \cat \qkw{|}\cat \keykey{k_m} \\[\NL]
\akw{propsOf}(\qpattProps: \JObj{ p_1 : S_1,\ldots,p_m : S_m }) &=& p_1 \cat \qkw{|}\cat\ldots \cat \qkw{|}\cat p_m \\[\NL]
\akw{propsOf}(K) &=& \ES \qquad\qquad\qquad\qquad \mbox{otherwise} \\[\NL]
\multicolumn{3}{l}{
\akw{propsOf}(\List{K_1,\ldots,K_n}) 
\ \  =\  \ \akw{propsOf}(K_1) \cat \qkw{|}\cat\ldots \cat \qkw{|}\cat \akw{propsOf}(K_n) 
}
\end{array}
$$

$\akw{propsOf}(\Kl)$ returns a pattern that matches all, and only, the property names matched by a $\qprops:O$ or
$\qpattProps:O$ keyword
that belongs to $\Kl$.
In the first line of the definition of $\akw{propsOf}$, we use $\keykey{k_i}$ to indicate a pattern that only matches $k_i$; for example,
$\keykey{address}$ is the pattern $\qkw{\!\!\NN\!\!{address}\$}$ (in POSIX notation);
$\emptyset$ is a pattern that matches no string.

\hide{Differently from $\qunProps$, the classical keyword $\qaddProps$ does not create problems with the and-merge operation, since it can be rewritten using
$\qpattProps$, as detailed in \cite{DBLP:journals/pvldb/AttoucheBCGSS22}; this is the reason why we defined an algorithm to rewrite the first one using the second one.}

The rule for the array keyword ${\qunIts}:S_u$ (Figure \ref{fig:valrules}) is similar to that for {\qunProps}:
Every item that
is not evaluated, directly or indirectly, by any adjacent keyword, is evaluated by ${\qunIts}:S_u$, hence it must satisfy $S_u$;
array items are directly evaluated by $\qprefIts$, $\qcont$, $\qits$, $\qunIts$, and are indirectly evaluated by the in-place keywords,
as happens for object properties. 

The keyword ${\qits:S_i}$ was called {\qaddIts} up to 
{\VerSeven}\footnote{Actually, up to {\VerSeven}, the function of the current
 {\fits} was distributed between {\fits} and {\faddIts}, see \cite{Draft04}.}
and it corresponds to {\qaddProps}: 
the items that are evaluated by an adjacent {\qprefIts} are regarded as ``evaluated'', 
but the schema $S_i$ is applied to all other items.

As a very artificial example, consider the two schemas below, both describing arrays. 
The first one imposes that the first element is a number, the second one is a string, 
that at least one number is present,
and that every item after the string is a number: $\qunIts\!\!: \afalse$ fails on every item that is not analyzed by any of the previous keywords,
including $\qany$ and $\qcont$. For example, $[3,\qkw{a},3]$ would be validated, and $[3,\qkw{a},\qkw{a}]$ would fail.

The second schema is stricter: it imposes the presence of a number, necessarily in the first position, since no other item would be accepted: $\qits\!\!: \afalse$
only ignores items that are evaluated by an adjacent $\qprefIts$.
For example, an array $[2,\qkw{a}]$ would not be validated by the second schema since the item {\qkw{a}}, even if it satisfies the $\qany$ keyword,
is analyzed again by $\qits:\afalse$.

\begin{Verbatim}[fontsize=\small, frame=lines]
{ "type": "array",
  "prefixItems": [ { "type": "number" } ],
  "anyOf": [ { "prefixItems": [ {}, { "type": "string" } ] } ],
  "contains": { "type": "number" },
  "unevaluatedItems": false  
}
\end{Verbatim}

\begin{Verbatim}[fontsize=\small, frame=lines]
{ "type": "array",
  "prefixItems": [ { "type": "number" } ],
  "anyOf": [ { "prefixItems": [ {}, { "type": "string" } ] } ],
  "contains": { "type": "number" },
  "items": false  
}
\end{Verbatim}


This limited expressive power of  {\qits} allows it to be executed without relying on annotations, but only on a statical inspection of the adjacent {\qprefIts} keyword,
if it exists: in the rule for {\qits}, the function $\key{prefLenOf}(\Kl)$ returns the length of the prefix-array of the
only {\qprefIts} keyword in $\Kl$, or zero in case no such keyword is present in $\Kl$. Apart from this, the rule exactly corresponds to that
for $\qaddProps$.

{
\infrule[\rits]
{
J=\JArr{J_1,...,J_n} \andalso
\KLJudgCJ{\Kl}\Ret{\rl}{(r,\pk)} \\[\NL]
\Pi = \SetST{(i,J_i)}{1\leq i \leq n \And\ i\not\in\SetTo{\key{prefLenOf}(\Kl)}} \\[\NL]
\forall \pi=(i,J_i)\in \Pi.\ \SJudgC{J_{i}}{S}\Ret{r_{\pi}}{(r_{\pi},\ps_{\pi})}
}
{\KLJudgCJ{(\Kl\plus\aits : S)}
\Ret{\rl\plus r}
{(r\And \BigAnd_{\pi\in\Pi}{r_\pi},\Set{1\ldots,n})}
}
}

\hide{
SHORTER ALTERNATIVE - WHAT IS BETTER, CONSIDERING SPACE ISSUES?

For an artificial example, both schemas below describe an array that contains up to one element, that is a number. In the first schema, we may substitute {\qunIts}
with
{\qaddIts}. In the second one, if we used {\qaddIts}, it would only accept empty arrays, since it would ignore what happens inside {\qall},
hence it would regard all items as unevaluated.

{\small
\begin{verbatim}
{ "type": "array", "prefixItems": [ "type": "number" ], "unevaluatedItems": false }
\end{verbatim}
}

{\small
\begin{verbatim}
{ "allOf":  [ {"type": "array"}, {"prefixItems": [ "type": "number" ]} ],
  "unevaluatedItems": false
}
\end{verbatim}
}
}

\paragraph{Deletion of annotations}
A fundamental property of annotations is the fact that, once a schema fails, 
all of the annotations generated by its keywords are deleted.\footnote{Differently
from failing schemas, failing keywords can produce annotations (see the rule for
$\fpattProps$); this detail is irrelevant for
validation, since a failing keyword forces the surrounding schema to fail hence the annotation
is not passed, but it is very important to reduce the number of error messages in some 
specific situations;
this is the only
reason why failing keywords generate annotations.}


This behavior is reflected by the rules that define the behavior of an object schema $\JObj{\Kl}$.
Specifically, while rule ($\rkw{objSchema-T}$) passes the annotation $\pk$ generated by $\Kl$, rule
($\rkw{objSchema-F}$) returns the pair ${(\bfalse,\ES)}$: the annotation $\pk$ generated
by the failing keyword list is forgotten.

\medskip
\shortrulefirst{\rkw{objSchema-T}}
{
\KLJudgCJ{\List{K_1,\ldots,K_n}}\Ret{\List{r_1,\ldots,r_n}}{(\btrue,\pk)} 
}
{\SJudgCJ{\JObj{K_1,\ldots,K_n}}
 \Ret{r}
 {(\btrue,\pk)}
}
\shortrule{\rkw{objSchema-F}}
{
\KLJudgCJ{\List{K_1,\ldots,K_n}}\Ret{\List{r_1,\ldots,r_n}}{(\bfalse,\pk)} 
}
{\SJudgCJ{\JObj{K_1,\ldots,K_n}}
 \Ret{r}
 {(\bfalse,\ES)}
}
\medskip

Deletion of annotations implies that a schema $\JObj{\anot: S}$ never generates
annotations: if it fails, rule ($\rkw{objSchema-F}$) suppresses the annotations of $\JObj{\anot: S}$; if it does not fail,  this means that $S$ fails, and, in this case, rule ($\rkw{objSchema-F}$) suppresses the annotations of $S$.
As a consequence, in {\mJS} the equivalence $\JObj{\anot: \JObj{\anot: S}}=S$ does not hold any longer, since the first schema returns no annotation,
while $S$ may generate annotations.

%
%

\begin{example}
Consider the following schema $S$.

\begin{Verbatim}[fontsize=\small,frame=lines]
{ "not":  { "properties": { "name" : { "type": "string" } },
            "minProperties" : 2
  }
}
\end{Verbatim}

Consider now the {\json} object $J=\JObj{\qkw{name}:\qkw{P}}$. 
This object fails the ${\qminP:2}$ keyword, hence the schema $\JObj{\qprops:{\ldots},\qminP:2}$ fails, 
hence that schema produces no annotation
because of rule $\rkw{objSchema-F}$ --- the annotations of $\qprops:{\ldots}$ are lost.
Due to the failure of that inner schema, the entire schema $\JObj{\qnot: \JObj{\ldots}}$
is successful, but it does not have an annotation to transmit. 
\end{example}

\paragraph{References and definitions}
In {\JS}, the $\qdref:u$ ``reference keyword'' allows one to validate an instance using a schema identified by the URI \key{u}.

Consider, for example, the following reformulation of the example from the Introduction.

\begin{Verbatim}[fontsize=\small,frame=lines]
{ "properties": { "from": { "$ref": "#/$defs/address" } },
  "anyOf": [ { "$ref": "#infoMsgS"}, { "$ref": "#errMsgS" } ],
  "required": [ "from" ],
  "additionalProperties": false,
  "$defs": {
     "infoMsg": { "$anchor": "infoMsgS",
                  "properties": { "class": { "const": "info" }, 
                                  "payload": { "type": "string" } },
                  "required": [ "class", "payload" ] },
     "errMsg":  { "$anchor": "errMsgS",
                  "properties": { "class": { "const": "error" }, 
                                  "errorCode": { "type": "integer" } },
                  "required": [ "class", "errorCode" ] },
     "address": { "type": "string" }
  }
}
\end{Verbatim}

The keyword $\qddefs:O$, generally used in the top-level schema, is a placeholder used to collect inside $O$ 
name-schema pairs $\qkw{nn}: S_{nn}$,
that are called \emph{definitions}, and we say that $S_{nn}$ is a \emph{{\ddefined} schema}, having $\qkw{nn}$ as its name.
A  {\ddefined} schema with name \qkw{nn} can be referred using either the path \qkw{\#/\$defs/nn}, or, in case 
it contains the  $\qda: aa$ keyword, also using the shorter form \qkw{\#aa}.


The rule for $\qdref:u$ is very simple: we retrieve the schema referenced by $u$ using the function $\key{deref}(u)$,
and we use the retrieved schema to validate $J$.
The URI $u$ has, in general, the shape $\key{absURI}\catHash\key{f}$, where $\key{absURI}$ identifies a resource and
$\key{f}$ identifies a fragment inside the resource.
We do not formalize $\key{deref}(u)$, because the association between $\key{absURI}$ and the resource is
implementation dependent.
In this paper, we will only use references with shape $\#\key{f}$ (empty $\key{absURI}$).
We use the formalism for local references as introduced in~\cite{DBLP:journals/pacmpl/AttoucheBCGSS24}. For the purposes of this paper, it suffices to
stipulate that validation is always performed with respect to an implicit current {\JS} document $S$, that $\key{deref}(\qkw{\#/\$defs/nn})$
returns the {\ddefined} schema with name \qkw{nn} in the implicit current document $S$, and that $\key{deref}(\qkw{\#aa})$ returns the only
subschema of the implicit current document $S$ that has  $\qda: aa$ as a top-level keyword.

\infrule[\rdref]
{
S' = \key{deref}(u) \andalso
\SJudg{C  + \key{absURI}\ }{J}{S'}\Ret{r}{(r,\ps)}
}
{
 \KJudg{C}{J}{\adref:u}
 \Ret{r}
 (r,\ps)
}

The validation behavior of the {\qddefs} keyword is formalized by rule (\rother) (see Figure \ref{fig:valrules}), which specifies that the
keyword is ignored
during validation; the schemas that {\qddefs} collects are only used when a $\qdref$ keyword that refers them is met. The same is true
for the {\qda} keyword: it is not used to validate instances, but only to guide the behavior of the $\key{deref}$ function.

{\VerTwenty} also defines the dynamic reference operator ${\addref: \key{absURI}\catHash\key{f}}$, 
where the schema that is retrieved depends on the ``dynamic context'', that is, the sequence of URIs that have
been retrieved before dereferencing $\key{absURI}\catHash\key{f}$.
We will not discuss it here, since it does not play a role in this paper; a detailed explanation of its behavior
can be found in the earlier work~\cite{DBLP:journals/pacmpl/AttoucheBCGSS24} by Attouche et al. 

A schema is \emph{closed} when all the reference keywords it contains are local and refer to a  {{\ddefined} schema}
that is actually present in the current schema --- in this paper we will assume that all the schemas with which we work are closed.
Of course, the lower bounds that we prove in this paper are not affected by the fact that we only use local references, provided that we
consider the size of any remote schema referenced by a schema $S$ as part of the size of $S$.

\hide{
In principle, the {\qddefs} and {\qda} keywords can appear anywhere in a schema, but in practice {\qddefs} usually appears in the top-level schema,
and  {\qda} usually appears at the top-level of a {{\ddefined} schema}, as in our example.
In principle, a path fragment \qkw{/p} can point everywhere in a schema, but in practice its most common form is 
\qkw{/\$defs/nn} where \qkw{nn} does not contain any other \qkw{/}.
In this paper we will assume that every schema uses {\qddefs}, {\qda} and {\qdref} in this way, without loss of generality, since,
when a schema is not in this form, we can just copy every referred schema that is placed in a non-standard place inside the top-level
{\qddefs} section and then rewrite the references to this schema in the \qkw{\$/defs/nn} format.
In the same way, we will assume that all references that can be found inside a schema or reached from a schema are local; since, in case they were not,
the referred schema can be downloaded and copied in the {\qddefs} section of the document, and the reference can then be renamed.
Thanks to this assumption, we distinguish between top-level schemas, whose structure is
$\JObj{K_1,\ldots,K_n,\qddefs : E }$, and inner schemas, which do not contain the {\qddefs} keyword, and we
use the notation $\TL{S}{E}$ to indicate a top-level schema whose non-{\qddefs} keywords are collected in $S$ and 
where $E$ is the argument of $\qddefs $ (the \emph{environment}),
so that $\TL{\JObj{K_1,\ldots,K_n}}{E}\eqdef\JObj{K_1,\ldots,K_n,\qddefs : E }$.
}

References in {\JS} may be recursive, but (informally) every time we are able to reach a reference from itself, we must traverse
at least one structural keyword; this constraint is called ``well-formedness''
in \cite{DBLP:conf/www/PezoaRSUV16,DBLP:conf/pods/BourhisRSV17}, 
``guarded recursion'' in \cite{DBLP:journals/pvldb/AttoucheBCGSS22}, and ``no infinite recursive nesting'' in the
specifications (\cite{specs2020}, Section 9.4.1). Guarded recursion can be defined as follows.
%

\begin{definition}[Guarded recursion constraint]\label{def:guarded}
In a closed schema $S$, a subschema $S'$ immediately unguardedly depends on $S''$ if
(a) $S'$ has a top-level boolean keyword that has $S''$ among its arguments or
(b) $S'$ has a $\qdref:u$ top-level keyword and $S''=\key{deref}(u)$.
The guarded recursion constraint says that the graph defined by this relation is acyclic.
\end{definition}

The specifications (\cite{specs2020}, Section 9.4.1) require that the repeated application of the validation rules
never produce an infinite loop; this is equivalent to Definition \ref{def:guarded}, because the boolean
operators and the reference operators are the only ones that are executed ``in-place'', that is,
which do not shift the attention on a strict subterm of the analyzed instance.
This constraint allows us to define the following measure that we will use in many inductive proofs; the
guarded recursion constraint could actually be rephrased as ``for any closed schema $S$, the in-place depth
of any subschema of $S$ is well defined''.

For technical reasons, rather than giving a direct definition of the in-place depth of $\qone:[S_1,\ldots,S_n]$,
we define it in terms of the  in-place depth of its boolean encoding, as follows.

\begin{definition}[$\key{booleanOneOf}(S_1,\ldots,S_n)$]\label{def:boolOneOf}
The schema $\key{booleanOneOf}(S_1,\ldots,S_n)$ is defined as the following encoding
of $\qone$ in terms of the other boolean operators:
$$
\begin{array}{llll}
\multicolumn{3}{l}{
\key{booleanOneOf}(S_1,\ldots,S_n) =
} \\[\NL]
\multicolumn{3}{l}{
\ \ \qany\!: 
}\\
\ \ \ &[&\{\ \qall\!:[\ S_1,\JObj{\qnot\!:S_{2}},\ldots,\JObj{\qnot\!:S_{n}}\ ]\ \},\\
&& \ldots,\\
&&\{\ \qall\!:[\ \JObj{\qnot\!:S_1},\ldots,\JObj{\qnot\!:S_{i-1}},S_i,\JObj{\qnot\!:S_{i+1}},\ldots,\JObj{\qnot\!:S_{n}}\ ]\ \},\\
&& \ldots,\\
&&\{\ \qall\!:[\ \JObj{\qnot\!:S_1},\ldots,\JObj{\qnot\!:S_{n-1}},{S_{n}}\ ]\ \} \\
&]
\end{array}
$$
\end{definition}

\begin{definition}[In-place depth]\label{def:ipd}
We define the in-place depth $ID$ of any subkeyword or subschema of a schema $S$ as follows
(we assume $\max(\ES)=0$):
$$\begin{array}{llllll}
 \!\!& ID(\JObj{K_1,\ldots,K_n}) \!\!&=\!\!& \max(ID(K_1),\ldots,ID(K_n))+1 \!\!& n \geq 0\\[\NL]
 \!\!& ID(\atrue/\afalse) \!\!&=\!\!& 0 \!\!& \\[\NL]
 \!\!& ID(\qall/\qany:[S_1,\ldots,S_n]) \!\!&=\!\!& \max(ID(S_1),\ldots,ID(S_n))+1 \!\!& n \geq 0\\[\NL]
 \!\!& ID(\qnot:S) \!\!&=\!\!& ID(S)+1 \!\!& \\[\NL]
 \!\!& ID(\qone:[S_1,\ldots,S_n]) \!\!&=\!\!& ID(\key{booleanOneOf}(S_1,\ldots,S_n)) + 1 \!\!& n \geq 0\\[\NL]
 \!\!& ID(\qdref:u) \!\!&=\!\!& ID(\key{deref}(u))+1 \!\!& \\[\NL]
 \!\!& ID(K) \!\!&=\!\!& 0 \!\!& \!\!\!\!\!\!\mbox{otherwise}\\[\NL]
\end{array}
$$
\end{definition}


%

In Figure~\ref{fig:valrules},  we report the rules for the non-terminal keywords that we use in the paper; for
a formalization of the entire language,
we refer to \cite{DBLP:journals/pacmpl/AttoucheBCGSS24}.

The rules for {\cJS} are identical to those for {\mJS} apart from (1) lack of the rules for the $\qunStar$ operators and
(2) since validation in {\cJS} does not depend on annotation, the Classical judgment just returns a boolean value $r$,
as in $\SJudgCJ{S}\Ret{}{r}$, rather than a value-annotation pair $(r,\ps)$;
some examples can be found in Section~\ref{sec:neg}.


\newlength{\RD}

\newif\iftwocols
\twocolstrue                                   
\iftwocols
\newcommand{\iftwo}[1]{#1}
\newcommand{\ifone}[1]{}
\newcommand{\maybeNL}{\\[\NL]}
\setlength{\RD}{17pt}
\else
\newcommand{\iftwo}[1]{}
\newcommand{\ifone}[1]{#1}
\newcommand{\maybeNL}{\andalso}
\setlength{\RD}{10pt}
\fi

\renewcommand{\SetTo}[1]{1..{#1}}

\newcommand{\Fontsize}{\tiny}
\renewcommand{\Fontsize}{\footnotesize}

\renewcommand{\rkw}[1]{\ensuremath{\mbox{\sf{\Fontsize #1}}}}
\renewcommand{\akw}[1]{\ensuremath{\mbox{\tt{\Fontsize #1}}}}
\renewcommand{\StackDash}[1]{\mathrel{\vdash\raisebox{.90ex}{\tt\kern-0.5em {\scalebox{0.5} {#1}\kern0.1em}}}}

\begin{figure*}
\Fontsize

\begin{multicols}{2}
\infrule[\rany]
{
\forall i\in \SetTo{n}.\ \ \SJudg{C}{J}{S_i}\Ret{r_i}{(r_i,\ps_i)}
}
{\KJudg{C}{J}{\aany:[S_1,...,S_n]}
 \Ret{r}
 (\BigOr_{i\in\SetTo{n}}{r_i},{\bigcup_{i\in\SetTo{n}}\ps_{i}} )
}
\columnbreak

\infrule[\rall]
{
\forall i\in \SetTo{n}.\ \SJudg{C}{J}{S_i}\Ret{r_i}{(r_i,\ps_i)} 
}
{\KJudgCJ{\aall:[S_1,...,S_n]}
 \Ret{r}
 (\BigAnd_{i\in\SetTo{n}}{r_i},{\bigcup_{i\in\SetTo{n}}\ps_{i}} )
}
\end{multicols}
\vspace{\RD}

\begin{multicols}{2}
\infrule[\rone]
{
\forall i\in \SetTo{n}.\ \SJudgC{J}{S_i}\Ret{r_i}{\ps_i} \andalso
r = (\ |\SetST{i}{r_i=\btrue}| = 1 \ )
}
{\KJudgCJ{\aone:[S_1,...,S_n]}
 \Ret{r}
 (r,{\bigcup_{i\in\SetTo{n}}}\ps_{i} )
}
\columnbreak

\infrule[\rnot]
{
\SJudg{C}{J}{S}\Ret{r}{(r,\ps)} 
}
{\KJudgCJ{\anot:S}
 \Ret{\Not r}
 (\Not r,\ps)
}
\end{multicols}
\vspace{\RD}



\begin{multicols}{2}
\newcommand{\aU}{aURI}

\infrule[\rdref]
{
S' = \key{deref}(u) \andalso  
\SJudg{C  + \key{\aU}\ }{J}{S'}\Ret{r}{(r,\ps)}
}
{
 \KJudg{C}{J}{\adref: u} 
 \Ret{r}
 (r,\ps)
}

\columnbreak 

\infrule[\rklist-IK]
{
K\in\key{IK} \andalso
\KLJudgCJ{\Kl}\Ret{\rl}{(r_l,\pk_l)} \andalso 
\KJudg{C}{J}{K}\Ret{r}{(r,\pk)} \\[\NL]
}
{\KLJudgCJ{(\Kl\plus K)}
\Ret{\rl\plus r}{(r_l \And r,\pk_l \cup \pk)}
}
\end{multicols}
\vspace{\RD}

\begin{multicols}{2}

\infax[\rklist-0]
{\KLJudgCJ{\List{}}
 \Ret{\List{}}{(\btrue,\ES)}
}

\columnbreak

\infrule[\rother]
{k \mbox{\ \ not defined in any other rule}}
{\KJudg{C}{J}{k{\Col}J'}
  \Ret{r}{(\btrue,\ES)}
}

\end{multicols}
\vspace{\RD}

\begin{multicols}{2}
\infax[\rtrueS]
{\SJudgCJ{\atrue}
 \Ret{\btrue}
 {(\btrue,\ES)}
}

\columnbreak

\infax[\rfalseS]
{\!\!\!\!\SJudgCJ{\afalse}
 \Ret{F}
 {(\bfalse,\ES)}
}

\end{multicols}
\vspace{\RD}

\begin{multicols}{2}

\infrule[\rkw{objSchema-T}]
{
\KLJudgCJ{\List{K_1,\ldots,K_n}}\Ret{\List{r_1,\ldots,r_n}}{(\btrue,\pk)} 
}
{\SJudgCJ{\JObj{K_1,\ldots,K_n}}
 \Ret{r}
 {(\btrue,\pk)}
}

\columnbreak

\infrule[\rkw{objSchema-F}]
{
\KLJudgCJ{\List{K_1,\ldots,K_n}}\Ret{\List{r_1,\ldots,r_n}}{(\bfalse,\pk)} 
}
{\SJudgCJ{\JObj{K_1,\ldots,K_n}}
 \Ret{r}
 {(\bfalse,\ES)}
}
\end{multicols}
\vspace{\RD}



%
%
%
%

\iftwo{\begin{multicols}{2}}
\infrule[\rpattProps]
{
J = \JObj{k_1{\Col}J_1,\ldots,k_n{\Col}J_n} \maybeNL
\Pi = \SetST{(k_i{\Col}J_i,p_{j}{\Col}S_j)}{k_{i} \in \rlan{p_{j}}} \\[\NL]
\forall \pi=(k_i{\Col}J_i,p_{j}{\Col}S_j) \in \Pi.\ 
\SJudgC{J_{i}}{S_{j}}\Ret{r_{\pi}}{(r_{\pi},\ps_{\pi})}  
}
{
\KJudgCJ{\apattProps: \JObj{ p_1 {\Col} S_1,\ldots,p_m {\Col} S_m }} \iftwo{\\[\NL]}
 \Ret{r}
 {(\BigAnd_{\pi\in\Pi}{r_\pi},\SetIIn{k_i}{(k_i{\Col}J_i,p_{j}{\Col}S_j)}{\Pi})}
}
\iftwo{\columnbreak}
\ifone{\vspace{\RD}}

\infrule[\rprops]
{
J = \JObj{k'_1{\Col}J_1,\ldots,k'_n{\Col}J_n} \maybeNL
\Pi = \SetST{(k'_i{\Col}J_i,k_{j}{\Col}S_j)}{k'_{i} = k_j} \\[\NL]
\forall \pi=(k'_i{\Col}J_i,k_{j}{\Col}S_j) \in \Pi.\ 
\iftwo{\ }
\SJudgC{J_{i}}{S_{j}}\Ret{r_{\pi}}{(r_{\pi},\ps_{\pi})}
}
{
\KJudgCJ{\aprops: \JObj{ k_1 {\Col} S_1,\ldots,k_m {\Col} S_m }}  \iftwo{\\[\NL]}
 \Ret{r}
 {(\BigAnd_{\pi\in\Pi}{r_\pi},\SetIIn{k'_i}{(k'_i{\Col}J_i,k_{j}{\Col}S_j)}{\Pi})}
}
\iftwo{\end{multicols}}
\vspace{\RD}

\iftwo{\begin{multicols}{2}}
\infrule[\runProps]
{
J=\JObj{k_1 : J_1,\ldots,k_n:J_n}  \andalso
\KLJudgCJ{\Kl}\Ret{\rl}{(r,\pk)} \\[\NL] 
\Pi = \SetST{(k_i{\Col}J_i)}{1\leq i \leq n \And\ k_i \not\in \pk}  \maybeNL 
\forall \pi=(k_i{\Col}J_i) \in \Pi.\ 
\SJudgC{J_{i}}{S}\Ret{r_{\pi}}{(r_{\pi},\ps_{\pi})}
}
{
\KJudgCJ{(\Kl\plus\aunProps: S)} \iftwo{\\[\NL]}
 \Ret{}
 {(r\And \BigAnd_{\pi\in\Pi}{r_\pi},\Set{k_{1}\ldots,k_{n}})}
}
\iftwo{\columnbreak}
\ifone{\vspace{\RD}}

\infrule[\raddProps]
{
J=\JObj{k_1 : J_1,\ldots,k_n:J_n}  \andalso
\KLJudgCJ{\Kl}\Ret{}{(r,\pk)} \\[\NL] 
\Pi = \SetST{(k_i{\Col}J_i)}{1\leq i \leq n \And\ k_i \not\in \rlan{\akw{propsOf}(\Kl)}} \maybeNL 
\forall \pi=(k_i{\Col}J_i) \in \Pi.\ 
\SJudgC{J_{i}}{S}\Ret{r_{\pi}}{(r_{\pi},\ps_{\pi})} 
}
{
\KJudgCJ{(\Kl\plus\aaddProps: S)}  \iftwo{\\[\NL]}
 \Ret{}
 {(r\And \BigAnd_{\pi\in\Pi}{r_\pi},\Set{k_{1}\ldots,k_{n}})}
}
\iftwo{\end{multicols}}
\vspace{\RD}

\iftwo{\begin{multicols}{2}}
\iftwo{\renewcommand{\rcont}{\rkw{cont}}}
\infrule[\rcont]
{
J=\JArr{J_1,...,J_n} \maybeNL
\forall i\in \SetTo{n}.\ \SJudgC{J_i}{S}\Ret{r_i}{(r_i,\ps_i)} 
\ifone{\andalso} \iftwo{\quad\ \ \ }
\pk_c = \SetST{i}{r_i = \btrue}
}
{\KJudgCJ{\acont : S}
\Ret{\rl\plus r}{(|\pk_c| \!\geq\! 1,\pk_c)}
}
\iftwo{\columnbreak}
\ifone{\vspace{\RD}}

\infrule[\rprefIts]
{
J = \JArr{J_1,\ldots,J_m} \maybeNL
\forall i\in \SetTo{\Min(n,m)}.\  \SJudgC{J_i}{S_i}\Ret{r_i}{(r_i,\ps_i)} 
}
{\KJudgCJ{\aprefIts: \JArr{S_1,\ldots,S_n}} \iftwo{\\[\NL]}
 \Ret{r}
 {(\BigAnd_{i\in\SetTo{\Min(n,m)}}{r_i},\Set{1,\ldots,\Min(n,m)})}
}
\iftwo{\end{multicols}}
\vspace{\RD}

\hide
{
$
\begin{array}{lllllllll}
\mbox{if}\ \ (\qprefIts: \JArr{ S_1,\ldots, S_m }) \in \Kl \ \ \mbox{for some}\ \  S_1,\ldots,S_n
& \mbox{then:} &
\akw{maxPrefixOf}({\Kl}) =   m 
& \mbox{otherwise:} & \akw{maxPrefixOf}({\Kl}) =  0
\end{array}
$
}

\iftwo{\begin{multicols}{2}}
\iftwo{\renewcommand{\runIts}{\rkw{unItems}}}
\infrule[\runIts]
{
J=\JArr{J_1,...,J_n} \andalso
\KLJudgCJ{\Kl}\Ret{\rl}{(r,\pk)} \maybeNL
\Pi = \SetST{(i,J_i)}{1\leq i \leq n \And\ i\not\in\pk} \\[\NL]
\forall \pi=(i,J_i)\in \Pi.\ \SJudgC{J_{i}}{S}\Ret{r_{\pi}}{(r_{\pi},\ps_{\pi})} 
}
{\KLJudgCJ{(\Kl\plus\aunIts : S)}\iftwo{\\[\NL]}
\Ret{\rl\plus r}
{(r \And \BigAnd_{\pi\in\Pi}{r_\pi},\Set{1\ldots,n})}
}
\iftwo{\columnbreak}
\ifone{\vspace{\RD}}

\infrule[\rits]
{
J=\JArr{J_1,...,J_n} \andalso
\KLJudgCJ{\Kl}\Ret{\rl}{(r,\pk)} \maybeNL
\Pi = \SetST{(i,J_i)}{1\leq i \leq n \And\ i\not\in\SetTo{\key{prefLenOf}(\Kl)}} \\[\NL]
\forall \pi=(i,J_i)\in \Pi.\ \SJudgC{J_{i}}{S}\Ret{r_{\pi}}{(r_{\pi},\ps_{\pi})}
}
{\KLJudgCJ{(\Kl\plus\aits : S)}\iftwo{\\[\NL]}
\Ret{\rl\plus r}
{(r \And \BigAnd_{\pi\in\Pi}{r_\pi},\Set{1\ldots,n})}
}
\iftwo{\end{multicols}}
\vspace{\RD}


\caption{Validation rules.}
\label{fig:valrules}
\end{figure*}

\renewcommand{\rkw}[1]{\ensuremath{\mbox{\sf{\small #1}}}}
\renewcommand{\akw}[1]{\ensuremath{\mbox{\tt{\small #1}}}}
\renewcommand{\StackDash}[1]{\mathrel{\vdash\raisebox{.90ex}{\tt\kern-0.5em {\tiny {#1}\kern0.5em}}}}

\subsection{Negation-completed {\cJS}}\label{sec:neg}

In \cite{DBLP:journals/tcs/BaaziziCGSS23}, the authors proved that negation can be eliminated from {\cJS}, if the
language is enriched by keywords:
${\qpattReq}: \JObj{ p : S}$, ${\qnotMof}: d$, 
${\qnotPatt : p}$, ${\qrepIts : b}$, $\qcontAft : S$.

These operators are defined by the following table and rules.
The rules specify that, if $J$ is an object, then ${\qpattReq}: \JObj{ p : S}$ requires the presence of at least one
field whose name matches $p$ and whose value satisfies $S$. 
The keyword $\qcontAft : S$,  if $J$ is an array, requires the presence of at least one
item that satisfies $S$ and whose position strictly follows the last position that is described by an adjacent 
$\qprefIts$;  $\qcontAft : S$ requires an item in the same range of positions that are constrained by $\qits: S$.
As happens for all the typed keywords,
any instance that is \emph{not} an object (respectively, an array) is validated by {\qpattReq} (respectively, by {\qcontAft}).

\medskip
\begin{tabular}{| l | l | l |}
\hline 
\textbf{assertion \key{kw{\Col}J'}} & \textbf{\TypeOf(\key{kw})} & \textbf{\rkw{cond}($J,\key{kw}{\Col}J'$)} \\ \hline \hline
\anotMof: d   &  \rnumber &  $\forall i\in \Int.\ J\neq (i\times d)$ \\ \hline
\anotPatt: p  &  \rstr &  $J \not\in \rlan{p}$ \\ \hline
\xrepIts: \xtrue  & \rarray &  $J=[J_1,\ldots,J_n] \And\ \exists i,j.\ i\neq j \And J_i= J_j  $  \\ \hline
\xrepIts: \xfalse  & \rarray & True  \\ \hline
\end{tabular}
\medskip

{
\newcommand{\SIZE}{\small}
\SIZE
\renewcommand{\rkw}[1]{\ensuremath{\mbox{\sf{\SIZE #1}}}}
\renewcommand{\akw}[1]{\ensuremath{\mbox{\tt{\SIZE #1}}}}
\renewcommand{\StackDash}[1]{\mathrel{\vdash\raisebox{.90ex}{\tt\kern-0.5em {\scalebox{0.5} {#1}\kern0.1em}}}}

\infrule[\rkw{patternRequired}]
{
J=\JObj{k_1:J_1,...,k_n: J_n}  \andalso
\forall k_i\in\rlan{p}.\ \SJudgC{J_i}{S}\Ret{r_i}{r_i} \\[\NL]
r = ( |\SetST{i}{k_i\in\rlan{p} \And r_i = \btrue}| \geq 1 )
}
{\KJudgCJ{\apattReq : \JObj{p:S}}
\Ret{r}{r}
}

%
%

\infrule[\rcontAft]
{
J=\JArr{J_1,...,J_n} \andalso
\KLJudgCJ{\Kl}\Ret{\rl}{r} \\[\NL]
\forall i\in \SetFromTo{\key{prefLenOf}(\Kl)+1}{n}.\ \SJudgC{J_i}{S}\Ret{r_i}{r_i} \andalso
r' = (|\SetST{i}{r_i = \btrue}| \geq 1)
}
{\KLJudgCJ{(\Kl\plus\acontAft : S)}
\Ret{\rl\plus r}
{r \And r'}
}

\renewcommand{\rkw}[1]{\ensuremath{\mbox{\sf{\small #1}}}}
\renewcommand{\akw}[1]{\ensuremath{\mbox{\tt{\small #1}}}}
\renewcommand{\StackDash}[1]{\mathrel{\vdash\raisebox{.90ex}{\tt\kern-0.5em {\tiny {#1}\kern0.5em}}}}
}


The resulting language is called Negation-Completed {\cJS}, has the same expressive power as
{\cJS}, but it enjoys ``negation elimination'': 
every schema in {\cJS} can be
rewritten into a schema in Negation-Completed {\cJS} whose size is polynomial 
in that of the original schema, and which is
``positive'', that is, it does not
contain neither negation nor $\qone$.
In the paper, we will use Positive Negation-Completed {\cJS} in the proof of the exponential blow-up for ${\qunProps}$
elimination (Section~\ref{sec:elimproof}), and in the corresponding proof for ${\qunIts}$.


\section{Elimination of $\qunProps$: The problem}\label{sec:problem}


In this paper, we prove that for any fixed schema, every instance of $\qunProps$ can be transformed into $\qaddProps$.
However, this process is not trivial.

In {\cJS}, most keywords distribute with respect to $\qany$ and $\qall$, that is, for any keyword $K\in IK$
we have\footnote{In case $K$ has the same name as a keyword in $\Kl_i$, instead of
$\JObj{\Kl_i,K}$ we must write $\JObj{\fall:[\JObj{\Kl_i},\JObj{K}]}$.}
$$
\begin{array}{lllll}
\JObj{\qany:[\JObj{\Kl_1},\ldots,\JObj{\Kl_n}],K} &\sim& \JObj{\qany:[\JObj{\Kl_1,K},\ldots,\JObj{\Kl_n,K}]} \\[\NL]
\JObj{\qall:[\JObj{\Kl_1},\ldots,\JObj{\Kl_n}],K} &\sim& \JObj{\qall:[\JObj{\Kl_1,K},\ldots,\JObj{\Kl_n,K}]} \\[\NL]
\end{array}
$$
where we use $S_1\sim S_2$ to indicate that schema $S_1$ is equivalent to schema $S_2$, meaning that they validate
the same instances.

Hence, the most natural way to eliminate $\qunProps$ would be to distribute it through the logical operators using this rule
until it reaches the leaves of a schema, where it could be rewritten as $\qaddProps$.

Consider, for example, the following schema: 
%
\begin{Verbatim}[fontsize=\small, frame=lines]
{ "anyOf":  [ { "$ref": "#sale" } , { "$ref": "#car" } ], 
  "unevaluatedProperties": false,
  "$defs": {
    "sale": { "$anchor": "sale", "properties": { "price": { "type": "integer" }}},
    "car":  { "$anchor": "car",  "properties": { "plate": { "type": "string" }}}
  }
}
\end{Verbatim}

The above schema
describes \emph{sales} with a \emph{price} and
\emph{cars} with a \emph{plate}, and considers their disjunction enriched with $\qunProps: \afalse$.
Unlike the example in the Introduction,  \emph{sales} and  \emph{cars} are not disjoint,
and this detail is crucial.

If we distribute $\qunProps$ through the disjunction, we obtain the following schema (we omit the $\qddefs$
keyword):
\begin{Verbatim}[fontsize=\small, frame=lines]
{ "anyOf":  [ { "$ref": "#sale", "unevaluatedProperties": false }, 
              { "$ref": "#car",  "unevaluatedProperties": false } ]
}
\end{Verbatim}

This new schema is more restrictive than the original one.
Consider the following instance:
$\JObj{ \qkw{price}{\Col}100, \qkw{plate}{\Col}\qkw{x111} }$.
It would be accepted by the original schema: it satisfies both \qkw{\#car} and
\qkw{\#sale}; its first property is evaluated by the first child of $\qany$, the second property by the second 
child, hence, in the original schema, it has no unevaluated property, 
and $\qunProps{\Col}\,\xfalse$ succeeds.
In the rewritten schema, the instance property $\qkw{plate}$ causes the $\qunProps{\Col}\,\xfalse$ in the
first $\qany$ branch to fail, since it is not evaluated 
by the keyword $\qdref{\Col}\qkw{\#sale}$, and the instance  property
$\qkw{price}$ causes the second branch to fail, since it is not evaluated 
by the keyword $\qdref{\Col}\qkw{\#car}$.
Hence, validation of the instance 
$\JObj{ \qkw{price}{\Col}100, \qkw{plate}{\Col}\qkw{x111} }$ fails.

%

To distribute $\qunProps$ through this $\qany$, we must distinguish three cases:
when only $\JObj{\qdref{\Col}\qkw{\#sale}}$ 
is satisfied, hence only $\qkw{price}$ is evaluated (if it is present);
 when only $\JObj{\qdref{\Col}\qkw{\#car}}$ 
is satisfied, hence only $\qkw{plate}$ is evaluated;
or when both are satisfied, hence both
$\qkw{price}$ and $\qkw{plate}$
are evaluated.
The resulting schema is as follows:
\begin{Verbatim}[fontsize=\small,frame=lines]
{ "anyOf":  [ { "$ref": "#sale", "not": {"$ref": "#car"}, 
                "unevaluatedProperties": false },
              { "$ref": "#car", "not": {"$ref": "#sale"}, 
                "unevaluatedProperties": false },
              { "allOf": [{"$ref": "#sale"}, {"$ref": "#car"}], 
                "unevaluatedProperties": false } ]  }
\end{Verbatim}

Now, we know exactly which properties are evaluated if present in each of the three cases, and $\qunProps$ can be safely rewritten as $\qaddProps$ as follows.
\begin{Verbatim}[fontsize=\small,frame=lines]
{ "anyOf":  [ { "$ref": "#sale", "not": {"$ref": "#car"}, 
                "properties": {"price":{}}, 
                "additionalProperties": false }, 
              { "$ref": "#car", "not": {"$ref": "#sale"}, 
                "properties": {"plate":{}}, 
                "additionalProperties": false },
              { "allOf": [{"$ref": "#sale"},  {"$ref": "#car"}], 
                "properties": {"plate":{}, "price":{}}, 
                "additionalProperties": false }  ] }
\end{Verbatim}

The operator $\qany$ is difficult to deal with since it may be satisfied in many different ways --- if it has $n$ different operands, it may be satisfied
by the satisfaction of any non-empty subset of them, but each different subset may correspond to a different set of evaluated properties,
and we will show with Corollary \ref{cor:expobj} that this problem cannot be avoided.

Although we will see later that $\qone$ and $\qall$ are somehow more tractable than $\qany$, neither of them enjoys distributivity.

Consider the operator $\qone$, and the following schema.
\begin{Verbatim}[fontsize=\small,frame=lines]
{ "oneOf":  [ { "$ref": "#sale" } , { "$ref": "#car" } ],
  "unevaluatedProperties": false 
}
\end{Verbatim}

If we distribute $\qunProps$ through the operator, we obtain the following schema.
\begin{Verbatim}[fontsize=\small,frame=lines]
{ "oneOf":  [ { "$ref": "#sale", "unevaluatedProperties": false }, 
              { "$ref": "#car",  "unevaluatedProperties": false } ] 
}
\end{Verbatim}

Again, the two schemas are not equivalent: this time, the rewritten schema is less restrictive.
The object $\JObj{ \qkw{price}:100 }$ would not be accepted by the original $\qone$ schemas: It satisfies both $\JObj{\qdref:\qkw{\#sale}}$ 
and $\JObj{\qdref:\qkw{\#car}}$, since no field is mandatory in the two schemas.\footnote{Actually, the original schema accepts no instance at all; this example is not meant to be
realistic, but only to illustrate failure of distributivity.}
However, this object would be accepted by the rewritten schema, since it satisfies the first branch, but it violates the
second one.
%

Consider now $\qall$, and the following schema.
\begin{Verbatim}[fontsize=\small,frame=lines]
{ "allOf":  [ { "$ref": "#sale" } , { "$ref": "#car" } ], 
  "unevaluatedProperties": false 
}
\end{Verbatim}
%

If we distribute $\qunProps$ through the conjunction, we obtain the following schema.
\begin{Verbatim}[fontsize=\small,frame=lines]
{ "allOf":  [ { "$ref": "#sale", "unevaluatedProperties": false }, 
              { "$ref": "#car",  "unevaluatedProperties": false } ] 
}
\end{Verbatim}

This new schema is more restrictive than the original one: for example, as happens in the
{\qany} case, both of its branches would refuse the
instance
$\JObj{ \qkw{price}:100, \qkw{plate}:\qkw{x111} }$,
which would instead be accepted by the original schema: It is both a sale and a car, and
each of its two fields is evaluated by the conjunction. Hence, the $\qunProps$ of the original schema would not create
problems.

These examples show that the elimination of $\qunProps$ is far from immediate.
In the next section, we give a formal proof of this fact.

\begin{remark}
Unlike the other boolean operators, rewriting $\qunProps$ in the presence of negation is very easy.
Consider the following schema.
%
\begin{Verbatim}[fontsize=\small,frame=lines]
{ "not": { "$ref": "#sale" } , "unevaluatedProperties": { "type": "string" } }
\end{Verbatim}

As we have discussed, the successful execution of $\qnot: S$ corresponds to a failure of~$S$, and~$S$ will not generate any annotation,
so $\qnot: S$ will not generate any annotation. Hence, $\qunProps: S$ can be rewritten as $\qaddProps:S$, since they are both applied to
every field of the object. Hence, the above schema can be rewritten as follows.
\begin{Verbatim}[fontsize=\small,frame=lines]
{ "not": { "$ref": "#sale" } , "additionalProperties": { "type": "string" } }
\end{Verbatim}
\end{remark}

\section{Elimination of $\qunProps$ and $\qunIts$ requires an exponential blow-up}\label{sec:blowup}

\subsection{Introduction}

In the previous section, we have seen that the elimination of $\qunProps$ is not a trivial task.
In this section, we prove that there exist families of schemas written in Static {\mJS} for which
the elimination can only be achieved through an exponential blow-up of the schema size, 
by exhibiting a family of schemas $S_n$ such that, for each schema $S_n$, the smallest corresponding schema written without using
$\qunProps$ has a size in $O(2^{|S_n|})$.
We then prove that the same property holds for $\qunIts$, even if we add the new operators
$\qminC$ and $\qmaxC$ to the target language.

\subsection{Exponentially of elimination of $\qunProps$}\label{sec:elimproof}

Consider the family of schemas where $S_n$ is defined as follows
\begin{Verbatim}[fontsize=\small,frame=lines]
{ "anyOf": [ { "required": [ "a1" ], "patternProperties": { "a1": true } },
             { "required": [ "a2" ], "patternProperties": { "a2": true } },
              ...
             { "required": [ "an" ], "patternProperties": { "an": true } }],
  "unevaluatedProperties": false
}
\end{Verbatim}

Recall that a pattern $\qkw{a1}$ matches any strings that includes $\qkw{a1}$, such as
$\qkw{a1}$ itself, or $\qkw{Ba1Ca2K}$.
The following property characterizes the instances that satisfy $S_n$.
\begin{propxrep}\label{pro:sn}
An instance $J$ satisfies $S_n$ if, and only if, it is not an object, or it is an object and:
\begin{compactenum}
\item $J$ contains at least one property whose name is exactly $\qkw{a}\cat i$, for some $i\leq n$;
\item for any property of $J$, the property name  matches one or more patterns $\qkw{a}\cat i$, with $i\leq n$;
\item for each property $k:J_k$ of $J$ whose name $k$ matches a set of $\qkw{a}\cat i$ patterns, for at least one of the $\qkw{a}\cat i$
     matched by $k$,
       there is a property of $J$ whose name is exactly $\qkw{a}\cat i$.
\end{compactenum}
\end{propxrep}

To understand this characterization, let us observe that the $i$-th branch of the $\qany$
evaluates a field $k$ iff the branch is successful, i.e.\ iff $J$ has a field named exactly $\qkw{a}\cat i$, and if $k$
matches $\qkw{a}\cat i$.
Hence, condition~(1) expresses the fact that $\qany$ needs at least one branch to
succeed. Conditions~(2) and~(3) express the fact that any field $k$ must be evaluated, otherwise the $\qunProps$ keyword
will fail, and a field is evaluated by the $i$-th branch iff conditions~(2) and~(3) hold for that field.

For example, $\JObj{\qkw{a1}: \xnull, \qkw{-a1-a3-}: \xnull}$ satisfies the schema, since
$\qreq:\qkw{a1}$ is satisfied, hence the first branch of $\qany$ evaluates the $\qkw{-a1-a3-}$ property; on the contrary,
 $\JObj{\qkw{a2}: \xnull, \qkw{-a1-a3-}: \xnull}$ does not satisfy the schema, since
no successful branch evaluates the property \qkw{-a1-a3-}.

Before proving the exponentiality of {\qunProps} elimination for this family of schemas, we introduce a definition and a lemma.
Before stating the definition, we recall that we use the term \emph{in-place keywords} for
${\qall:A}$,${\qany:A}$, ${\qone:A}$, ${\qnot:S}$, and ${\qdref:u}$, and \emph{structural keywords}
for any other keyword.

\begin{definition}\label{def:skof}
For any schema $S$ and for any instance $J$, we use $\SOf{J}$ to denote
all subschemas of $S$ that are satisfied by $J$,
$\KOf{J}$ to denote
all subkeywords of $S$ that are satisfied by $J$,
and $\SKOf{J}$ to denote the set of all structural subkeywords of $S$ that are satisfied by $J$, so that $\SKOf{J}\subseteq\KOf{J}$.
\end{definition}

\begin{lemma}[Monotonicity]\label{lem:monotonicity}
For any positive schema $S$ in Negation-Completed {\cJS} and for any pair of instances $J_1$ and $J_2$, 
$\SKOf{J_1}\subseteq \SKOf{J_2}$ implies that $\KOf{J_1}\subseteq \KOf{J_2}$
and that $\SOf{J_1}\subseteq \SOf{J_2}$.
\end{lemma}

\begin{appendixproof}
We want to prove that (1) $\SKOf{J_1}\subseteq \SKOf{J_2}$ and $\SJudgC{J_1}{S'}$ implies
$\SJudgC{J_2}{S'}$ and (2)  $\SKOf{J_1}\subseteq \SKOf{J_2}$ and $\KJudgC{J_1}{K}$ implies
$\KJudgC{J_2}{K}$ for all subschemas $S'$ and subkeywords $K$ of $S$, by mutual induction on the in-place depth
(Definition \ref{def:ipd}).
(1) is immediate for  $S=\atrue/\afalse$, and is proved by induction on  for $S=\JObj{K_1,\ldots,K_n}$.
(2)  is proved by induction when $K$ is boolean or is $\qdref:u$, and follows from 
$\SKOf{J_1}\subseteq \SKOf{J_2}$ otherwise.
\end{appendixproof}

\begin{corollary}\label{cor:inplace}
For any positive schema $S$ in Negation-Completed {\cJS} and for any pair of instances $J_1$ and $J_2$, 
if $J_1$ satisfies $S$, $J_2$ does not satisfy $S$, and 
$\SKOf{J_2}\supseteq (\SKOf{J_1} \setminus \Cal{A}^{-})$, then 
$\Cal{A}^{-}$ is not empty.
\end{corollary}

We now use Corollary \ref{cor:inplace} to prove the exponentiality result.

\begin{theoxrep}\label{the:expobj}
For any schema $S_n$ of the above family, 
any equivalent schema $S$ that is expressed using Positive Negation-Completed {\cJS}
as defined in Section~\ref{sec:neg}, has a dimension
that is bigger than $2^n$.
\end{theoxrep}

\begin{proofsketch}
\renewcommand{\InSketch}[1]{}
\renewcommand{\InAppendix}[1]{#1}

Consider any $n$ and any schema $S$, written in Positive Negation-Completed {\cJS},
such that $S\sim S_{n}$.
\InSketch{In this sketch we ignore the $\qminP$, $\qmaxP$, and $\qconst$ keywords, which are described in the
full proof.}

Consider all the subkeywords of $S$ that are either $\qprops:O$,
$\qpattProps:O$,
$\qaddProps:S$, 
or $\qpropN:S$ --- we call them the \emph{prop-keywords}.
In our proof, we will only use objects where every property has a $\anull$ value.
For this reason we define, for each prop-keyword $\pK$, its \emph{accepted language} $\rlan{\pK}$ as 
the set of  property names $k$ such that an object $\JObj{k:\anull}$ satisfies that prop-keyword.
Since all the prop-keywords validate an object if and only if they validate all of its fields regarded one at a time,
an object $J$ where the value of each property is $\anull$ is validated by a prop-keyword $\pK$ if,
and only if, all names of all properties of $J$ belong to the \emph{accepted language} of $\pK$.

 
Of course, the accepted language of a keyword $\qaddProps: S$ depends on the adjacent 
keywords.
For example, consider the following schema with two prop-keywords:
$$
\begin{array}{llll}
\JObj{\pK_1, \pK_2} = \JObj{\qpattProps:\JObj{\qkw{a1}{\Col}\{\}, \qkw{\PP{a2}}{\Col}\{\}}, \qaddProps:\afalse}
\end{array}
$$
Since ${\pK_1}$ never fails, its accepted language is $\Sigma^*$: $\rlan{\pK_1}=\Sigma^*$.
$\pK_2$ \emph{fails} over $\JObj{k:\anull}$ if, and only if, $k$ is not evaluated by $\pK_1$, that is,
using the $\akw{propsOf}$ function defined in the rules, if $k\not\in\rlan{\akw{propsOf}(\pK_1)}$,
that is, iff $k\not\in\rlan{\qkw{a1|\PP{a2}}}$, hence the \emph{accepted} language of $\pK_2$ is $\rlan{\qkw{a1|\PP{a2}}}$.


For each prop-keyword $\pK$ we say that its \emph{pattern signature} is the set of all
patterns $\qkw{a}\cat i$ in $\Set{\qkw{a1},\ldots,\qkw{an}}$ such that 
$\rlan{\qkw{a}\cat i}\subseteq \rlan{\pK}$;
for example, if $\rlan{\pK}=\rlan{\qkw{a1|\PP{a2}}}$;
then its \emph{pattern signature} is $\Set{\qkw{a1}}$; the string $\qkw{a2}$
is not in the signature, since $\rlan{\qkw{a2}}$ is not completely contained in $\rlan{\qkw{a1|\PP{a2}}}$.

\InAppendix{Finally, we collect all subkeywords of $S$ with shape $\qmaxP : m$ or $\qconst: J_o$,
and we use $\key{BIG}$ to denote a number that is greater of each $m$ parameter and of the length $|J_o|$
of each $J_o$ parameter.}

We now show that, for any non-empty subset $A$ of $\Set{\qkw{a}\cat 1,\ldots,\qkw{a}\cat n}$,
there exists one prop-keyword inside $S$ that has exactly $A$ as its pattern signature.
Fixed $A=\Set{\qkw{a}\cat{i_1},\ldots,\qkw{a}\cat{i_l}}$,
we consider an object $J$ that contains the properties $\Set{\qkw{a}\cat{i_1},\ldots,\qkw{a}\cat{i_l}}$
(with a null value);
then,
for each prop-keyword $\pK$ such that there exists $\qkw{a}\cat{i_j}$ in $A$ that is not in the 
pattern signature of~$\pK$, we add a null property $\key{PNot}(j,\pK)$ whose name belongs to
$\rlan{\qkw{a}\cat{i_j}}\setminus \rlan{\pK}$\InSketch{.}\InAppendix{;
finally, we add a set of $\key{BIG}$ extra null properties $\key{PExtra}(1)$,\ldots, $\key{PExtra}(\key{BIG})$, where the name of 
$\key{PExtra}(q)$ is obtained by concatenating $q+1$ copies of $\qkw{a}\cat{i_1}$.}

By construction, the object $J$ belongs to $S_n$: Consider Property \ref{pro:sn};
(1): 
the object $J$ contains one property $\qkw{a}\cat i$
(since $A$ is not empty); (2 and 3):
every property with shape  $\key{PNot}(j,\pK)$ matches the property $\qkw{a}\cat{i_j}$
that is present in the object, and every property $\key{PExtra}(q)$  matches $\qkw{a}\cat{i_1}$
that is present in the object.
We consider now the set $\SKOf{J}$ of all structural
keywords inside $S$ (prop-keywords or other structural keywords) that are satisfied by this $J$
to show that some of them are prop-keywords whose pattern signature is exactly $A$.

Observe that the pattern signature of each prop-keyword in 
$\SKOf{J}$ is a superset
of $A$: if the  pattern signature of a prop-keyword $\pK'$ is not a superset of $A$,
then the object $J$ contains, by construction, a property $\key{PNot}(j,\pK')$ that is not in the accepted language of $\pK'$,
hence such a keyword $\pK'$ is not satisfied by $J$, hence this $\pK'$ cannot be in $\SKOf{J}$;
we use (*) to refer to this fact later on.

We consider now the set $\Cal{PK}^{A}$ of those prop-keywords that belong to $\SKOf{J}$ and 
whose pattern signature is exactly $A$, with the aim of proving that $\Cal{PK}^{A}$ it is not empty.
To this aim, we build a variant $J'$ of $J$ that satisfies all keywords in $\SKOf{J}\setminus\Cal{PK}^{A}$ but 
does not satisfy $S$, as follows.

%

We consider an enumeration $\Set{\qkw{a}\cat {c_1},\ldots,\qkw{a}\cat {c_o}}$ of
 the complement of $A$, defined as $Co(A)=\Set{\qkw{a}\cat 1,\ldots,\qkw{a}\cat n}\setminus A$.
We build $J'$ from  $J$ by adding 
a 
property with name $k^+=\qkw{a}\cat {c_1}\cat\ldots\cat\qkw{a}\cat {c_o}\cat\qkw{\_}$.
The object $J'$ violates either condition (2) or condition (3) of Property \ref{pro:sn}:
if $o=0$, then $K^+=\qkw{\_}$, which violates condition (2); if $o\geq 1$, then $k^+$ matches 
the patterns of $Co(A)$, but it does not match any of the patterns in $A$.
In other terms, the field $k^+$ will fail the $\qunProps:\afalse$ keyword,
hence $J'$ does not satisfy $S_n$, hence it does not satisfy $S$.

We prove now that $J'$ satisfies all keywords $\pK$ in the set $\SKOf{J}\setminus\Cal{PK}^{A}$.
If $\pK$ is a prop-keyword, then, by (*), its pattern signature is a superset of $A$; since $\pK\not\in\Cal{PK}^{A}$,
the inclusion is strict, hence its pattern signature includes at least one pattern $\qkw{a}\lcat c_l$ of $Co(A)$; by construction,
the new property of $J'$ matches all patterns in $Co(A)$, hence it matches $\qkw{a}\lcat c_l$,
hence the new property is in $\rlan{\pK}$, hence it is accepted by $\pK$; 
since all other properties of $J'$ belong to $J$ hence are accepted by $\pK$,
then $J'$ satisfies the prop-keyword $\pK$. For the other structural keywords in $\SKOf{J}\setminus\Cal{PK}^{A}$, we reason as follows.
\InAppendix{$J$ has strictly more than $\key{BIG}$ properties, hence no keyword in $\SKOf{J}$ is a 
$\qmaxP:m$ keyword or a $\qconst: J_o$ keyword.
Since $J'$ has more properties than $J$, then it satisfies all the $\qminP:n$ keywords in $\SKOf{J}$.}
Since we did not remove any property from $J$, then $J'$ satisfies all $\qreq:A$ and $\qpattReq:O$ that are
in $\SKOf{J}$.
No other structural keyword is able to distinguish one object from another, hence all other
keywords in $\SKOf{J}\setminus\Cal{PK}^{A}$ are satisfied by $J'$.

Since $S$ is a positive schema, since $J$ satisfies $S$ and $J'$ does not, 
since $J'$ satisfies all keywords in  $\SKOf{J}\setminus\Cal{PK}^{A}$, 
we can conclude by Corollary \ref{cor:inplace}  that $\Cal{PK}^{A}$ is not empty, that is, that there
exists at least one keyword whose pattern signature is exactly $A$.
Hence, we have proved that for every non-empty subset $A$ of $\Set{\qkw{a1},\ldots,\qkw{an}}$ there is one 
subkeyword of $S$ that has $A$ as its pattern signature.
Since $\Set{\qkw{a1},\ldots,\qkw{an}}$ has $2^n-1$ different non-empty subsets, this implies that $S$ has at least 
$2^n-1$ subkeywords.

\end{proofsketch}

We have proved that $S_n$ needs at least $2^n-1$ keywords in order to be represented by a positive
Negation-Completed {\cJS} schema.
The exponentiality result follows for general {\cJS} since it is has been proved in \cite{DBLP:journals/tcs/BaaziziCGSS23} that
negation can be eliminated from {\cJS} with a polynomial expansion.

\begin{corollary}\label{cor:expobj}
There exists a family of Static {\mJS} schemas $S_n$  whose only {\mJS} operator is
one instance of $\qunProps$,
such that, for each family of schemas $S'_n$ expressed in {\cJS}, if each $S'_i$ is equivalent
to $S_i$, then $S'_n$ has a size in $\Omega(2^n)$.
\end{corollary}

\subsection{Exponentiality of elimination of $\qunIts$}

\newcommand{\San}{S^{\!{\mathcal  A}}_{\!n}}

We now present a family ${\San}$ of array schemas that has the same property of the $S_n$ family of object schemas defined
in the previous section.

As a technical tool, we also define, for each $n$, the family
$\Cal{T}_n=\Set{\qkw{T}\cat{1},\ldots,\qkw{T}\cat{n}}$
of object schemas, with the property that, for any set
$\Cal{T'}=\Set{\qkw{T}\cat{i_1},\ldots,\qkw{T}\cat{i_l}}\subseteq \Cal{T}_n$, the object $\JObj{\qkw{a}\cat{i_1}:\anull,\ldots,\qkw{a}\cat{i_l}:\anull}$
satisfies all schemas in $\Cal{T'}$ and no schema in $\Cal{T}_n\setminus\Cal{T'}$.
\begin{definition}[${\San}$, $\qkw{T}\cat{i}$]\label{def:san}
The schema ${\San}$ is defined as follows; we will also use $\qkw{T}\cat{i}$ to denote the
schema with the anchor $\qkw{T}\cat{i}$.
{\small
\begin{verbatim}
{ "anyOf": [
    { "prefixItems": [{"$ref":"#T1"}], "minItems":1, "contains":{"$ref" "#T1"} },
    { "prefixItems": [{"$ref":"#T2"}], "minItems":1, "contains":{"$ref":"#T2"} },
    ...
    { "prefixItems": [{"$ref":"#Tn"}], "minItems":1, "contains":{"$ref":"#Tn"} }
  ],
  "unevaluatedItems": false,
  "$defs": { "T1": { "$anchor": "T1", "required": [ "a1" ] },
             "T2": { "$anchor": "T2", "required": [ "a2" ] },
             ...
             "Tn": { "$anchor": "Tn", "required": [ "an" ] } }
}
\end{verbatim}
}
\end{definition}

Each  $\qany$ branch requires that the first element of the array satisfies the schema $\qkw{T}\lcat i$, and, when this condition holds,
the branch evaluates all array elements that satisfy $\qkw{T}\cat i$; observe that when $\qprefIts$ and $\qminIt$ are both satisfied, then
$\qcont$ holds
as a consequence, but its presence ensures that all other items in the array that satisfy $\qkw{T}\lcat i$ are also evaluated,
so that $\qunIts: \afalse$ does not apply to them.
In summary, an array satisfies this schema if, and only if: 
\begin{compactenum}
\item its first element exists and it satisfies at least one of the $\qkw{T}\cat i$ schemas;
\item every other element satisfies at least one of the $\qkw{T}\cat i$ schemas that are satisfied by the first item.
\end{compactenum}

We now prove that the encoding of this family of schemas requires an exponential blow-up.

\begin{remark}
This family resembles the one that we used for objects, but there are some important differences, especially in the fact that
{\qprops} evaluates all fields whose name matches a given string, independently from their value, while
{\qcont} evaluates the items whose value matches a given schema, independently from their position.
The object keywords {\qprops}, {\qaddProps}, and {\qunProps} are all uniformly defined in terms of property names.
On the other side,
the array keywords {\qprefIts}, {\qits} and {\qunIts} are defined in terms of item positions, but
{\qcont} depends on item content. For this reason, one may expect that {\qunProps} and {\qunIts} should enjoy different
properties with respect to the elimination of {\qunStar} keywords, but this is not really the case.
\end{remark}

%
%
%
%
\begin{propxrep}\label{pro:expits}
For any schema ${\San}$ of the above family, 
any equivalent schema $S$ that is expressed using Positive Negation-Completed {\cJS},
has a dimension
that is bigger than~$2^n$.
\end{propxrep}

\begin{proofsketch}
Consider any schema $S$ written in Positive Negation-Completed {\cJS} and equivalent to ${\San}$.
Consider the set $\Cal{T}_n=\Set{\qkw{T}\cat 1,\ldots,\qkw{T}\cat n}$ of the object schemas
that appear inside ${\San}$.
For each keyword $\iK$ with shape $\qits : S^{\prime}$ (an \emph{items-keyword}), we say that its 
\emph{tail signature} $\TLF$ is the set of all
schemas $\qkw{T}\cat i$ in $\Cal{T}_n$ such that 
$S^{\prime}$ is satisfied by every element of $\qkw{T}\cat i$.\footnote{We call it \emph{tail}
signature since, in case  $\fits : S^{\prime}$ is adjacent to a $\fprefIts : \JArr{S_1,\ldots,S_n}$ keyword, then 
$\fits : S^{\prime}$ only affects the elements in the tail of an instance, that is, those whose position is greater than
$n+1$; differently from the pattern signature of the previous proof,
the tail signature of $\fits: S'$ does not depend on the adjacent keywords.}
For example, the tail signature of an items-keyword $\qits: \JObj{\qany:[\qkw{T1},\qkw{T3}]}$
is $\Set{\qkw{T1},\qkw{T3}}$.

We now show that, for any non-empty subset $A$ of $\Cal{T}_n$,
there exists one items-keyword $\iK$ somewhere inside $S$ that has $A$ as its tail signature.
Assume, towards a contradiction, that there exists a subset $A$ of 
$\Cal{T}_n$ such that no items-keyword inside $S$ has $A$ as its tail signature.
We fix an integer $\key{BIG}$ that is strictly bigger than the maximum $m$ such that
$\qmaxIt: m$ 
appears in $S$, is strictly bigger than the maximum length of the argument $J_A$
of each subkeyword of S that is ${\qprefIts:J_A}$ or ${\qconst:J_A}$, and is strictly bigger than 1.
Fixed a non-empty set $A=\Set{\qkw{T}\cat{i_1},\ldots,\qkw{T}\cat{i_l}}\subseteq \Cal{T}_n$ such that no 
items-keyword has $A$ as its tail signature,
we choose two different objects, $J_1$ and $J_2$, whose 
properties are exactly ${\qkw{a}\cat{i_1},\ldots,\qkw{a}\cat{i_l}}$, so that both $J_1$
and $J_2$ satisfy all  the schemas in $A$ and  no schema in $\Cal{T}_n\setminus A$.
We build an array $J$ of length at least $\key{BIG}$ as follows:
\begin{compactitem}
\item its first item is $J_1$;
\item its next \key{BIG}-1 items, which exist since $\key{BIG}>1$, are equal to $J_2$,
   so that this array  violates every  $\quniqIts:\atrue$ keyword  inside $S$ and, 
    having at least $\key{BIG}$ items, also violates every $\qmaxIt:m$  and $\qconst:J$ keyword inside $S$;
\item then,  for every items-keyword {\iK} with keyword $\qits: S_{\iK}$ whose tail signature $\TLF$ does not satisfy $A\subseteq \TLF$, we
choose a schema $\qkw{T}\cat i_{\iK}$ in $A\setminus \TLF$ and an element $J_{\iK}$ of $\qkw{T}\cat i_{\iK}$ that does not satisfy $S_{\iK}$, which exists
since $\qkw{T}\cat i_{\iK}$ is not in the tail signature $\TLF$, 
     and we add $J_{\iK}$ to the array $J$; this ensures that $J$ does not satisfy  $\qits: S_{\iK}$, so that every items-keyword that 
     belongs to $\SKOf{J}$ (i.e., that
     validates $J$) has a tail signature $\TLF$ such that $A\subseteq \TLF$; since we assumed that no $\iK$ has 
     $\TLF=A$, we conclude that $\iK\in\SKOf{J} \Implies \TLF\supset A$.
\end{compactitem}

The array $J$ satisfies ${\San}$, since its first item satisfies all $\qkw{T}\cat i$ that belongs to $A$ and
all its items ($J_1$, $J_2$, and all the different $J_{\iK}$'s) satisfy some $\qkw{T}\cat i\in A$, hence all its items are evaluated by a successful
branch of ${\San}$, hence none is passed to $\qunIts:\afalse$.
Now, towards a contradiction, we build an array $J^+$ that satisfies $S$ but does not satisfy ${\San}$.
To this aim, we consider an enumeration $\Set{\qkw{T}\cat{j_1},\ldots,\qkw{T}\cat{j_p}}$
of the complement of $A$, $Co(A)=\Cal{T}_n\setminus A$,  we let  $J_a$ be an object whose field names 
are exactly $\Set{\qkw{a}\cat{j_1},\ldots,\qkw{a}\cat{j_p}}$,
and we observe that $J_a$ satisfies all schemas in $Co(A)$ and no schema of $A$.
We build the array $J^+$ by adding  $J_a$ after the last element of the array $J$.

$J^+$ satisfies all keywords in $\SKOf{J}$ because:
\begin{compactitem}
\item by point (2), no satisfied keyword is either a $\qmaxIt:m$ or a $\quniqIts:\atrue$ keyword
  or a $\qconst:A$ keyword;
\item keywords $\qcont:S$, $\qcontAft:S$, $\qminIt:m$, and $\qrepIts:b$ that were already satisfied by $J$ cannot be 
invalidated by the addition of one element;
\item a  keyword $\qprefIts:A$ satisfied by $J$ cannot be violated by $J^+$ since $J$ is longer than $\key{BIG}$ that
  is bigger than the argument of any $\qprefIts:A$ in $S$, hence the added element is not in the prefix;
\item the schema $S'$ of an item-keyword $\iK=\qits:S'$ is satisfied by all items of $J^+$ that belong to $J$ because $\iK\in\SKOf{J}$,
   and $S'$ is also satisfied by $J_a$ because, by (3) above, we have $\TLF\supset A$, hence $\TLF$ contains at least one schema of
   $\qkw{T}\cat{j_p}\in Co(A)$, and $J_a$ satisfies that schema by construction, hence $J_a$ satisfies $S'$, 
   hence $J^+$ satisfies $\qits:S'$;
\item any other structural keyword consider any two arrays to be equivalent.
\end{compactitem}

Hence, $J$ satisfies $S$ and $J^{+}$  satisfies all keywords of $\SKOf{J}$ 
hence, since $S$ is written in negation-free
JSON Schema, then $J^{+}$  satisfies $S$.
However, $J^{+}$ does not satisfy ${\San}$: since the first item of $J^{+}$ is $J_1$, then
$J^{+}$ satisfies all and only the {\qany} branches that correspond to the elements of $A$;
since the item $J_a$ satisfies no schema of $A$, then it is not evaluated by
any successful branch of the $\qany$ keyword, hence it causes {\qunIts} to fail; this contradicts the equivalence between $S$ and ${\San}$.
Hence, we have proved that, for any $A$ non-empty subset of $\Cal{T}_n$, we must have in $S$ at least 
one items-keyword whose tail signature is $A$.
Since $\Cal{T}_n$ has $2^n-1$ different non-empty subsets, this implies that $S$ has at least 
$2^n-1$ different keywords.
\end{proofsketch}

\begin{corollary}\label{cor:exparr}
There exists a family of Static {\mJS} schemas ${\San}$ whose only {\mJS} operator is
one instance of $\qunIts$,
such that, for each family of schemas $S'_n$ expressed in {\cJS}, if each $S'_i$ is equivalent
to $S_i$, then~$S'_n$ has a size in $\Omega(2^n)$.
\end{corollary}

\subsection{Adding {\qminC} and {\qmaxC}}

{\VerEight} of {\JS} introduced the operators {\qminC} and {\qmaxC}.
When ${\qminC:m}$ is adjacent to a ${\qcont:S}$ keyword, it forces the presence of at least $m$ items that satisfy 
$S$; 
when ${\qmaxC:q}$ is adjacent to a ${\qcont:S}$ keyword, it limits the number of items that satisfy $S$ to be less
than $q$, as expressed by the following rule.

\infrule[\rcont]
{
J=\JArr{J_1,...,J_n} \andalso
\forall i\in \SetTo{n}.\ \SJudgC{J_i}{S}\Ret{r_i}{(r_i,\ps_i)} \andalso
\pk_c = \SetST{i}{r_i = \btrue} 
}
{\KJudgCJ{\acont : S, \aminC:m, \amaxC:q}
\Ret{\rl\plus r}{(m \!\leq\! |\pk_c| \!\leq\! q,\pk_c)}
}

The addition of these operators to {\cJS} has already been studied: they are considered in the algorithm to verify satisfiability and
inclusion presented in \cite{DBLP:journals/pvldb/AttoucheBCGSS22}, and are included in Negation-Completed {\cJS} in~\cite{DBLP:journals/tcs/BaaziziCGSS23}.

Since these operators add a counting ability to {\JS}, it is natural to ask whether the exponentiality result would
still hold if we added $\qminC:m$ and $\qmaxC:q$ to the target language. 
We prove here that this is still the case --- these operators do not help making the translation more compact.

{\mJS} allows any combination of $\acont : S$, $\aminC:m$, and $\amaxC:q$ keywords in the same schema and gives a
meaning to each combination.
If we allow the $\Inf$ value for the $q$ parameter of $\amaxC:q$, the meaning of any combination where some
operator is missing can be defined by the following normalization
procedure: if $\acont : S$ is missing, then the two other keywords are deleted; otherwise, a missing
$\aminC:m$ can be inserted as $\aminC:1$, and  a missing
$\amaxC:q$ can be inserted as $\amaxC:\Inf$. 
In our proof, we assume that every schema is normalized in this way, so that
we consider the triple $\acont : S, \aminC:m, \amaxC:q$ to be a single keyword, defined by the above rule.

%
%
\begin{propxrep}
For any schema ${\San}$ of Definition \ref{def:san},
if $S$ is equivalent to ${\San}$ and is expressed using Positive Negation-Completed {\cJS}
enriched with the {\qminC} and {\qmaxC} keywords, then $S$ has a dimension
that is bigger than $2^n$.
\end{propxrep}

\begin{proofsketch}
Consider any schema written in Positive Negation-Completed {\cJS}
enriched with the {\qminC} and {\qmaxC} keywords, and equivalent to ${\San}$.
Assume that any  $\qmaxC: q$ keyword of $S$ has $q \geq 1$; this can
be obtained by rewriting every triple $\qcont : S^{\prime}$, $\qminC: 0$, $\qmaxC: 0$ 
as
$ \qits : \JObj{ \qnot: S^{\prime}}$ and then performing not-elimination, while
triples with $\qminC: n$, $\qmaxC: 0$, and $n\neq 0$ can be rewritten as $\xfalse$.
For this schema, we repeat the same construction as in the proof of Property \ref{pro:expits},
but this time we choose $\key{BIG}$ so that it is also strictly bigger than any $q$ such that
$\qmaxC: q$ appears in $S$.

Once we have fixed a non-empty $A=\Set{\qkw{T}\cat{i_1},\ldots,\qkw{T}\cat{i_l}}\subseteq \Cal{T}_n$ 
such that no items-keyword has $A$ as its tail signature,
we choose, as in the previous proof, two different objects, $J_1$ and $J_2$, whose 
properties are exactly ${\qkw{a}\cat{i_1},\ldots,\qkw{a}\cat{i_l}}$, so that both $J_1$
and $J_2$ satisfy all  the schemas in $A$ and  no schema in $\Cal{T}_n\setminus A$.
Then, as in the previous proof, for every items-keyword {\iK} with keyword $\qits: S_{\iK}$ whose tail signature $\TLF$ does not satisfy $A\subseteq \TLF$, we
choose a schema $\qkw{T}\cat i_{\iK}$ in $A\setminus \TLF$ and an element $J_{\iK}$ of $\qkw{T}\cat i_{\iK}$ that does not satisfy $S_{\iK}$, which exists
since $\qkw{T}\cat i_{\iK}$ is not in $\TLF$.
At this point, we build an array $J$ that contains $\key{BIG}$ copies of $J_1$, followed by $\key{BIG}$ copies of $J_2$,
and, for each $\iK$ whose tail signature $\TLF$ does not satisfy $A\subseteq \TLF$, we add $\key{BIG}$ copies of the item $J_{\iK}$ that ensures 
that  $\iK\not\in\SKOf{J}$.
Hence, as in the previous proof, we have that $J$ satisfies ${\San}$, and that every items-keyword in $\SKOf{J}$ satisfies $\TLF\supset A$.

We now build the array $J^+$ by adding an element $J_a$ at the end of $J$ as in the previous proof
and we prove that $J^+$ satisfies all keywords in $\SKOf{J}$.
For all keywords in $\SKOf{J}$ that are different from $\qcont : S_C, \qminC:m, \qmaxC:q$, we reason as in the previous proof.

For every keyword 
$\qcont : S_C, \qminC:m, \qmaxC:q$,
since $J$ satisfies the $\qminC:m$ requirement, then $J^{+}$, which contains all items of $J$ plus one more, also
satisfies $\qminC:m$ as well.

For the $\qmaxC:q$ keyword, if $q=\Inf$, then it is trivially satisfied.
If $q$ is finite, it is strictly smaller then $\key{BIG}$, hence none of the $J_1$, $J_2$, and $J_{\iK}$ elements of $J$ satisfies $S_C$, since 
all of them are repeated at least $\key{BIG}$ times and $\key{BIG}>q$, hence, if they satisfied $S_C$, they would violate $\qmaxC:q$. Hence, the only
item in $J^{+}$ that may satisfy $S_C$ is $J_a$ but, since $q\geq 1$, a single item is not sufficient to violate $\qmaxC:q$;
hence, $J^{+}$ satisfies $\qmaxC:q$.

As in the previous proof, we have proved that $J^{+}$ satisfies every keyword in $\SKOf{J}$ but it does not satisfy ${\San}$, 
hence we have a contradiction, hence we have proved that for every $A$ there is a keyword inside $S$ whose tail signature is $A$, 
hence we have an exponential number of keywords.
As a conclusion, the exponentiality proof holds even if $S$ is allowed to use {\qminC} and {\qmaxC}.
\end{proofsketch}

As proved in \cite{DBLP:journals/tcs/BaaziziCGSS23}, negation can be eliminated from {\cJS} with a polynomial size expansion even in the presence of {\qminC} and {\qmaxC}, hence we have the following corollary.

\begin{corollary}\label{cor:exparr}
There exists a family of Static {\mJS} schemas ${\San}$ whose only {\mJS} operator is
one instance of $\qunIts$,
such that, for each family of schemas $S'_n$ expressed in {\cJS}, 
enriched with the {\qminC} and {\qmaxC} keywords,
if each $S'_i$ is equivalent
to $S_i$, then $S'_n$ has a size in $\Omega(2^n)$.
\end{corollary}

\section{Eliminating $\xunProps$  using normalization}\label{sec:elimination}

\subsection{Introduction}

We have shown in Section \ref{sec:problem} that, in general, it is not sound to just push {\xunStar} keywords through a logical operator, and
in Section \ref{sec:blowup} we have proved that, in general, the elimination of {\xunStar} keywords may require an exponential increase of 
the schema size.

In this section, we first introduce a baseline technique to eliminate {\xunStar} keywords, based on \emph{eXclusive Disjunctive Normal Form} (XDNF), which is simple
and general, but not practical.
We then use the XDNF intuition to introduce our approach, based on the notions of \emph{statically characterized schemas},
\emph{cover closure}, and \emph{Evaluation Normal Form} (ENF),
which yields schemas that can be much smaller than the baseline.

\subsection{Pushing $\xunProps$ through eXclusive DNF}\label{subsec:pushex}


A schema $S$ is in \emph{DNF} (\emph{Disjunctive Normal Form}) when
$S=\qanyOf{S_1,\ldots,S_n}$, and each $S_i$ combines structural keywords
$K^i_{j}$ using conjunctive operators:
$S_i=\JObj{K^i_{1},\ldots,K^i_{m_{i}}}$ or $S_i=\qallOf{\JObj{K^i_{1}},\ldots,\JObj{K^i_{m_{i}}}}$.

We have seen in Section~\ref{sec:problem} that {\qunProps} cannot be pushed through $S=\qanyOf{S_i,\ldots,S_n}$ in general, but this is possible when 
 $S$ is in \emph{eXclusive DNF} (\emph{XDNF}), that is, it is  in DNF and all conjunctions $S_i$ are mutually exclusive.
 
Consider, for example, the following schema that has 
a core definition \qkw{\#/\$defs/extensible}
 that is extended in different ways; the core definition is ``extensible'', but, after the extension,
the result is closed by adding $\qunProps: \xfalse$; this is a common use case of this operator.

\begin{Verbatim}[fontsize=\small,frame=lines]
{ "anyOf": [
    { "$ref": "#/$defs/extensible",
      "properties": { "kind": {"const": "A"}, "address": {"type": "string"} },
      "required": ["kind"] },
    { "$ref": "#/$defs/extensible",
      "properties": { "kind": {"const": "M"}, "model": {"type": "string"} },
      "required": ["kind"] }
  ],
  "unevaluatedProperties": false,
  "$defs": {
       "extensible": { "type": "object",  "properties": { "name": "string"} } 
  }
}
\end{Verbatim}

%

The first branch is satisfied by objects having a mandatory $\qkw{kind}$ property with a constant value $\qkw{A}$ as well as optional properties $\qkw{address}$ and $\qkw{name}$ of type string, but it does not impose constraints on other properties; on the other hand, the second branch is satisfied by objects having a mandatory $\qkw{kind}$ property with a constant value $\qkw{M}$ as well as optional properties $\qkw{model}$ and $\qkw{name}$ of type string, but, as for the previous branch, it does not impose further constraints on other properties.

Since the two branches of $\qany$ are mutually exclusive, because of the disjoint values allowed for property $\qkw{kind}$, then, for each instance $J$, when one branch is successful,
the other branch fails, hence it does not evaluate any property; hence,  ${\qunProps} : S_u$ can be pushed
through $\qany$.

Now we observe that we can statically compute
the properties $p_1,\ldots,p_n$ that are evaluated by each branch,
and rewrite it as 
$$
\begin{array}{llllllll}
  \JObjOpen  K^i_{1},\ldots,K^i_{m_{i}},
  \qall: [  \JObj{ \qprops : \JObj{p_1{\Col}\JObj{},...,p_n{\Col}\JObj{}}, \qaddProps : S_u} ] 
  \JObjClose
  \end{array}
  $$
where the ``useless'' unary $\qall$ is necessary since one of the $K^i_j$ is $\qprops:\JObj{}$,
and we cannot have two different
$\qprops$ adjacent in the same schema.\footnote{We could of course merge the two keywords into one,
but we use $\fall$ to keep the approach simpler.}

Hence, the above schema can be rewritten as follows.
\begin{Verbatim}[fontsize=\small,frame=lines]
{ "anyOf": [
    { "$ref": "#/$defs/extensible",
      "properties": { "kind": {"const": "A"}, "address": {"type": "string"} },
      "required": ["kind"],
      "allOf": [ { "properties": { "name": {}, "kind": {}, "address": {} } ,
                   "additionalProperties": false } ]
    },
    { "$ref": "#/$defs/extensible",
      "properties": { "kind": {"const": "M"}, "model": {"type": "string"} },
      "required": ["kind"],
      "allOf": [ { "properties": { "name": {}, "kind": {}, "model": {} },
                   "additionalProperties": false } ] 
    }
  ],
  "$defs": { "extensible": { "type": "object", 
                             "properties": { "name": {"type": "string"} } }
  }
}
\end{Verbatim}

This method can be always applied: we can rewrite every schema as an XDNF schema, by 
first rewriting the schema in DNF, and then 
by substituting the $n$ arguments of the disjunction $\qany: [ S_1, \ldots, S_n ]$ with $O(2^n)$ mutually exclusive conjunctions:
for each non-empty subset $\Cal{S}$ of $\Set{S_1, \ldots, S_n}$, we consider the conjunction where all elements of
$\Cal{S}$ are required to hold and all those in the complement of $\Cal{S}$ are negated, so that every possible
combination of validity/non-validity of the schemas in $\Set{S_1, \ldots, S_n}$ that validates $\qany: [ S_1, \ldots, S_n ]$
is satisfied by exactly one of these mutually exclusive conjunctions. \hide{(for details, see Remark \ref{rem:XDNF}).}
Since the result is in XDNF, we can push $\qunProps$ through the $\qany$ inside each branch.
Since each branch of an XDNF is a conjunction of structural operators, we can always statically compute which properties
are evaluated by that branch, and hence substitute $\qunProps$ with $\qaddProps$.


It is easy to prove that the size of this translation is bounded by $O(2^N)$, hence it is asymptotically ``optimal'', according to the result of
Corollary \ref{cor:expobj}.
However, the $O(2^N)$ lower bound of Corollary \ref{cor:expobj} is a worst-case scenario that one hopes to avoid in most
practical cases, while this XDNF
approach generates an exponential explosion from any disjunction.

A second problem of this approach is the fact that reduction to DNF requires not-elimination, which is not trivial in {\mJS}: while 
operators such as
$\qall$ and $\qany$, or $\qminP$ and $\qmaxP$, are each the De Morgan dual of the other; other keywords, such as $\qpattProps$, do not have a De Morgan dual.
This issue can be managed through the techniques introduced in \cite{DBLP:journals/tcs/BaaziziCGSS23} for {\cJS},
but the problem in {\mJS} is a bit more complex, since we
need to define a negation dual for $\qunProps$ and a negation dual for $\qnot$, as discussed in Remark \ref{rem:demorgan}

We present here an algorithm that overcomes these two problems: it generalizes the condition of ``exclusiveness'' to a weaker
condition of ``cover-closure'' that takes advantage from a static characterization
of the properties evaluated by each subschema; differently from DNF and XDNF,
the cover-closure condition can be reached without any kind of not-elimination, and, in many
practical cases, adding a limited number of extra cases to the original disjunction.

Our experiments, described in Section \ref{sec:experiments}, show that this algorithm behaves very well on real-world schemas.

\begin {remark}\label{rem:demorgan}
In {\cJS}, $\qnot$ is self-dual,
which means that $\qnotOf{\qnotOf{S}}$ can always be rewritten as $S$. Unfortunately, this is not the case in {\mJS}, since $\qnotOf{\qnotOf{S}}$
yields the same boolean result as $S$, but returns no annotation; this fact complicates the not-elimination of the
$\qnot$ operator.

\hide{
When we arrive at the leaves, we use the $\qpattReq$ keyword defined in \cite{DBLP:journals/tcs/BaaziziCGSS23} in order to dualize $\qpattProps$; this keyword can then be
translated back to {\cJS} as described in \cite{DBLP:journals/tcs/BaaziziCGSS23}. The double negation of $\qpattProps: S$ requires a double $\qnot$ since we want to retain its boolean
behavior but not its annotations; this is not needed with $\qreq : [p]$, since this keyword produces no annotation.
$$
\begin{array}{lllll}
\XNOT( \qany : [ S_1,\ldots, S_m] ) &=& \qall : [ \XNOTOf{S_1},\ldots, \XNOTOf{S_m}]) \\
\XNOT( \qall : [ S_1,\ldots, S_m] ) &=& \qany : [ \XNOTOf{S_1},\ldots, \XNOTOf{S_m}]) \\
\XNOT (\qnotOf {S}) &=& \XNNOT (S) \\
\XNOTOf{\qpattProps : S} &=& \qpattReq : \JObj{ \XNOTOf { S} } \\
\XNOTOf{\qreq : [p] } &=& \qprops : \JObj { p : \XNOTOf{S}} \\[3\NL]
\XNNOT( \qany : [ S_1,\ldots, S_m] ) &=& \qany : [ \XNNOTOf{S_1},\ldots, \XNNOTOf{S_m}]) \\
\XNNOT( \qall : [ S_1,\ldots, S_m] ) &=& \qall : [ \XNNOTOf{S_1},\ldots, \XNNOTOf{S_m}]) \\
\XNNOT (\qnotOf {S}) &=& \XNOT (S) \\
\XNNOTOf{\qpattProps : S} &=&\qnotOf{\qnotOf{ \qpattProps : S} } \\
\XNNOTOf{\qreq : [p] } &=&  \JObj { \qreq : [p] } \\
\end{array}
$$
}

Consider now $\qunProps: S$; it fails when there exists an unevaluated property that satisfies $\qnotOf{S}$,
hence, in order to rewrite $\qnot:\JObj{\qunProps: S}$, we need
an operator $\qunReq: S$ that \emph{requires} the presence of an unevaluated property that
satisfies $S$, in the same way as we had to define $\qpattReq:S$ in order to eliminate 
$\qnot:\JObj{\qpattProps:S}$.
Moreover, this new $\qunReq: S$ operator would be annotation-dependent, hence we would then
need to eliminate it as we do with the other $\qunStar$ operators.


All of this is doable, but is quite complex, and expensive; an approach that avoids not-elimination should be
preferred.
\end{remark}

\subsection{Statically characterizing the properties and items that are evaluated by a schema}

Our algorithm takes advantage of those situations where one can statically characterize the properties
and the items that a schema evaluates, which happens, for example, for all schemas that do not depend
on an $\qany$ or $\qone$ operator.
We now formalize this notion.

For example, observe that $S=\qallOf{\qprefIts:[S_1,S_2],\qcont:S_c}$ ``evaluates'' an item $J_i$ at position $i$ of an array $J$ if, and only if,
(1) $J$ satisfies $S$ and  (2) either $i\leq 2$ or $J_i$ satisfies $S_c$. We will express this fact by saying that
``the pair $(2,S_c)$ characterizes the item evaluation of $S$'' (Definition \ref{def:pair}.2 and Definition \ref{def:sta}.4-5-6).
However, if you consider a schema $S=\qanyOf{S_1,S_2}$, where $S_1$ is characterized by $(1,S^1_c)$ and
$S_2$ is characterized by $(2,S^2_c)$, you can only have a lower bound and an upper bound of what is evaluated.
The lower bound is $(1,\qallOf{S^1_c,S^2_c})$: if $\SJudgCJ{S}$,
whichever is the $\qany$ branch that succeeds, if an element has position 1, or if it satisfies $\qallOf{S^1_c,S^2_c}$, then the
element is evaluated.
The upper bound is $(2,\qanyOf{S^1_c,S^2_c})$: if an item of a $J$ such that $\SJudgCJ{S}$ is evaluated by $S$, then
either its position is $\leq 2$ or it satisfies either $S^1_c$ or $S^2_c$. We now formalize all of this.

\begin{definition}[``$k$ matches $\Set{p_1,\ldots,p_n}$'', ``$J$ satisfies $(h,S)$'']\label{def:pair} \ \\
\begin{compactenum}
\item For a set of patterns $P=\Set{p_1,\ldots,p_n}$, we define $\rlan{P}=\bigcup_{i_\in\SetTo{n}}\rlan{p_i}$, and we say that
$k$ matches $P$ when $k\in\rlan{P}$.

\item Given an array $\JArr{J_1,\ldots,J_n}$ we say that the item $J_i$ at position $i$ satisfies the pair $(h,S)$ with
$h\in\Nat$, if either $i \leq h$ or $J_i$ satisfies $S$.
\end{compactenum}
\end{definition}

\begin{definition}[``$\Set{p_1,\ldots,p_n}$/$(h,S)$ is an upper bound/is a lower bound/statically characterizes'']\label{def:sta}\ \\
\begin{compactenum}
\item A set of patterns $\Set{p_1,\ldots,p_n}$ \emph{is an upper bound for property evaluation} of a schema $S$ iff,
for any object $J$, if $J$ satisfies $S$, 
then every evaluated property matches $\Set{p_1,\ldots,p_n}$.

\item  A set of patterns $\Set{p_1,\ldots,p_n}$ \emph{is a lower bound for property evaluation} of a schema $S$ iff,
for any object $J$,  if $J$ satisfies $S$, then every property of $J$ that matches $\Set{p_1,\ldots,p_n}$
is evaluated.

\item A set of patterns $\Set{p_1,\ldots,p_n}$ \emph{characterizes the property evaluation} of a schema $S$ iff it is both an upper bound and a lower bound.

\item A pair $(h,S')$ \emph{is an upper bound for item evaluation} of a schema $S$ iff,
for any array $J$, if $J$ satisfies $S$, 
then every item evaluated by $S$ satisfies $(h,S')$.

\item A pair $(h,S')$ \emph{is a lower bound for item evaluation} of a schema $S$ iff,
for any array $J$, if $J$ satisfies $S$, 
then every item of $J$ that satisfies $(h,S')$ is evaluated by $S$.

\item A pair $(h,S')$ \emph{characterizes the item evaluation} of a schema $S$ iff it is both an upper bound and a lower 
bound.

\item A schema $S$ is \emph{statically characterized} iff there exist $\Set{p_1,\ldots,p_n}$ that \emph{characterizes its property evaluation}  and a pair $(h,S')$ that \emph{characterizes its item evaluation}.

\end{compactenum}

\end{definition}

\hide{
\begin{example}\label{ex:uplow}
Consider again the schema of Section \ref{subsec:pushex} and consider the first branch of $\qany$. 
 This branch is satisfied by any object having a mandatory $\qkw{kind}$ property, optional $\qkw{address}$ and $\qkw{name}$ properties of type string, as well as any other property. Therefore, any object satisfying this branch \emph{must} have a $\qkw{kind}$ property, which is evaluated by the $\qkw{properties}$ keyword inside the branch; $\qkw{address}$ and $\qkw{name}$, if any, are evaluated respectively by the $\qkw{properties}$ keyword inside the branch and by the $\qkw{properties}$ keyword inside the referred schema. Any other property is left unevaluated. 

\GG{Wrong example, the first branch of the ${\qany}$ evaluates all properties. }
As a consequence, $\Set{\exactly{kind}, \exactly{address}, \exactly{name}}$ is an upper bound for property evaluation for the first branch of the ${\qany}$. It is easy to see that the same set of patterns is also a lower-bound.
\end{example}

\DC{should we add an example for the array counterpart of the definition?}
}

We now define functions $\MinEP$, $\MaxEP$, $\MinEI$, $\MaxEI$,
which compute a lower bound and an upper bound for the properties and the items that are evaluated 
by a schema.
They are defined in Tables~\ref{tab:EP} and~\ref{tab:EI}.

\begin{table}[htb]
\caption{Functions $\MinEP(S)$ and $\MaxEP(S)$.}\label{tab:EP}
\label{tab:exP}
\begin{tabular}{|l|l|l|c|}
\hline
\textbf{Schema}   & $\MinEP(S)$ & $\MaxEP(S)$ \\ \hline\hline
 $ \JObj{K_1,\ldots,K_n}\ \ n \geq 2$ & $\bigcup_{i\in \SetTo{n}}{\MinEPK{K_i}}$ & $\bigcup_{i\in \SetTo{n}}{\MaxEPK{K_i}}$ \\ \hline
 $\JObj{\qprops: \JObj{p_1:S_1,...,p_m:S_m}}$  & $\Set{\exactly{p_1},\ldots,\exactly{p_m}}$  & $\Set{\exactly{p_1},\ldots,\exactly{p_m}}$ \\ \hline
 $\JObj{\qpattProps: \JObj{p_1:S_1,...,p_m:S_m}}$  & $\Set{{p_1},\ldots,{p_m}}$  & $\Set{{p_1},\ldots,{p_m}}$ \\ \hline
 $\JObj{\qaddProps: S_a}$ &  $\Set{\qkw{.*}}$  &  $\Set{\qkw{.*}}$  \\ \hline
 $\JObj{\qunProps: S_u}$ &  $\Set{\qkw{.*}}$  &  $\Set{\qkw{.*}}$  \\ \hline
 $\JObj{\qdref: u}$ & $\MinEP(\key{deref}(u))$ & $\MaxEP(\key{deref}(u))$  \\ \hline
 $\JObj{\qany: \JArr{S_1,\ldots,S_n}}$ & $\bigcap_{i\in \SetTo{n}}{\MinEP(S_i)}$ & $\bigcup_{i\in \SetTo{n}}{\MaxEP(S_i)}$ \\ \hline
 $\JObj{\qone: \JArr{S_1,\ldots,S_n}}$ & $\bigcap_{i\in \SetTo{n}}{\MinEP(S_i)}$ & $\bigcup_{i\in \SetTo{n}}{\MaxEP(S_i)}$ \\ \hline
 $\JObj{\qall: \JArr{S_1,\ldots,S_n}}$ & $\bigcup_{i\in \SetTo{n}}{\MinEP(S_i)}$ & $\bigcup_{i\in \SetTo{n}}{\MaxEP(S_i)}$ \\ \hline
any other schema 
      & $\Set{}$  & $\Set{}$ \\ \hline
\end{tabular}
\end{table}

\begin{table}[htb]
\caption{Functions $\MinEI(S)$ and $\MaxEI(S)$.}
\label{tab:EI}

\begin{tabular}{|m{14em}|m{13em}|m{13em}|}
\hline
\textbf{Schema}   & $\MinEI(S)$ & $\MaxEI(S)$ \\ \hline\hline
$ \JObj{K_1,\ldots,K_n}\ \ n \geq 2$
& $(\max_{i\in \SetTo{n}}(h_i),$ 
$\ \mbox{\ \ } \JObj{\qany : [S'_1,\ldots,S'_n]})$ \mbox{\ \ \ }
where $(h_i,S'_i)=\MinEIK{K_i}$  
& $(\max_{i\in \SetTo{n}}(h_i),$
$\ \mbox{\ \ } \JObj{\qany : [S'_1,\ldots,S'_n]})$ \mbox{\ \ \ }
where $(h_i,S'_i)=\MaxEIK{K_i}$ 
\\ \hline
$\JObj{\qkw{prefixItems}:[S_1,\ldots,S_n]}$  & $(n,\xfalse)$ & $ (n,\xfalse)$  \\ \hline
$\JObj{\qkw{contains}: S_a}$ &  $(0,S_a)$  &  $(0,S_a)$  \\ \hline
$\JObj{\qits: s} $  & $ (\Inf,\xtrue)$ & $ (\Inf,\xtrue)$  \\ \hline
$\JObj{\qunIts: S_u}$ &  $ (\Inf,\xtrue)$  & $ (\Inf,\xtrue)$   \\ \hline
$\JObj{\qdref: u}$ & $\MinEI(\key{deref}(u))$ & $\MaxEI(\key{deref}(u))$  \\ \hline
$
\!\!\!\begin{array}{lllllllllllll}
\JObj{\qany: \JArr{S_1,\ldots,S_n}}  \ \mbox{or}   \\
\JObj{\qone: \JArr{S_1,\ldots,S_n}}
\end{array}
$
& $(\min_{i\in \SetTo{n}}(h_i),$
$\ \mbox{\ \ } \JObj{\qall : [S'_1,\ldots,S'_n]})$ \mbox{\ \ \ }
where $(h_i,S'_i)=\MinEI(S_i)$  
& $(\max_{i\in \SetTo{n}}(h_i),$
$\ \mbox{\ \ } \JObj{\qany : [S'_1,\ldots,S'_n]})$ \mbox{\ \ \ }
where $(h_i,S'_i)=\MaxEI(S_i)$  
\\ \hline
$\JObj{\qall: \JArr{S_1,\ldots,S_n}}$ 
& $(\max_{i\in \SetTo{n}}(h_i),$ 
$\ \mbox{\ \ } \JObj{\qany : [S'_1,\ldots,S'_n]})$ \mbox{\ \ \ }
where $(h_i,S'_i)=\MinEI(S_i)$  
& $(\max_{i\in \SetTo{n}}(h_i),$
$\ \mbox{\ \ } \JObj{\qany : [S'_1,\ldots,S'_n]})$ \mbox{\ \ \ }
where $(h_i,S'_i)=\MaxEI(S_i)$  
\\ \hline
any other schema 
 & $(0,\xfalse)$  & $(0,\xfalse)$ \\ \hline
\end{tabular}\label{tab:exI}

\end{table}

\begin{propxrep}\label{pro:minmax}
For any schema $S$, $\MinEP(S)$ is a lower bound for its property evaluation, and $\MaxEP(S)$ is an upper bound.

For any schema $S$, $\MinEI(S)$ is a lower bound for its item evaluation, and $\MaxEI(S)$ is an upper bound.

%
%
\end{propxrep}

\begin{proof}
We prove, by induction, what follows.
For any schema $S$, $\MinEP(S)$ is a lower bound for its property evaluation, and $\MaxEP(S)$ is an upper bound.
For any schema $S$, $\MinEI(S)$ is a lower bound for its item evaluation, and $\MaxEI(S)$ is an upper bound.
For any keyword $K$, $\MinEPK{K}$ is a lower bound for the property evaluation of schema $\JObj{K}$, 
and $\MaxEPK{K}$ is an upper bound for the same schema.
For any keyword $K$, $\MinEIK{K}$ is a lower bound for the item evaluation of schema $\JObj{K}$, 
and $\MaxEIK{K}$ is an upper bound for the same schema.

We start with $\MinEP$ and $\MaxEP$.

Cases ${\qprops}$ and ${\qpattProps}$ are trivial.\ 
Cases {\qunProps} is trivial as well -  a successful schema $\JObj{\qunProps}$ evaluates
all properties. Of course, when $\qunProps$ is not the only keyword it does not evaluate all properties
but only those that are not evaluated by the other properties, but we deal with this aspect in the inductive
proof for case $\JObj{K_1,\ldots,K_n}$. The same holds for $\qaddProps$.

Case $\qdref:u$ is trivial.

$\MinEP(\JObj{K_1,\ldots,K_n})=\bigcup_{i\in \SetTo{n}}{\MinEPK{K_i}}$:
We want to prove that 
for any object $J$,  if $J$ satisfies $\JObj{K_1,\ldots,K_n}$, then every property of $J$ that matches 
${\bigcup_{i\in \SetTo{n}}{\MinEPK{K_i}}}$ is evaluated.
If a property $p$
matches ${\bigcup_{i\in \SetTo{n}}{\MinEPK{K_i}}}$, then it matches $\MinEPK{K_l}$ for some $l$.
Since $\JObj{K_1,\ldots,K_n}$ is satisfied, then we know that $K_l$ is satisfied,
hence, by induction, property $p$ is evaluated by the schema $\JObj{K_l}$.
If $K_l$ is an independent keyword, we conclude that $p$ is evaluated by the keyword $K_l$, hence 
is evaluated by the schema.
If $K_l$ is either $\qunProps$ or $\qaddProps$, then we know that $p$ is either evaluated by $K_l$ or by a keyword that precedes $K_l$,
hence it is evaluated by the schema.

$\MaxEP(\JObj{K_1,\ldots,K_n})=\bigcup_{i\in \SetTo{n}}{\MaxEPK{K_i}}$:
We want to prove that 
for any object $J$,  if a property $p$ of $J$ is evaluated by $\JObj{K_1,\ldots,K_n}$, then $p$ matches 
${\bigcup_{i\in \SetTo{n}}{\MaxEPK{K_i}}}$.
If $p$ is evaluated by $\JObj{K_1,\ldots,K_n}$ then $p$ is evaluated by at least
one keyword $K_l$; if $K_l$ is an independent property, then $p$ is also evaluated by
$\JObj{K_l}$, hence, by induction, $p$ is in $\MaxEPK{K_l}$, hence it is in ${\bigcup_{i\in \SetTo{n}}{\MaxEPK{K_i}}}$.
If $K_l$ is $\qunProps:S$, then it is possible that $p$ is not evaluated by $\JObj{K_l}$ since $p$ may be
evaluated by a keyword that precedes $K_l$, but still $p$ is in ${\bigcup_{i\in \SetTo{n}}{\MaxEPK{K_i}}}$
since $\MaxEPK{K_l}=\qkw{.*}$, and $\qkw{.*}$ matches any property. The same proof holds for
$\qaddProps:S$.

$\MinEP(\qone: \JArr{S_1,\ldots,S_n}) = \bigcap_{i\in \SetTo{n}}{\MinEP(S_i)}$:
We want to prove that 
for any object $J$,  if $J$ satisfies $\JObj{(\qone: \JArr{S_1,\ldots,S_n})}$, then every property of $J$ that matches 
$\bigcap_{i\in \SetTo{n}}{\MinEP(S_i)}$ is evaluated.
Assume that  $J$ satisfies $\JObj{(\qone: \JArr{S_1,\ldots,S_n})}$. Then, $J$ satisfies $S_l$ for some $l$, hence,
by induction, every property of $J$ that matches $\MinEP(S_l)$ is evaluated.
If a property of $J$  matches $\bigcap_{i\in \SetTo{n}}{\MinEP(S_i)}$, then it matches $\MinEP(S_l)$,
and the thesis follows.

$\MaxEP(\qone: \JArr{S_1,\ldots,S_n}) = \bigcup_{i\in \SetTo{n}}{\MaxEP(S_i)}$:
We want to prove that 
for any object $J$,  if a property $p$ of $J$ is evaluated by $\qone: \JArr{S_1,\ldots,S_n}$, then $p$ matches 
${\bigcup_{i\in \SetTo{n}}{\MaxEP(S_i)}}$.
If a property $p$ of $J$ is evaluated by a successful $\qone: \JArr{S_1,\ldots,S_n}$, then it is evaluated by the only
branch of $\qone$ that succeeds; let us call it $S_l$. 
By induction, property $p$ matches $\MaxEP(S_l)$, hence it matches ${\bigcup_{i\in \SetTo{n}}{\MaxEP(S_i)}}$.

$\MinEP(\qany: \JArr{S_1,\ldots,S_n}) = \bigcap_{i\in \SetTo{n}}{\MinEP(S_i)}$:
We want to prove that 
for any object $J$,  if $J$ satisfies $\JObj{(\qany: \JArr{S_1,\ldots,S_n})}$, then every property of $J$ that matches 
$\bigcap_{i\in \SetTo{n}}{\MinEP(S_i)}$ is evaluated.
Assume that  $J$ satisfies $\JObj{(\qany: \JArr{S_1,\ldots,S_n})}$. Then, $J$ satisfies $S_l$ for some $l$, hence,
by induction, every property of $J$ that matches $\MinEP(S_l)$ is evaluated.
If a property of $J$  matches $\bigcap_{i\in \SetTo{n}}{\MinEP(S_i)}$, then it matches $\MinEP(S_l)$,
and the thesis follows.

$\MaxEP(\qany: \JArr{S_1,\ldots,S_n}) = \bigcup_{i\in \SetTo{n}}{\MaxEP(S_i)}$:
We want to prove that 
for any object $J$,  if a property $p$ of $J$ is evaluated by $\qany: \JArr{S_1,\ldots,S_n}$, then $p$ matches 
${\bigcup_{i\in \SetTo{n}}{\MaxEP(S_i)}}$.
If a property $p$ of $J$ is evaluated by a successful $\qany: \JArr{S_1,\ldots,S_n}$, then it is evaluated by some
branches of $\qany$ that succeed; let us consider anyone of them, and call it $S_l$. 
By induction, property $p$ matches $\MaxEP(S_l)$, hence it matches ${\bigcup_{i\in \SetTo{n}}{\MaxEP(S_i)}}$.

$\MinEP(\qall: \JArr{S_1,\ldots,S_n}) = \bigcup_{i\in \SetTo{n}}{\MinEP(S_i)}$:
We want to prove that 
for any object $J$,  if $J$ satisfies $\JObj{(\qall: \JArr{S_1,\ldots,S_n})}$, then every property of $J$ that matches 
$\bigcup_{i\in \SetTo{n}}{\MinEP(S_i)}$ is evaluated.
Assume that  $J$ satisfies $\JObj{(\qall: \JArr{S_1,\ldots,S_n})}$. Then, $J$ satisfies every $S_l$ hence,
by induction, every property of $J$ that matches $\MinEP(S_l)$ is evaluated, for each $l$,
hence every property in $\bigcup_{i\in \SetTo{n}}{\MinEP(S_i)}$ is evaluated.

$\MaxEP(\qall: \JArr{S_1,\ldots,S_n}) = \bigcup_{i\in \SetTo{n}}{\MaxEP(S_i)}$:
We want to prove that 
for any object $J$,  if a property $p$ of $J$ is evaluated by $\qall: \JArr{S_1,\ldots,S_n}$, then $p$ matches 
${\bigcup_{i\in \SetTo{n}}{\MaxEP(S_i)}}$.
If a property $p$ of $J$ is evaluated by $\qall: \JArr{S_1,\ldots,S_n}$, then it is evaluated by some $S_l$. 
By induction, property $p$ matches $\MaxEP(S_l)$, hence it matches ${\bigcup_{i\in \SetTo{n}}{\MaxEP(S_i)}}$.

Any other keyword is either a structural keyword different from the ones listed above or is the $\qnot$ keyword.
The structural keywords different from the ones listed above do not evaluate any property of the instance under validation.
A successful $\qnot:S$ keyword is applied to a schema $S$ that does not validate the instance, and a schema that does not
validate an instance does not return any validated property.
Schemas {\xtrue} and {\xfalse} do not evaluate any properties.

We move to $\MinEI$ and $\MaxEI$.

We say that an item $J_i$ of an array 
$J=\JArr{J_1,\ldots,J_n}$, satisfies the first component of a pair $(l,S)$, if $i\leq l$, and that is satisfies
the second component if $J_i$ satisfies $S$.
We say that $J_i$ satisfies the pair $(l,S)$ if it satisfies either the first or the second component.

$\MinEI/\MaxEIK{\qits:S}=(\Inf,\xtrue)$: a schema $\JObj{ \qits: S}$ evaluates all items of any evaluated array,
and every item of an array satisfies the pair $(\Inf,\xtrue)$.

$\MinEI/\MaxEIK{\qprefIts:\JArr{S_1,\ldots,S_n}}=(n,\xtrue)$: a schema $\JObj{ \qprefIts:\JArr{S_1,\ldots,S_n}}$ 
evaluates all and only the items with position $\leq n$, which are exactly the items that satisfy $(n,\xfalse)$.

$\MinEI/\MaxEIK{\qcont: S}=(0,S)$: a schema $\JObj{ \qcont: S}$ 
evaluates all and only the items that satisfy $S$, which are exactly the items that satisfy $(0,S)$.

All other cases are analogous to those for $\MinEP/\MaxEP$, where $\min$ on the first component and
$\qall$ on the second one take the place of $\bigcap$, and $\max$ on the first component and
$\qany$ on the second one take the place of $\bigcup$.
We report a couple of cases as examples.

$\MinEI(\JObj{K_1,\ldots,K_n})=(\max_{i\in \SetTo{n}}(h_i), \JObj{\qany : [S'_1,\ldots,S'_n]})$
where $(h_i,S'_i)=\MinEI(S_i)$:
We want to prove that 
for any object $J$,  if $J$ satisfies $\JObj{K_1,\ldots,K_n}$, then every item of $J$ that satisfies the pair 
$\MinEI(\JObj{K_1,\ldots,K_n})$ is evaluated.
If an item $J_i$ satisfies the pair, then either its position is less than $\max_{i\in \SetTo{n}}(h_i)$,
or it satisfies $\JObj{\qany : [S'_1,\ldots,S'_n]}$.
In the first case, its position is less then $h_l$ for some $l$; in the second case, it satisfies $S'_l$ for some $l$;
hence, in both cases, $J_i$ satisfies a pair $(h_l,S'_l)$ for some $l$.
Since $(h_l,S'_l)=\MinEIK{K_l}$,
then, by induction, item $J_i$ is evaluated by the schema $\JObj{K_l}$.
If $K_l$ is an independent keyword, we conclude that $J_i$ is evaluated by the keyword $K_l$, hence 
is evaluated by the schema.
If $K_l$ is either $\qits$ or $\qunIts$, then we know that $J_i$ is either evaluated by $K_l$ or by a keyword that precedes $K_l$,
hence it is evaluated by the schema.

$\MaxEI(\JObj{K_1,\ldots,K_n})=(\max_{i\in \SetTo{n}}(h_i), \JObj{\qany : [S'_1,\ldots,S'_n]})$
where $(h_i,S'_i)=\MaxEI(S_i)$:
We want to prove that 
for any object $J$,  if an item $J_i$ of $J$ is evaluated by $\JObj{K_1,\ldots,K_n}$, then $J_i$ satisfies the pair 
$\MaxEI(\JObj{K_1,\ldots,K_n})$.
If $J_i$ is evaluated by $\JObj{K_1,\ldots,K_n}$ then $J_i$ is evaluated by at least
one keyword $K_l$. If $K_l$ is an independent item, then $J_i$ is also evaluated by
$\JObj{K_l}$, hence, by induction, $J_i$ satisfies $\MaxEIK{K_l}=(h_l,S'_l)$, hence, either its position is
less then $h_l$, or it satisfies $S'_l$. In the first case its position is less then $\max_{i\in \SetTo{n}}(h_i)$,
in the second case it satisfies $ \JObj{\qany : [S'_1,\ldots,S'_n]}$, hence it satisfies the pair.
If $K_l$ is $\qunProps:S$, then it is possible that $J_i$ is not evaluated by $\JObj{K_l}$ since $J_i$ may be
evaluated by a keyword that precedes $K_l$, but still $J_i$ satisfies $ \JObj{\qany : [S'_1,\ldots,S'_n]}$
since $\MaxEIK{K_l}=(\Inf,\xtrue)$, hence $S'_l=\xtrue$ (a second reason why $J_i$ satisfies the pair
is that  $h_l=\Inf$ hence the position of $J_i$ satisfies  $\max_{i\in \SetTo{n}}(h_i)$).
The same reasoning holds when $K_l$ is $\qits:S$.

$\MinEI(\qone: \JArr{S_1,\ldots,S_n}) =(\min_{i\in \SetTo{n}}(h_i), \JObj{\qall : [S'_1,\ldots,S'_n]})$
where $(h_i,S'_i)=\MinEI(S_i)$:
We want to prove that 
for any object $J$,  if $J$ satisfies $\JObj{(\qone: \JArr{S_1,\ldots,S_n})}$, then every item $J_i$
of $J$ that satisfies 
the pair $\MinEI(\qone: \JArr{S_1,\ldots,S_n})$ is evaluated.
If an item $J_i$ satisfies the pair, then either its position is less than $\min_{i\in \SetTo{n}}(h_i)$,
or it satisfies $\JObj{\qall : [S'_1,\ldots,S'_n]}$.
In the first case, its position is less then $h_l$ for every $l$; in the second case, it satisfies $S'_l$ for every $l$;
hence, in both cases, $J_i$ satisfies every pair $(h_l,S'_l)$, that is, it satisfies $\MinEI(S_l)$ for every $l$.
Since  $J$ satisfies $\JObj{(\qone: \JArr{S_1,\ldots,S_n})}$, then, $J$ satisfies $S_o$ for some $o$, hence,
by induction, every item of $J$ that satisfies $\MinEI(S_o)$ is evaluated, and we have just proved that $J_i$
satisfies every pair  $\MinEI(S_l)$, including  $\MinEI(S_o)$, hence $J_i$ is evaluated.

$\MaxEI(\qone: \JArr{S_1,\ldots,S_n}) =(\max_{i\in \SetTo{n}}(h_i), \JObj{\qany : [S'_1,\ldots,S'_n]})$
where $(h_i,S'_i)=\MaxEI(S_i)$:
We want to prove that 
for any object $J$,  if an item $J_i$ of $J$ is evaluated by $\qone: \JArr{S_1,\ldots,S_n}$, then $J_i$ satisfies the pair
$\MaxEI(\qone: \JArr{S_1,\ldots,S_n}) $.
If an item $J_i$ of $J$ is evaluated by a successful $\qone: \JArr{S_1,\ldots,S_n}$, then it is evaluated by the only
branch of $\qone$ that succeeds; let us call it $S_o$. 
By induction, item $J_i$ satisfies $\MaxEI(S_o)$, hence it either has a position that is smaller than $h_o$
or it satisfies $S'_o$. In the first case it satisfies the first component $\max_{i\in \SetTo{n}}(h_i)$,
in the second case it satisfies the second component $\JObj{\qany : [S'_1,\ldots,S'_n]}$.

\end{proof}

\hide{
The property specifies that $\MinEP(S)$ is a lower bound, but not that it is the best possible 
lower bound; for example, if a branch $S_i$ of $\qanyOf{S_1,\ldots,S_n}$ were unsatisfiable, we could exclude
that branch from the computation, and obtain, in general, a greater, hence more precise, lower bound,
and the same holds for the other three functions.
Our algorithm uses four functions for some crucial optimizations; stricter bounds produce more
effective optimizations, but there is a cost-efficacy trade-off.
Property~\ref{pro:minmax} ensures that our specific definition is sound, and our experiments are reassuring about the
cost-efficacy trade-off, but all the theory that we develop would still work with a different definition of this
function, provided it satisfies the soundness property.
}

When $\MinEP(S)$ and $\MaxEP(S)$ denote the same language, then
$\MinEP(S)$ (or, equivalently, $\MaxEP(S)$) statically characterizes the property evaluation of $S$,
and our algorithm can use this fact to produce an optimized (smaller) encoding
(as we will see in Section \ref{sec:computing}). 
However, deciding whether
$\MinEP(S)$ and $\MaxEP(S)$ are equivalent is computationally expensive, and therefore
different implementations can devote a different effort to this optimization.
In order to express this fact, we define a function $\ExEP_{eq}(S)$,
which is parametrized on the function $eq$ that is used to compare two sets of patterns,
so that  $\ExEP_{eq}(S)=\Set{p_1,\ldots,p_n}$ when 
$eq$ is able to prove that $\MinEP(S)$ and $\MaxEP(S)$ are both equivalent to some
$\Set{p_1,\ldots,p_n}$, and $\ExEP_{eq}(S)=\uparrow$ otherwise.
Parametrizing $\ExEP_{eq}$ over $eq$ allows us to discuss the effect of different choices for function~$eq$.

\begin{definition}[$\ExEP_{eq}(S)$, $\ExEI_{eq}(S)$, $\ExEP_{eq}(S)\Down/\Up$, $\ExEI_{eq}(S)\Down/\Up$]\label{def:eq}
Assume that $\key{eq}$ is a binary relation such that 
$$
\begin{array}{lllllllllllllll}
eq(\Set{p_1,\ldots,p_n},\Set{p'_1,\ldots,p'_m}) &\Implies& \rlan{\Set{p_1,\ldots,p_n}}=\rlan{\Set{p'_1,\ldots,p'_m}}\\[\NL]
eq((n,S),(m,S')) &\Implies& (n=m) \And (\forall J.\  (J \text{ satisfies } S) \Iff (J \text{ satisfies } S'))
\end{array}
$$
The functions $\ExEP_{eq}(S)$ and  $\ExEI_{eq}(S)$ are defined as follows.
$$
\begin{array}{rlllllllllllllll}
\Not eq(\MinEP(S),\MaxEP(S))   &\Implies& \ExEP_{eq}(S)&=&\uparrow   & &   \\[\NL]
 eq(\MinEP(S),\MaxEP(S))   &\Implies& \ExEP_{eq}(S)&=&\MinEP(S)   & &   \\[\NL]
\Not eq(\MinEI(S),\MaxEI(S))   &\Implies& \ExEI_{eq}(S)&=&\uparrow   & &   \\[\NL]
 eq(\MinEI(S),\MaxEI(S))   &\Implies& \ExEI_{eq}(S)&=&\MinEI(S)   & &   
\end{array}
$$
We use $\ExEP_{eq}(S)\Down$ to indicate that $\ExEP_{eq}(S)\neq\,\uparrow$,
and similarly for $\ExEI_{eq}(S)\Down$.
We use $\ExEP_{eq}(S)\Up$ to indicate that $\ExEP_{eq}(S)=\uparrow$,
and similarly for $\ExEI_{eq}(S)\Up$.
\end{definition}

For any $eq$ that enjoys the properties of Definition \ref{def:eq}, the following property holds.

\begin{propxrep}[$\ExEP_{eq}(S)$, $\ExEI_{eq}(S)$, characterize property/item evaluation]\label{pro:ex}
For any schema $S$, if $\ExEP_{eq}(S)\Down$, then $\ExEP_{eq}(S)$ characterizes the property evaluation of $S$.
For any schema $S$, if $\ExEI_{eq}(S)\Down$, then $\ExEI_{eq}(S)$ characterizes the item evaluation of $S$.
\end{propxrep}

From now on, we assume that the function $eq$ is fixed and that it includes at least syntactical equality, so that
$eq(P,P)$ and $eq((n,S),(n,S))$ hold.

\begin{example}
Consider the following schema.
\begin{Verbatim}[fontsize=\small,frame=lines]
{ "anyOf": [ 
    { "properties": { "a": { "type":"string" } }, "additionalProperties":false },
    { "properties": { "b": { "type":"string" } }, "additionalProperties":false }
  ]
}
\end{Verbatim}

\newcommand{\KeyKey}[1]{\qkw{\PPP{#1}}}
The first branch $S_1$ of $\qany$ has $\MinEP(S_1)=\MaxEP(S_1)=\ExEP_{eq}(S_1)=\Set{\KeyKey{a},\qkw{.*}}$;
the second branch $S_2$ has $\MinEP(S_1)=\MaxEP(S_1)=\ExEP_{eq}(S_1)=\Set{\KeyKey{b},\qkw{.*}}$.
Our algorithm may rewrite both of them as just $\Set{\qkw{.*}}$, but let us assume it does not.
Then, we have $\MinEP(\qany:[S_1,S_2])=\Set{\KeyKey{a},\qkw{.*}}\cap\Set{\KeyKey{b},\qkw{.*}}=\Set{\qkw{.*}}$,
and $\MaxEP(\qany:[S_1,S_2])=\Set{\KeyKey{a},\qkw{.*}}\cup\Set{\KeyKey{b},\qkw{.*}}=\Set{\KeyKey{a},\KeyKey{b},\qkw{.*}}$.

Now, the value of $\ExEP_{eq}(\qany:[S_1,S_2])$ depends on the function $eq$.
If $eq$ is just set equality, that we indicate as $=$, then we have that $\neg eq(\MinEP(\qany:[\ldots]),\MaxEP(\qany:[\ldots]))$,
hence $\ExEP_{=}(\qany:[S_1,S_2])\Up$.
The function $eq$ that we have implemented, however, regards any two sets that contain $\qkw{.*}$ as equivalent, ignoring any other pattern
in this case; let us call this function $eq^{+}$; in this case, $\ExEP_{eq^{+}}(\qany:[S_1,S_2])=\Set{\qkw{.*}}$.
\end{example}

We now show that $\ExEP_{eq}(S)$ is undefined only when $S$ depends on either an $\qany$ or a $\qone$ keyword.
We first define the notion of \emph{\AAF}.

\begin{definition}[$K$ {\AAF}  $K'$]
For a fixed schema $S$, we say an occurrence of a subkeyword $K$ is {\AAF} an occurrence a subkeyword $K'$ if there is a path
from $K$ to $K'$ that only crosses boolean keywords or references.
Formally,  $K$ is immediately {\AAF} $K'$ if either (1) $K$ is a boolean keyword $\qall/\qany/\qone:[S_1,\ldots,S_k]$, and
$K'$ is a top-level keyword in one of the $S_i$ arguments of $K$, or (2) $K$ is a reference keyword $\qdref: u$, and
$K'$ is a top-level keyword in $\key{deref}(u)$.
 $K$ is {\AAF} $K'$ if there exists a sequence $K_1,\ldots,K_n$ of keyword occurrences such that either $n=1$, or,
 for each $i\in\SetTo{n-1}$, $K$ is immediately {\AAF} $K'$.
 An occurrence of a subschema $S'$ of $S$ is {\AAF} $K'$ if one of the top-level keywords of $S'$ is {\AAF}
  $K'$.
\end{definition}

The following property implies that  $\ExEP_{eq}(S)$ is always well defined unless $S$ is {\AAF}
an $\qany$ keyword or a $\qone$ keyword that evaluates different properties in its different branches,
and similarly for  $\ExEI_{eq}$.

\begin{propxrep}[ $\ExEP_{eq}$,  $\ExEI_{eq}$, and $\qany$/$\qone$]\label{pro:sta}
$\MinEP(S)=\MaxEP(S)$ holds  for any schema $S$ which is not {\AAF}
any $\qany$ or $\qone$
keyword $K'$ such that $\MinEP(\JObj{K'})\neq\MaxEP(\JObj{K'})$.
The same property holds if we substitute $\MinEP/\MaxEP$ with $\MinEI/\MaxEI$.
\end{propxrep}

\begin{proof}
To follow the fine details of the proof,
please remember that $\MinEP$ is always applied to schemas, such as $\JObj{K}$, and
never directly to keywords, such as $K$, which forces us to add/remove curly brackets 
in some points of the proof.

Assume that $S$ is a schema that is not {\AAF} any $\qany$ or $\qone$
keyword $K'$ with $\MinEPK{K'}\neq\MaxEPK{K'}$.
For any line in Table \ref{tab:EP} we show that $\MinEP(S)$ and $\MinEP(S)$ are equal,
by induction on the in-place depth of $S$. 

All lines different from $\qone$, $\qany$, $\qdref$, $\qall$, $\JObj{K_1,\ldots,K_n}$ are immediate
since they have $\MinEP=\MaxEP$.

In case $S=\qallOf{K_1,\ldots,K_n}$, since  $S$ is a schema that is not {\AAF} any $\qany$ or $\qone$
keyword $K'$ with $\MinEPK{K'}\neq\MaxEPK{K'}$, the same property holds for
$\JObj{K_1},\ldots,\JObj{K_2}$. Hence all these schemas enjoy $\MinEP(\JObj{K_i})=\MaxEP(\JObj{K_i})$,
hence we have $\bigcup_{i\in\SetTo{n}}\MinEP(\JObj{K_i})=\bigcup_{i\in\SetTo{n}}\MaxEP(\JObj{K_i})$,
i.e., $\MinEP(S)=\MaxEP(S)$.
The same proof holds for $\qall$.
In case $S=\JObj{\qdref:u}$,  since  $S$ is a schema that is not {\AAF} any $\qany$ or $\qone$
keyword $K'$ with $\MinEPK{K'}\neq\MaxEPK{K'}$, the same property holds for $\key{deref}(u)$.
By induction on the in-place depth, we have that $\MinEP(\key{deref}(u))=\MaxEP(\key{deref}(u))$,
hence $\MinEP(S)=\MaxEP(S)$.

In case $S=\JObj{K_a}=\qanyOf{S_1,\ldots,S_n}$, if we had $\MinEP(\JObj{K_a})\neq\MaxEP(\JObj{K_a})$
then we would violate the hypothesis that $S$ is not {\AAF} any $\qany$ or $\qone$
keyword $K'$ with $\MinEPK{K'}\neq\MaxEPK{K'}$, hence we must have $\MinEP(\JObj{K_a})=\MaxEP(\JObj{K_a})$,
i.e., $\MinEP(S)=\MaxEP(S)$.
The same proof holds for $\qone$.

The same proof presented for $\ExEP_{eq}$ holds for  $\ExEI_{eq}$.
\end{proof}

\subsection{Pair covering and Evaluation Normal Form}\label{sec:pcenf}

We now describe our generalization of the XDNF and our algorithm for the elimination
of {\qunStar}.

Consider the following schema.

\begin{Verbatim}[fontsize=\small,frame=lines]
{ "anyOf": [
    { "properties": { "address":{} } },
    { "properties": { "model":{} } }
  ],
  "unevaluatedProperties": false
}
\end{Verbatim}

As we have seen, an object with an address and a model is validated by this schema, since it satisfies both branches,
and the address is evaluated by the first branch and the model by the second one:
the two branches cooperate to evaluate all the fields that must be evaluated in order to 
bypass the {\qunStar} constraint. Hence, we cannot push
{\qunProps: \afalse} through the branches of {\qany}, since, after pushing, the different branches cannot cooperate any more, and
hence the object with both fields would not be validated by the new schema.
Consider now the following schema (that is equivalent):
\begin{Verbatim}[fontsize=\small,frame=lines]
{ "anyOf": [
    { "properties": { "address":{} } },
    { "properties": { "model":{} } }
    { "properties": { "address":{}, "model":{} } },
  ],
  "unevaluatedProperties": false
}
\end{Verbatim}

It is not in XDNF, but, if we push {\qunProps: \afalse} through the branches of {\qany}, any $J$ that was accepted before is still accepted.
Consider the schema after pushing.
\begin{Verbatim}[fontsize=\small,frame=lines]
{ "anyOf": [
   { "properties": { "address":{} }, "unevaluatedProperties": false },
   { "properties": { "model":{} }, "unevaluatedProperties": false },
   { "properties": { "address":{}, "model":{} }, "unevaluatedProperties": false }
  ]
}
\end{Verbatim}

An object with both fields is still accepted by the third branch.
This time, it was possible to push {\qunProps: \afalse} through {\qany} because, for any two branches $S_1$ and
$S_2$ that can cooperate to evaluate the fields of an instance $J$,
there is a branch $S_c$ that ``\Cs'' them, i.e., that, if applied to a $J$ that satisfies both $S_1$ and $S_2$, 
then $S_c$ is satisfied, and it
evaluates all fields that are evaluated either by $S_1$ or by $S_2$.
This condition is qualified by ``if applied to a $J$ that satisfies both $S_1$ and $S_2$'', hence it 
holds trivially when $S_1$ and $S_2$ are disjoint, so that, when $S_1$ and $S_2$ are disjoint,
then any schema ($S_1$, for example) {\Cs} the two, but we
will see that there are many other situations when two schemas are {\Cd} by one of the two, or by a third schema.
This means that reaching a disjunctive form where all branches are {\Cd} by a branch in the disjunction is much easier than
reaching an XDNF; this is the basis of our algorithm.

We now formally define the notion of {\PCs} and the corresponding notion of {\ICs}, used to analyze array evaluation.

\begin{definition}[$S_c\ \PCs (S_1,S_2), S_c\ \ICs (S_1,S_2)$]
We say that $S_c\ \PCs (S_1,S_2)$ iff, for any $J$ that satisfies both $S_1$ and $S_2$:
\begin{compactenum}
\item $J$ satisfies $S_c$;
\item every property of $J$ that is evaluated by $S_1$ is also evaluated by $S_c$;
\item every property of $J$ that is evaluated by $S_2$ is also evaluated by $S_c$.
\end{compactenum}
The definition of $S_c\ \ICs (S_1,S_2)$ is obtained by substituting ``property'' with ``item''.
\end{definition}

In an XDNF all branches are mutually exclusive, and we generalize that with cover-closure.
A disjunction is cover-closed if, for each pair of schemas,
it contains a schema that {\PCs} both and a schema that {\ICs} both. For simplicity, from now on we only focus on
the {\PCs} relation, but everything that we say can be immediately transferred to the {\ICs} relation.

For any two schemas $S_1$ and $S_2$, the schema $\JObj{\qall : [ S_1, S_2]}$ {\PCs} the pair,
so we can make a disjunction cover-closed by adding that schema to the list of the disjunction arguments,
but we have many interesting special cases when one of the two schemas covers both, so that we have
cover-closure at no cost.
For example, when two schemas $S_1$ and $S_2$ are mutually exclusive, then any schema (such as $S_1$ itself)
trivially {\PCs} both, so that cover-closure generalizes exclusiveness.
Moreover,
it is easy to check that, if $\MinEP(S_1)\supseteq \MaxEP(S_2)$, then $S_1$ {\PCs} $(S_1,S_2)$, so that, as a consequence:
\begin{compactenum}
\item if two schemas $S_1$ and $S_2$ are characterized by the same set of patterns, then each of them {\PCs} the pair
        $(S_1,S_2)$;
\item one schema $S_1$ that evaluates all properties (since it contains $\qunProps$, for example) {\PCs} any pair $(S_1,S_2)$
   formed with another arbitrary schema $S_2$;
\item given a schema $S_2$ that evaluates no properties, any arbitrary schema $S_1$ {\PCs} $(S_1,S_2)$.
\end{compactenum}

Hereafter, given a list of schemas $L=\List{S_1,\ldots,S_n}$, we use $\qany : L$
to indicate the keyword $\qany: [S_1,\ldots,S_n]$, and similarly for $\qone$ and $\qall$.

In order to eliminate $\qunStar$ operators from a disjunction, we must (1) push the operators
through the branches and (2) in any branch, rewrite the pushed operator using its ``annotation independent''
counterpart $\qaddProps$ or $\qits$.
An XDNF allows for (1) because all arguments are pairwise disjoint and (2) because
all arguments are conjunctions of structural keywords, which allows us to compute which fields/items
are evaluated.
We generalize the idea by defining the ENF, which allows (1) because all arguments are {\Cd} by another one 
and (2) because
all arguments are statically characterized.

This is its formal definition.

\begin{definition}[{evaluation normal form} --- ENF]\label{def:ENF}
We say that a list of schemas $L=\List{S_1,\ldots,S_n}$ is cover-closed  when every
pair of schemas $(S_i,S_j)$ of $L$ is {\Cd} by a schema in $L$ (hence, the empty list and
every singleton are cover-closed).
We say that a list of schemas $L=\List{S_1,\ldots,S_n}$ is \emph{{\ESC}}  when every
$S_i$ is statically characterized.
We say that a keyword $\qany: L$ or a schema $\JObj{\qany: L}$ is cover-closed (or {\ESC}) when the
list $L$ is cover-closed
(or, respectively, {\ESC}).

We say that a schema is in \emph{evaluation normal form} ENF when:
\begin{compactenum}
\item it is a $\qany$ single-keyword schema, that is, it is equal to $\JObj{\qany:L}$ for some $L$;
\item the list $L$ is cover-closed;
\item the list $L$ is  {\ESC}.
\end{compactenum}
\end{definition}

Every schema $\JObj{\qany: [S_1,\ldots,S_n]}$ in XDNF is also in ENF, since every two schemas
$S_i$, $S_j$ that are disjoint are {\PCd} by $S_i$ (and by $S_j$), and since every product
$S_i$ is statically-characterized, by Property \ref{pro:sta}, since $S_i$ is not {\AAF} any $\qany$ or $\qone$.

A schema $\JObj{\qany: [S_1,\ldots,S_n]}$ in ENF is not necessarily in XDNF, nor even in DNF:
for example, each of the $S_i$ may contain any nesting of boolean operators and of references, as long as it is 
 statically-characterized, for example because it has shape $\qnotOf{S}$; hence, the ENF is much less
 restrictive than the XDNF.
 
 \subsection{Computing the ENF}\label{sec:computing}

An arbitrary schema can be transformed in ENF using the algorithm of Definition~\ref{def:funENF};
from now on, we assume that $eq$ is fixed, hence we write  $\ExEP$ rather than $\ExEP_{eq}$.

The first two lines specify that the function $\ENF(S)$ is, essentially, the identity function when~$S$ is statically characterized
(case $\ExEP(S)\Down$).
These first two lines are crucial in keeping the size of $\ENF(S)$ similar to the size of $S$.

By Property \ref{pro:sta}, for any keyword $K$ that is different from $\qall$, $\qany$, $\qone$, or $\qdref$,
we have that $\ExEP_{eq}(\JObj{K})\Down$ even with the simplest $eq$,
hence, the only cases where we can have $\ExEP(S)\Up$ are those of multi-keyword schemas,
which we immediately reduce to single-keyword schemas (case 3 in the definition),
and single-keyword schemas whose keyword is either $\qall$, $\qany$, $\qone$, or $\qdref$,
and where $\neg eq(\MinEP(\JObj{K}),\MaxEP(\JObj{K}))$ (cases 4 to 7).

Observe that also these in-place keywords may satisfy  $\ExEP_{eq}(\JObj{K})\Down$. 
For example, when~$K$ is $\qone{\Col}A$ or $\qany{\Col}A$, we have $\ExEP_{eq}(\JObj{K})\Down$ 
as soon as $eq(\MinEP(\JObj{K}),\MaxEP(\JObj{K}))$.
As another example,  when $K$ is $\qall{\Col}A$ or $\qdref{\Col}u$, we have
$\ExEP_{eq}(\JObj{K})\Down$ when they are not {\AAF}
any $\qone{\Col}A$ or $\qany{\Col}A$.
In all these cases, the trivial rules of case $\ExEP(S)\Down$ are applied (cases 1 and 2).

We now analyze all the lines in the table.

When $S = \qallOf{S_1,\ldots,S_m}$, the algorithm
first normalizes all of the subschemas $S_i$, hence obtaining a conjunction
$\JObj{\qall : [ \JObj{\qany: L_1},\ldots,\JObj{\qany: L_m}]}$
where every $L_i$ is cover-closed. Then, it uses an auxiliary function $\LAnd$ with the following properties
(we use $S_1\sim S_2$ to indicate that the two schemas validate the same instances):
\begin{compactenum}
\item  $\JObj{\qany : \LAnd(\List{L_1,\ldots,L_m})} \sim \JObj{\qall : [ \JObj{\qany: L_1},\ldots,\JObj{\qany: L_m}]}$;
\item if every $L_i$ is cover-closed, then $\LAnd(L_1,\ldots,L_m)$ is cover-closed.
\end{compactenum}
The auxiliary function is defined in Definition~\ref{def:andor};
it just computes a sort of ``Cartesian product'' of the lists $L_1,\ldots,L_m$ using the recursive function
$\LAnd$, and we will prove that it enjoys the two desired properties.
Observe that $\LAnd(\Cal{L})$  transforms a list of lists of schemas ($\Cal{L}=\List{L_1,\ldots,L_m}$) into a flat list of schemas ($L$).

A multi-keywords schema $\JObj{K_1,\ldots,K_n}$ is treated as $\qallOf{\JObj{K_1},\ldots,\JObj{K_n}}$.

When $S$ is a single-keyword schema $\JObj{\qany:{[S_1,\ldots,S_m]}}$, the algorithm
first normalizes all subschemas, obtaining a disjunction
$\JObj{\qany : [ \JObj{\qany: L_1},\ldots,\JObj{\qany: L_m}]}$.
Then, it uses an auxiliary function $\LOr$ with the property that
 \begin{compactenum}
\item  $\JObj{\qany : \LOr(\List{L_1,\ldots,L_m}} \sim \JObj{\qany : [ \JObj{\qany: L_1},\ldots,\JObj{\qany: L_m}]}$;
\item if every $L_i$ is cover-closed, then $\LOr(L_1,\ldots,L_m)$ is cover-closed.
\end{compactenum}
$\LOr(\Cal{L})$ is a recursive function that transforms a list of lists of schemas into a flat list of schemas using the auxiliary
function $\LClose(L',L'')$, which computes a list of schemas that is big enough to ensure that
$L'\LC L''\LC \LClose(L',L'')$ is cover-closed.

$\LClose(L',L'')$ is defined by a lower bound and an upper bound.
The upper bound just collects, for every $S'\in L'$ and $S''\in L''$, their conjunction $\JObj{\qall: [ S',S'']}$, 
which ensures that the two are {\Cd} in the result.
The lower bound specifies that an implementation is allowed to omit a conjunction
from the result when the pair is {\Cd} by a schema in $L'\LC L''$;
any implementation is free to decide how aggressively to aim for the lower bound, which yields a smaller ENF
but requires a bigger computational effort.\footnote{
In our implementation, we just check whether $(S',S'')$ is {\Cd} by either $S'$ or by $S''$ 
because $\MinEP(S_1)\supseteq \MaxEP(S_2)$ or $\MinEP(S_2)\supseteq \MaxEP(S_1)$
(as discussed in Section \ref{sec:pcenf}), 
otherwise we ignore the optimization possibility and insert $\JObj{\qall: [ S',S'']}$ into $\LClose(L',L'')$.}

When $S = \qoneOf{S_1,\ldots,S_m}$, the algorithm first rewrites $S$ as a disjunction of
$m$ conjunctions 
$\JObj{\qall:[\ \qnotOf{S_1},\ldots,\qnotOf{S_{i-1}},\ S_i,\ \qnotOf{S_{i+1}},\ldots,\qnotOf{S_{m}}\ ]}$, 
accordingly with the boolean semantics of this operator.
Then, it normalizes each conjunction,
hence obtaining a disjunction
$\qany : [ \JObj{\qany: L_1},\ldots,\JObj{\qany: L_m}]$. At this point, each $L_i$ is cover-closed and no $J$ may satisfy both
a schema in $L_i$  and a schema in $L_j$ with $i\neq j$, hence the list 
$ L_1\LC\ldots\LC L_m$ is cover-closed, hence 
 $\qany : [ \JObj{\qany: L_1},\ldots,\JObj{\qany: L_m}]$ can be rewritten
 as
 $\qany : L_1\LC\ldots\LC L_m$ (see Property \ref{pro:ENFcorrect} for a formal proof).

Finally, when $S$ is a reference, the reference is substituted by its definition, which is recursively normalized.
This recursive step can never generate an infinite loop,
thanks to the guarded recursion constraint,
since normalization only expands boolean operators and 
references but never crosses the structural operators.\footnote{In the implementation, the definition $\key{deref}(u)$ of 
each reference $\fdref:u$ is normalized the first time the reference is met, 
and the result is
memorized to be reused when needed, so that $\ENF(\key{deref}(u))$ is only computed once even if there are many occurrences of $\fdref:u$.}

\begin{definition}[$\LAnd(\Cal{L}), \LClose(L',L''), \LOr(\Cal{L})$]\label{def:andor}
$$\begin{array}{llllllll}
\LAnd(\List{}) &=\!\!\!\!\!& \List{\xtrue} \\[\NL]
\LAnd(L :: \Cal{L}) &=\!\!\!\!\!& \ListST{\JObj{\qall: [ S',S'']}}{ S'\!\in\! L,\ S'' \!\in\! \LAnd(\Cal{L})} \\[\NL]
\LClose(L',L'') \!\!\!\!\!&\subseteq\!\!\!\!\!& \ListST{\JObj{\qall: [ S',S'']}}{ S'\!\in\! L',\ S'' \!\in\! L'' } \\[\NL]
\LClose(L',L'') \!\!\!\!\!&\supseteq\!\!\!\!\!& \ListST{\JObj{\qall: [ S',S'']}}{ S'\!\in\! L',\ S'' \!\in\! L'',\ \not \exists S\!\in\!(L'\LC L'').\ S\ \Cs\ (S',S'') } \\[\NL]
\LOr(\List{}) & =\!\!\! \!\!& \List{} \\[\NL]
\LOr(L:: \Cal{L}) & =\!\!\!\!\! & L \LC \LOr(\Cal{L}) \LC \LClose(L,\LOr(\Cal{L})) \\[\NL]
\end{array}$$
\end{definition}

\begin{definition}[$\ENF(S)$]\label{def:funENF}

Function $\ENF(S)$ is defined as follows:

\vspace{10pt}
\begin{tabular}{|l|l|l|l|c|}

\multicolumn{4}{l}{\textbf{If $\ExEP(S)\Down$   and  $\ExEI(S)\Down$} \ \ \ \textbf{{(e.g., $\qnotOf{S}$, or $\JObj{\qprops: S}$, or \ldots)}}:} \\[\NL] \hline
1&$\ENF(S)$ & =& $S   \qquad\qquad\qquad\qquad\ \mbox{\ \ if }S = \JObj{\qany:L}$\\
\hline
2&$\ENF(S)$ & =& $\JObj{\qany: [S]}  \qquad\mbox{\ \ if }S \neq  \JObj{\qany:L}$\\
\hline 
\multicolumn{4}{l}{}\\
\multicolumn{4}{l}{
\textbf{If $\ExEP(S)\Up$  or  $\ExEI(S)\Up$ \ \   {($\JObj{K_1,\ldots,K_m}$, $\JObj{\qany:L}$, $\JObj{\qone:L}$, $\JObj{\qall:L}$, $\JObj{\qdref:u}$)}:}}\\[\NL] \hline
3&$\ENF(\JObj{K_1,\ldots,K_m} )$\ with $m>1$ &=&
                                                $\JObj{\qany : \LAnd(\List{L_1,\ldots,L_m})}$  \\
                                                &&&$\key{where}\ \ENF(\JObj{K_i}) = \JObj{\qany : L_i}$ \ \ 
                                                 $\key{for}\ i\in\SetTo{m}$\\ \hline
4&$\ENF(\JObj{\qall : [ S_1,\ldots, S_m]} )$ &=&
                                                $\JObj{\qany : \LAnd(\List{L_1,\ldots,L_m})}$  \\
                                               &&&$\key{where}\ \ENF(S_i) = \JObj{\qany : L_i}$ \ \ 
                                                 $\key{for}\ i\in\SetTo{m}$\\ \hline
5&$\ENF(\JObj{ \qany : [ S_1,\ldots, S_m]} )$ &=&
                                                $\JObj{\qany : \LOr(\List{L_1,\ldots,L_m})} $ \\
                                                &&&$\key{where}\ \ENF(S_i) = \JObj{\qany : L_i}$ \ \ 
                                                 $\key{for}\ i\in\SetTo{m}$\\ \hline
6&$\ENF(\JObj{ \qone : [ S_1,\ldots, S_m]} )$ &=&
                                                $\JObj{\qany : L_1\LC\ldots\LC L_m}$  \\
                                              &&&$\key{where}\ \ENF(\{\qall:[\qnotOf{S_1},\ldots,\qnotOf{S_{i-1}},S_i,$\\[\NL]
                                              &&&$\qquad\qnotOf{S_{i+1}},\ldots,\qnotOf{S_{m}}]\}) = \JObj{\qany : L_i}$ \ \ \\[\NL]
                                               &&& $\key{for}\ i\in\SetTo{m}$\\ \hline
7&$\ENF(\JObj{\qdref : u})$ &=&  $\ENF(\key{deref}(u))$ \\ \hline
\end{tabular}
\end{definition}

\vspace{10pt}
The function $\ENF(S)$ is well-defined; this is essentially a consequence of the guarded recursion constraint.

\begin{propxrep}\label{pro:welldefined}
The function $\ENF(S)$ is well-defined.
\end{propxrep}

\begin{proofsketch}
In cases 3 to 7, $\ENF(S)$ is defined in terms of a set of applications of $\ENF$ to a set of schemas $S_i$
(just one in case 7),
but in all of these cases the in-place depth (Definition \ref{def:ipd}) of each $S_i$ is
smaller than the in-place depth of $S$; in the case for $\qone$ this happens thanks to the use of 
$\key{booleanOneOf}$ in the definition of in-place
depth for $\qone$.
\end{proofsketch}

We now prove that $\LAnd$ and $\LOr$ enjoy the properties that we promised in the text.

\begin{propxrep}[$\LAnd(\Cal{L}), \LOr(\Cal{L})$]\label{pro:andor}
For any list of lists of schemas $\Cal{L}$ such that every list $L\in\Cal{L}$ is cover-closed
and is {\ESC}, and for any two lists
$L'$ and $L''$ that are  cover-closed and {\ESC}:
\begin{compactenum}
\item \label{andor:i} $\LAnd(\Cal{L})$ is cover-closed and is {\ESC};
\item \label{andor:ii} $L' \LC L'' \LC \LClose(L',L'') $ is cover-closed and is {\ESC};
\item \label{andor:iii} $\LOr(\Cal{L})$ is cover-closed and is {\ESC};
\item \label{andor:iv} $\qany : \LAnd(L_1,\ldots,L_n)$ is equivalent to $\qall : [ \JObj{\qany: L_1},\ldots,\JObj{\qany: L_n}]$;
\item \label{andor:v} $\qany: L' \LC L'' \LC \LClose(L',L'') $ is equivalent to $\qany: L' \LC L''  $;
\item \label{andor:vi} $\qany : \LOr(L_1,\ldots,L_n)$ is equivalent to $\qany : [ \JObj{\qany: L_1},\ldots,\JObj{\qany: L_n}]$;
\item \label{andor:vii} If each list $L_i$ is cover-closed and is {\ESC}, and, for any $i\neq j$, $\JObj{\qany:{L_i}}$ and $\JObj{\qany:{L_j}}$ 
    are disjoint, then $L_1 \LC \ldots \LC L_n$ is cover-closed and is {\ESC}.
\end{compactenum}
\end{propxrep}

\begin{appendixproof}
All lists that are described in points \ref{andor:i}, \ref{andor:ii}, \ref{andor:iii}, and \ref{andor:vii},are built by choosing elements from lists that are {\ESC}
and by adding elements with shape $\qallOf{S_1,S_2}$ where $S_1$ and $S_2$ are statically characterized; all such elements
are hence statically characterized, hence, all these lists are {\ESC}.

(\ref{andor:i}) $\LAnd(\Cal{L})$ is cover-closed: by induction.
Case $\LAnd(\List{})$: holds since every singleton is cover-closed.
Case $\LAnd(L :: \Cal{L}) =\ListST{\JObj{\qall: [ S',S'']}}{ S'\in L,\ S'' \in \LAnd(\Cal{L})}$.
Assume $S_1$ and $S_2$ belong to 
$\ListST{\JObj{\qall: [ S',S'']}}{ S'\in L,\ S'' \in \LAnd(\Cal{L})}$; then, $S_1=\JObj{\qall: [ S'_1,S''_1]}$
and $S_2=\JObj{\qall: [ S'_2,S''_2]}$ with $S'_1\in L$, $S''_1\in L$, $S'_2\in \LAnd(\Cal{L})$, $S''_2\in \LAnd(\Cal{L})$.
By hypothesis, there exist $S'_3\in L$ that covers $(S'_1,S'_2)$, and $S''_3\in \LAnd(\Cal{L})$ that covers $(S''_1,S''_2)$.
The schema $S_3=\JObj{\qall: [ S'_3,S''_3]}$ belongs to $\LAnd(L :: \Cal{L})$ by construction. If $J$ satisfies both $S_1$ and
$S_2$, then it satisfies all of $S'_1,S''_1, S'_2,S''_2$, hence it satisfies both $S'_1,S'_2$ and $S''_1,S''_2$, hence it satisfies
both $S'_3$ and $S''_3$, hence is satisfies $S_3$. The successful evaluation of $S_3$ evaluates all properties and items evaluated by
$S'_3$ and by $S''_3$, hence it evaluates all items evaluated by $S_1=\JObj{\qall: [ S'_1,S''_1]}$ and those evaluated by 
$S_2=\JObj{\qall: [ S'_2,S''_2]}$.

(\ref{andor:ii}) $L' \LC L'' \LC \LClose(L',L'')$ is cover-closed:
every two elements of $L'$ or of $L''$ have a cover in $L'$ or in $L''$, by construction.
Any two elements $S'\in L'$ and $S''\in L''$ are either {\Cd} by an element of $L'\LC L''$ or by an 
element $\JObj{\qall: [ S',S'']}$ in $\LClose(L',L'')$, by the lower-bound condition of $\LClose$.
 Consider now $S_1\in L'$ and $S_2\in\LClose(L',L'')$, where $S_2= \JObj{\qall: [ S',S'']}$,
 with $S'\in L'$ and $S''\in L''$.
 $S_1$ and $S'$ are {\Cd} by an element $S'_1$ of $L_1$; $S'_1$ and $S''$, by construction,
 are either {\Cd} by an element of $L'\LC L''$ or by an 
element $\JObj{\qall: [ S'_1,S'']}$ in $\LClose(L',L'')$.
The same proof holds for $S_1\in L''$ and $S_2\in\LClose(L',L'')$.
Finally, consider two elements of $\LClose(L',L'')$; by the upper-bound condition they can be written as
$\JObj{\qall: [ S'_1,S''_1]}$ and $\JObj{\qall: [ S'_2,S''_2]}$.
The pair $(S'_1,S'_2)$ is {\Cd} by $S'_3$ in $L'$, and the pair $(S''_1,S''_2)$  is {\Cd} by $S''_3$ in $L''$.
The pair $(S'_3,S''_3)$ is either {\Cd} by  an element of $L'\LC L''$ or by an 
element $\JObj{\qall: [ S'_3,S''_3]}$ in $\LClose(L',L'')$. This element is satisfied by any $J$ that satisfies 
$\JObj{\qall: [ S'_1,S''_1]}$ and $\JObj{\qall: [ S'_2,S''_2]}$, and it evaluates all properties and items that are evaluated
by these two terms.

(\ref{andor:iii}) $\LOr(\Cal{L})$ is cover-closed: by induction.
Case $C(\List{})$: holds since the empty list is cover-closed.
Case $\LOr(L:: \Cal{L}) = L \LC \LOr(\Cal{L}) \LC \LClose(L ,\LOr(\Cal{L}))$.
The list $\LOr(\Cal{L})$ is cover-closed by induction, and the thesis follows from the closure of 
$L' \LC L'' \LC \LClose(L',L'') $  (point \ref{andor:ii}).

(\ref{andor:iv}) $\qany : \LAnd(L_1,\ldots,L_n)$, is equivalent to $\qall : [ \JObj{\qany: L_1},\ldots,\JObj{\qany: L_n}]$:
by induction on $n$.

Case $n=0$ is trivial, since both sides are satisfied by any $J$.

Case $n+1$: we want to prove that 
$\qany :  \LAnd(L_1,\ldots,L_{n+1})$, is equivalent to\\
$\qall : [ \JObj{\qany: L_1},\ldots,\JObj{\qany: L_{n+1}}]$:\\
We use $\Cal{L}$ to indicate $\List{L_2,\ldots,L_{n+1}}$, and we have:\\
$\LAnd(L_1,\ldots,L_{n+1}) = \ListST{\JObj{\qall: [ S',S'']}}{ S'\in L,\ S'' \in \LAnd(\Cal{L})}$, hence $J$ satisfies\\
$\qany :  \LAnd(L_1,\ldots,L_{n+1})$ iff \\
$\exists S'\in L,\ S'' \in \LAnd(\Cal{L}).\ J$ satisfies ${S'}{}$, $J$ satisfies ${S'}{}$, hence\\
$\qany :  \LAnd(L_1,\ldots,L_{n+1})$ is equivalent to\\
$\qall : [ \JObj{\qany: L_1} ,    \JObj{\qany: \LAnd(\Cal{L})}         ]$, equivalent, by induction, to:\\
$\qall : [ \JObj{\qany: L_1}, \JObj{\qall : [ \JObj{\qany: L_2},\ldots,\JObj{\qany: L_{n+1}}]}      ]$, equivalent, by associativity, to:\\
$\qall : [ \JObj{\qany: L_1}, \JObj{\qany: L_2},\ldots,\JObj{\qany: L_{n+1}}      ]$, qed.

(\ref{andor:v}) $\qany: L' \LC L'' \LC \LClose(L',L'') $ is equivalent to $\qany: L' \LC L''  $. 
Every $J$ that satisfies the right hand side is satisfied by the corresponding element of the left hand side.
In the other directions, every $J$ that satisfies an element of $L'$ or $L''$ satisfies the corresponding
element of the right hand side. If $J$ satisfies an element $\JObj{\qall: [ S',S'']}$ of $\LClose(L',L'') $,
then it satisfies both $S'$ in $L'$ and $S''$ in $L''$, hence it satisfies the left hand side.

(\ref{andor:vi}) $\qany : \LOr(L_1,\ldots,L_{n})$, is equivalent to $\qany : [ \JObj{\qany: L_1},\ldots,\JObj{\qany: L_{n}}]$:
by induction on $n$.

Case $n=0$ is trivial, since both sides are equal to $\qany : []$.

Case $n+1$: we want to prove that 
$\qany : \LOr(L_1,\ldots,L_{n+1})$, is equivalent to\\
$\qany : [ \JObj{\qany: L_1},\ldots,\JObj{\qany: L_{n+1}}]$.

We let $\Cal{L}=\List{L_2,\ldots,L_n}$\\
$\qany : \LOr(L_1,\ldots,L_{n+1})$ is, by definition:\\
$\qany : (L_1 \LC \LOr(\Cal{L}) \LC \LClose(L_1,\LOr(\Cal{L})))$; 
by the property (\ref{andor:v}), is equivalent to:\\
$\qany : (L_1 \LC \LOr(\Cal{L}) )$; by associativity:\\ 
$\qany : (L_1 \LC \JObj{ \qany: \LOr(\Cal{L})})$; by induction:\\
$\qany : (L_1 \LC \JObj{\qany : [ \JObj{\qany: L_2},\ldots,\JObj{\qany: L_{n+1}}]}$; we conclude by associativity.

(\ref{andor:vii})  If each list $L_i$ is cover-closed and, for any $i\neq j$, $\JObj{\qany:{L_i}}$ and $\JObj{\qany:{L_j}}$ are disjoint,
            then $L_1 \LC \ldots \LC L_n$ is cover-closed.

The proof is immediate: given two schemas $S_1$ and $S_2$  from $L_1 \LC \ldots \LC L_n$, if they belong to the same
list $L_i$, then they are {\Cd} by a schema in $L_i$, since $L_i$ is cover-closed.
If they belong to two different lists $L_i$ and $L_j$ than there exists no
$J$ that satisfies both $S_1$ and $S_2$ since that $J$ would satisfy both $\JObj{\qany:{L_i}}$ and $\JObj{\qany:{L_j}}$, contradicting the 
hypothesis that they are disjoint, hence $S_1$ and $S_2$ are disjoint hence the pair $(S_1,S_2)$ is
trivially {\Cd} by $S_1$ (and by $S_2$ as well).
\end{appendixproof}

\begin{propxrep}\label{pro:ENFcorrect}
For any $S$, $\ENF(S)$ is equivalent to $S$ and is in ENF.
\end{propxrep}

\begin{appendixproof}
%

The cases when $\ExEP(S)\Down$ and  $\ExEI(S)\Down$ are trivial.
All other cases are proved by induction on the in-place depth.

Case $\JObj{K_1,\ldots,K_m} $\ with $m>1$: \\
equivalence: by definition, $\JObj{K_1,\ldots,K_m}$ is equivalent to 
$\JObj{\qall : [ \JObj{K_1},\ldots, \JObj{K_m}]}$;
by induction, this is equivalent to 
$\JObj{\qall : [ \JObj{\qany:L_1},\ldots, \JObj{\qany:L_m}]}$,
by Property \ref{pro:andor} (4) this is equivalent to 
$\JObj{\qany : \LAnd(L_1,\ldots,L_m)}$.\\
ENF: by the inductive hypothesis, every $L_i$   is {\ESC} and is cover-closed,
hence, by Property \ref{pro:andor} (1), $\LAnd(L_1,\ldots,L_m)$ is {\ESC} and is cover-closed,
hence $\JObj{\qany : \LAnd(L_1,\ldots,L_m)}$ is in ENF.

Case $\JObj{\qall:[S_1,\ldots,S_m]}$: essentially the same.

Case $\JObj{\qany:[S_1,\ldots,S_m]}$: the same, using Property \ref{pro:andor} (6) and Property \ref{pro:andor} (3).

Case $\JObj{\qone:[S_1,\ldots,S_m]}$: \\
equivalence: by definition, $\JObj{\qone:[S_1,\ldots,S_m]}$ is equivalent to 
$\JObj{\qany : [ S'_1,\ldots, S'_m]}$
where $S'_i= \JObj{\qall:[\qnotOf{S_1},\ldots,\qnotOf{S_{i-1}},S_i,\qnotOf{S_{i+1}},\ldots,\qnotOf{S_{m}}]}$;
by induction, this is equivalent to 
$\JObj{\qany : [ \JObj{\qany:L_1},\ldots, \JObj{\qany:L_m}]}$,
by associativity, this is equivalent to 
$\JObj{\qany :  L_1\LC\ldots\LC L_m}$.\\
ENF: by the inductive hypothesis, every $L_i$   is {\ESC} and is cover-closed,
hence, by Property \ref{pro:andor} (7), $ L_1\LC\ldots\LC L_m$ is {\ESC} and is cover-closed,
hence $\JObj{\qany : L_1\LC\ldots\LC L_m}$ is in ENF.

Case $\JObj{\qdref:u}$: 
equivalence: by definition, $\JObj{\qdref:u}$ is equivalent to $\key{deref}(u)$, which is equivalent to
$\ENF(\key{deref}(u))$ by induction on the in-place depth.
ENF: $\ENF(\key{deref}(u))$ is in ENF by induction on the in-place depth.
\end{appendixproof}

During the normalization process, we eliminate all duplicates from the arguments of all $\qall$ and $\qany$ keywords that we build;
this allows us to ensure that the final size of the schema that we get cannot grow more then $O(2^{|S|})$.

\begin{propxrep}
$\ENF(S)$ generates a schema whose size is in  $O(2^{|S|})$.
\end{propxrep}

\begin{appendixproof}
Consider a closed schema $S$ of size $N$.
For every $S'$ that is a subschema of $S$, we say that $S'$ is a source-node for $S$, 
and that $\qnotOf{S'}$ is a source-node for $S$, 
so that the number of different source-nodes for $S$ is $O(N)$.
For every $n$-uple $S_1,\ldots,S_n$ of source-nodes, we say that $\qallOf{S_1,\ldots,S_n}$  is a 
source-conjunction,
and we also consider any source-node $S'$ as a source-conjunction.
We will also assume that, whenever our algorithm computes the conjunction of two source-conjunctions, it 
flattens them, so that the result is still a source-conjunction.

We prove that:
\begin{quote}
when $S$ is a source-conjunction, then $\ENF(S)=\JObj{\qany: L}$,
where $L$ is a list of source-conjunctions. 
\end{quote}
We will then use this fact in order to establish a size bound.

Case (1):
we have $S=\JObj{\qany:L}$ and $\ENF(S)=S$. By hypothesis, $S$ is a source
conjunction; since it is neither a conjunction nor a negation, then it is a subschema of the initial schema, 
hence $L$ is a list of subschemas, hence $L$ is a list of source-conjunctions. 

Case (2):
here the result is $\qanyOf{S}$, where the only element of the list $[{S}]$
is $S$, which is a source-conjunction by hypothesis.

%

Cases for
$\ExEP(S)\Up \text{ or } \ExEI(S)\Up$ 
are proved by induction on the in-place depth, and by cases, as follows.

$$
\begin{array}{lllll}
(3)\ \ENF(\JObj{K_1,\ldots,K_n} ) &=&
                                                \JObj{\qany : \LAnd(\List{L_1,\ldots,L_m})}  \\
                                                &&\key{where}\ \ENF(\JObj{K_i}) = \JObj{\qany : L_i} \ \ 
                                                 \key{for}\ i\in\SetTo{m}
\end{array}
$$
By induction, every $L_i$ is a list of source-conjunctions.
The result follows since $\LAnd$ returns a list where every element is a conjunction of elements from the input lists, and the
conjunction of many source-conjunctions is still a source-conjunction.

$$
\begin{array}{lllll}
(4)\ \ENF(\JObj{\qall : [ S_1,\ldots, S_m]} ) &=&
                                                \JObj{\qany : \LAnd(\List{L_1,\ldots,L_m})}  \\
                                               &&\key{where}\ \ENF(S_i) = \JObj{\qany : L_i} \ \ 
                                                 \key{for}\ i\in\SetTo{m}\\[\NL]
\end{array}
$$
Since $\JObj{\qall : [ S_1,\ldots, S_m]}$ is a source-conjunction, then every $S_i$ is a source-node of the
original schema, hence every $S_i$ is a source-conjunction, hence, by induction, every $L_i$ is a list of source-conjunctions.
The result follows since $\LAnd$ builds all elements of its result as conjunctions of the elements of the input lists.

$$
\begin{array}{lllll}
(5)\ \ENF(\JObj{ \qany : [ S_1,\ldots, S_m]} ) &=&
                                                \JObj{\qany : \LOr(\List{L_1,\ldots,L_m})}  \\
                                                &&\key{where}\ \ENF(S_i) = \JObj{\qany : L_i} \ \ 
                                                 \key{for}\ i\in\SetTo{m}\\[\NL]
\end{array}
$$
Since $\JObj{ \qany : [ S_1,\ldots, S_m]}$ is a source-conjunction that is neither a $\qnot$ nor a $\qall$ keyword, then it is a subschema of the original
schema, hence every $S_i$ is a subschema of the original schema, 
hence it is a source-conjunction, hence,
by induction, every $L_i$ is a list of source-conjunctions.
The result follows since $\LOr$ concatenates elements that are in the input lists with conjunctions of such elements.

$$
\begin{array}{lllll}
(6)\ \ENF(\JObj{ \qone : [ S_1,\ldots, S_m]} ) &=&
                                                \JObj{\qany : L_1\LC\ldots\LC L_m}  \\
                                              &&\key{where}\ \ENF(\{\qall:[\qnotOf{S_1},\ldots,\qnotOf{S_{i-1}},S_i,\\[\NL]
                                              &&\qquad\qnotOf{S_{i-1}},\ldots,\qnotOf{S_{m}}]\}) = \JObj{\qany : L_i} \ \ \\[\NL]
                                               && \key{for}\ i\in\SetTo{m}\\[\NL]
\end{array}
$$
As in case (5), every  $S_i$ is a subschema of the original schema, hence every
$$\{\qall:[\qnotOf{S_1},\ldots,\qnotOf{S_{i-1}},S_i,ì\qnotOf{S_{i-1}},\ldots,\qnotOf{S_{m}}]\}$$
is a source-conjunction, hence we can apply induction on the in-place depth
and conclude that every $L_i$ is a source-conjunction;
list concatenation preserves this property.

$$
\begin{array}{lllll}
(7)\ \ENF (\JObj{\qdref : u}) &=& \ENF(\key{deref}(u)))  \\
\end{array}
$$
By induction on in-place depth, $\ENF(\key{deref}(u))$ returns  a schema with shape $\JObj{\qany: L}$ where $L$ is a list of 
source-conjunctions.

Hence, the algorithm returns a schema $\qanyOf{C_1,\ldots,C_n}$, such that every $C_i$ is a source conjunction
$C_i=\qallOf{S^i_1,\ldots,S^i_{m_i}}$.
By associativity, commutativity, and idempotence of $\qall$, we can systematically eliminate all duplicates from the 
arguments of the $\qall$ of $C_i$.
If we define a {\ddefined} schema for each source-node, every source-conjunction can be represented as a list of 
$N$ references or negated references, where $N$ is the number of
subschemas of the initial schema\footnote{just $N$, and not $2N$, since when both a reference and its negation are present we 
can rewrite the conjunction as \emph{false}}.

When we manipulate a list $L$ of source-conjunctions, in every step of our algorithm we can substitute it with a sublist $L'$ provided
that $\qany: L'$ is equivalent to $\qany: L$ (this can be proved by induction).
For this reason, whenever a list $L$ of source-conjunctions contains two source-conjunctions $\qall: [S_1,\ldots,S_n]$ and 
$\qall: [S'_1,\ldots,S'_n]$
which are pseudo-duplicates, that is, which only differ in the order of their elements, we can eliminate one of the two.
Hence, we can ensure that our algorithm only manipulates lists of source-conjunctions such that the 
source-conjunctions contain no duplicates and the list contains no pseudo-duplicates.
Since we only have $2^N$ different source-conjunctions --- if we regard pseudo-duplicates as equal --- every list of 
conjunctions that we manipulate contains less then $2^N$ source-conjunctions, hence has size smaller than $2^N \times N$.
Hence, for every schema $S$ smaller than $N$,
the size of $\ENF(S)$ is smaller than $2^N \times N$.
\end{appendixproof}

\subsection{Eliminating  $\qunProps$ and $\qunIts$}

We can finally use the function  $\ENF(S)$ to
define a function $\UElim(S)$ that eliminates the keywords
$\qunProps$ and $\qunIts$ from every schema $S$.

We say that a schema that contains $\qunStar:S_u$ as a top-level keyword is an
uneval-schema;
we apply $\UElim(S)$, hence $\ENF(S)$, to every uneval-schema.
In order to keep the size of the result in $O(2^N)$, we do not
want to apply $\ENF$ to a subschema $S'$, get an exponential blow-up,
 and then apply $\ENF$ again to a schema
that contains the result of $\ENF(S')$ as a subschema; for this reason, we do not want to have an
uneval-schema nested inside another uneval-schema.
For this reason, as a preliminary step,
we \emph{unnest} the source schema by (1) defining a new
definition $u_{S'}:S'$ for each occurrence of an uneval-subschema $S'$ of $S$  
that is not a {\ddefined} schema in the {\qddefs} section
and (2) by substituting that occurrence with ${\qdref:u_{S'}}$.
Hence, after unnesting, every uneval-schema is a {\ddefined} schema in the {\qddefs} section,
which does not contain any other uneval-schema, but only references to such schemas.
The unnested schema has size $O(N)$, where~$N$ is the size of the schema before unnesting.

We introduce the algorithm through an example, before introducing the formal definition
of the involved functions $\UElim(S)$, $\PUP{S_u}{\_}$, $\PUPe{S_u}{\_}$.
Consider the following schema.
\begin{Verbatim}[fontsize=\small,frame=lines]
{ "items": { "anyOf":  [ { "$ref": "#sale" } , { "$ref": "#car" } ], 
             "unevaluatedProperties": false },
  "$defs": {
    "sale": { "$anchor": "sale", "properties": { "price": { "type": "integer" }}},
    "car":  { "$anchor": "car",  "properties": { "plate": { "type": "string" }}}
  }
}
\end{Verbatim}

After unnesting, we obtain the following schema, where the only uneval-schema
is the {\ddefined} schema $\qkw{\#/\$defs/r}$.
\begin{Verbatim}[fontsize=\small,frame=lines]
{ "items": { "$ref":  "#/$defs/r" },
  "$defs": {
    "sale": { "$anchor": "sale", "properties": { "price": { "type": "integer" }}},
    "car":  { "$anchor": "car",  "properties": { "plate": { "type": "string" }}},
    "r":    { "anyOf":  [ { "$ref": "#sale" } , { "$ref": "#car" } ], 
              "unevaluatedProperties": false }
  }
}
\end{Verbatim}

After the schema has been unnested, we apply $\UElim(S)$ to every {\ddefined} schema;
if the schema has a shape $\JObj{K_1,\ldots,K_n,\qunStar:S_u}$,  $\UElim(S)$
computes $\ENF(\JObj{K_1,\ldots,K_n})$, and then pushes $S_u$ through all branches of the ENF
using $\PUP{S_u}{\ENF(\JObj{K_1,\ldots,K_n})}$.

For example, in this case, the ENF of the $\JObj{\qany:A}$ part of the schema identified by 
\qkw{\#/\$defs/r} would produce a {\qany} schema with three arguments:
$$
\begin{array}{lllllllll}
\multicolumn{3}{l}{
\UElim(\JObj{\qany:[\JObj{\qdref:\qkw{\#sale}},\JObj{\qdref:\qkw{\#car}}],\qunProps:\afalse} ) 
} \\[\NL]
\multicolumn{3}{l}{
= \PUP{\afalse}{\ENF(\JObj{\qany:[\JObj{\qdref:\qkw{\#sale}},\JObj{\qdref:\qkw{\#car}}]})} 
} \\[\NL]
= \PUPkey({\afalse},\{\qany:[ & \JObj{\qdref:\qkw{\#sale}},\\[\NL]
              & \JObj{\qdref:\qkw{\#car}},\\[\NL]
              &\qallOf{\JObj{\qdref:\qkw{\#sale}},\JObj{\qdref:\qkw{\#car}}} \\[\NL]
              &\!\!\!\!\!\!\!\!\! ]\}) \\[\NL]
\end{array}
$$
At this point, the function $\PUP{S_u}{\qanyOf{S_1,\ldots,S_n}}$ uses 
$\PUPe{S_u}{S_i}$ to push $S_u$ to each branch $S_i$ of the ENF.
$\PUPe{S_u}{S_i}$ returns
$$\qallOf{S_i,\JObj{\qpattProps:\JObj{p_1:\{\},\ldots,p_n:\{\}},\qaddProps:S_u}}$$
where $\qpattProps:\JObj{p_1:\{\},\ldots,p_n:\{\}}$ cites all properties of $\ExEP(S_i)$, so that
the keyword $\qaddProps:S_u$ applies $S_u$ to the other properties. 
$$
\begin{array}{lllllllll}
\multicolumn{2}{l}{
    \PUPkey({\afalse},\{\qany:[ 
    }& \JObj{\qdref:\qkw{\#sale}},\\[\NL]
              && \JObj{\qdref:\qkw{\#car}},\\[\NL]
              &&\qallOf{\JObj{\qdref:\qkw{\#sale}},\JObj{\qdref:\qkw{\#car}}} \\[\NL]
              &&\!\!\!\!\!\!\!\!\! ]\}) \\[\NL]
= \{\qany:[ & \multicolumn{2}{l}{\PUPe{\afalse}{\JObj{\qdref:\qkw{\#sale}}},}\\[\NL]
              &  \multicolumn{2}{l}{\PUPe{\afalse}{\JObj{\qdref:\qkw{\#car}}},}\\[\NL]
              & \multicolumn{2}{l}{\PUPe{\afalse}{\qallOf{\JObj{\qdref:\qkw{\#sale}},\JObj{\qdref:\qkw{\#car}}}}} \\[\NL]
              &\!\!\!\!\!\!\!\!\! ]\} \\[\NL]
= \{\qany:[ \\[\NL] 
       \qquad           \{\qall: &\multicolumn{3}{l}{ [ \JObj{\qdref:\qkw{\#sale}}, } \\[\NL]
    & \multicolumn{3}{l}{
                             \ \ \JObj {\qpattProps\JObj{\qkw{price}:\{\}}, \qaddProps:\afalse}  ]\},
              }\\[\NL]  
       \qquad           \{\qall: &\multicolumn{3}{l}{ [ \JObj{\qdref:\qkw{\#car}}, }\\[\NL]
    & \multicolumn{3}{l}{
                             \ \ \JObj {\qpattProps\JObj{\qkw{model}:\{\}}, \qaddProps:\afalse}  ]\},
              }\\[\NL]
    
       \qquad           \{\qall: &\multicolumn{3}{l}{ 
                                                         [ \qallOf{\JObj{\qdref:\qkw{\#sale}},\JObj{\qdref:\qkw{\#car}}},}\\[\NL]
    & \multicolumn{3}{l}{
                             \ \ \{\ \qpattProps\JObj{\qkw{price}:\{\}, \qkw{model}:\{\}}, 
                             }\\[\NL]
    &  \multicolumn{3}{l}{
                             \ \ \ \ \        \qaddProps:\afalse \}  ]\}
                              }\\[\NL]
              &\!\!\!\!\!\!\!\!\! ]\} \\[\NL]
\end{array}
$$


We now give the formal definition of $\PUP{S_u}{S}$ and $\PUPe{S_u}{S}$;
the full definition of  $\UElim(S)$ is only in Definition \ref{def:elim}, since we have to discuss arrays
before.


\begin{definition}[$\key{pProps}(\Set{p_1,\ldots,p_n}$]\label{def:props}
$\key{pProps}(\Set{p_1,\ldots,p_n})$ stands for
 $\qpattProps: \JObj {p_1: \JObj{},\ldots,p_n: \JObj{}}$.
\end{definition}

\begin{definition}[$\PUP{S_u}{S}$]\label{def:PUP}

Let $S=\qanyOf{ S_1,   \ldots, S_n}$ be a schema where every $S_i$ is
statically characterized; 
$\PUP{S_u}{S}$ and $\PUPe{S_u}{S_i}$ are defined as follows:
$$
\begin{array}{llllllll}
\PUPe{S_u}{ S_i } 
\!\!& = & 
\!\!\JObj{ \qall: [ S_i, \JObj{\key{pProps}(\ExEP(S_i)), \qaddProps: S_u }]} 
\\[2\NL]
\multicolumn{3}{l}{
\PUP{S_u}{ \qanyOf{ S_1,   \ldots, S_n}} 
}
\\ & = & 
 \!\!\JObjOpen\ \qany:[\ \PUPe{S_u}{ S_1 },\ldots,\PUPe{S_u}{ S_n}\ ]\ \JObjClose
\end{array}
$$
\end{definition}

\hide{
Observe that, when $\ExEP(S_i)$ matches every string, the schema $\JObj{\qall: [S_i,\JObj{\ldots}] }$
in the above definition can be substituted with $S_i$.

When $\rlan{\ExEP(S_n)} = \Sigma^*$ we can substitute $\JObj{\qall : [ S_i, \JObj{props(\ExEP(S_i), \qaddProps : S }]}$
in the above definition
with $S_i$, and when $propsOf(S_i)=\ExEP(S_i)$ we can substitute $\JObj{\qall : [ S_i, \JObj{props(\ExEP(S_i), \qaddProps : S }]}$
with $S_i$ enriched with the extra keyword $\qaddProps :S$.
}

The approach to push and eliminate $\qunIts$ is very similar.
In this case, we compute, for each branch $S_i$ of the ENF, the two components
$(h,S_e)$ of $\ExEI(S_i)$, which we indicate, respectively, with
$\ExEI_1(S_i)$ and $\ExEI_2(S_i)$.
We use $\key{prefIts}(h)$ to indicate a keyword $\qprefIts: [ \xtrue^1,\ldots,\xtrue^h]$ that evaluates
the first $h$ items of an array. Hence the following schema $S^i$ ensures that all items
after $\ExEI_1(S_i)$ that do not satisfy $\ExEI_2(S_i)$ satisfy $S_u$: 
$$
S^i \ =\ \JObj{\key{prefIts}(\ExEI_1(S_i)), \qits : \qanyOf{S_u,\ExEI_2(S_i)}}
$$
Hence, we can eliminate $\qunIts:S$ from a schema in ENF by adding $S^i$ to each $S_i$ conjunct of the
ENF $\qany : [ S_1,   \ldots, S_n  ]$ as follows, provided that $\ExEI_1(S_i)$ is finite;
when $\ExEI_1(S_i)=\Inf$, or $\ExEI_2(S_i)=\xtrue$, then all items of any array are evaluated, and, hence, 
 $S^i=\atrue$, and $\PUIe{S_u}{ S_i }=S_i$: when all items have been evaluated, then
 $\qunIts:S_u$ is trivially satisfied.

\begin{definition}[$\PUI{S_u}{S}$]\label{def:PUI}

Let $S=\qanyOf{ S_1,   \ldots, S_n}$ be a schema where every $S_i$ is
statically characterized; 
$\PUI{S_u}{S}$ and $\PUIe{S_u}{S_i}$ are defined as follows:
$$
\begin{array}{llllllllll}
  \ExEI_1(S_i) = \Inf \Or \ExEI_2(S_i) = \atrue: 
  \\[\NL]
\  \ \ \PUIe{S_u}{ S_i }  =\ \ S_i 
\\[3\NL]

  \ExEI_1(S_i) \neq \Inf \And \ExEI_2(S_i) \neq \atrue: \\[\NL]
\ \ \ \PUIe{S_u}{ S_i } \\
\qquad\qquad=\ \JObjOpen\ \qall :  [\ S_i, \ \  \!\!\JObj{\key{prefIts}(\ExEI_1(S_i)), \qits : \qanyOf{S_u,\ExEI_2(S_i)} }  \ ]\ \JObjClose 
 \\[3\NL]
\PUI{S_u}{ \qanyOf{ S_1,   \ldots, S_n}}  \\
\qquad\qquad=\ 
\JObjOpen\ \qany :[\ \PUIe{S_u}{ S_1 },\ldots,\PUIe{S_u}{ S_n}\ ]\ \JObjClose
\end{array}
$$

\end{definition}

We are finally ready for the formal definition of $\UElim(S)$.
The function $\UElim(S)$ does not need to recursively descend $S$ since,
thanks to unnesting, all the occurrences of $\qunStar:S_u$ are at the top-level
of a {\ddefined} schema.

\begin{definition}[$\UElim(S)$]\label{def:elim}
$\UElim(S)$ is defined by the first rule that can be applied to $S$ among the four rules that follow.
$$
\begin{array}{lllll}
\multicolumn{3}{l}{ \UElim(\JObj{K_1,\ldots,K_n, \qunProps: S_p, \qunIts: S_i }) }  \\[\NL]
\multicolumn{3}{l}{ \ \ = \{\,\qall{\Col}[\,\PUP{S_p}{\ENF (\JObj{K_1,\ldots,K_n})} ,\PUI{S_i}{\ENF (\JObj{K_1,\ldots,K_n})}\,]\,\} } \\[\NL]
\multicolumn{3}{l}{ \UElim(\JObj{K_1,\ldots,K_n, \qunProps: S_p})  =   \PUP{S_p}{\ENF (\JObj{K_1,\ldots,K_n})}} \\[\NL]
\multicolumn{3}{l}{ \UElim(\JObj{K_1,\ldots,K_n, \qunIts: S_i})  =   \PUI{S_i}{\ENF (\JObj{K_1,\ldots,K_n})}} \\[\NL]
\UElim(S)  &=& S\\[3\NL]
\end{array}
$$
\end{definition}

We now prove that $\UElim(S)$ is equivalent to $S$ and eliminates all instances of $\qunStar$.

\begin{propxrep}\ \\[-10pt]
\label{pro:uelim}
\begin{compactenum}
\item For any $S$ in ENF, $J$ satisfies $\PUP{S_u}{S}$ iff it satisfies $S$ and all the properties that are not evaluated by $S$ satisfy $S_u$.
\item For any $S$ in ENF, $J$ satisfies $\PUI{S_u}{S}$ iff it satisfies $S$ and all the items that are not evaluated by $S$ satisfy $S_u$.
\item For any $J$ and $S$, $J$ satisfies $\UElim(S)$ iff $J$ satisfies $S$.
\end{compactenum}
\end{propxrep}

\begin{appendixproof}
In this proof, we say that a schema $S\,\PCs\,\Cal{S}$, where $\Cal{S}$ is a non-empty set of schemas, iff:
(1) every $J$ that satisfies every schema in $\Cal{S}$ also satisfies $S$ and (2) every property that is evaluated by a schema
in $\Cal{S}$ is also evaluated by $S$.
We prove by induction that if a set of schemas $\Cal{S}$ is cover-closed, then any subset $\Cal{S}'$ of $\Cal{S}$ is $\PCd$ by a schema in $\Cal{S}$.
Case $|\Cal{S}'|=1$ is trivial.\ Case $\Cal{S}' = \Set{S_1,\ldots,S_{n+1}}$: by induction, $\Set{S_1,\ldots,S_{n}}$ is $\PCd$ by an element
$S_l$ of $\Cal{S}$, and, since $\Cal{S}$ is cover-closed, there is an element $S_m$ of $\Cal{S}$ that $\PCs$ the pair $(S_l,S_{n+1})$;
we prove that $S_m$ {\PCs} $\Cal{S}'$:
every $J$ that satisfies all elements of $\Cal{S}'$ satisfies both $S_l$ and $S_{n+1}$, hence it satisfies $S_m$; if a property is evaluated
by an element of $\Cal{S}'$ then it is evaluated either by $S_l$ or by $S_{n+1}$, hence it is evaluated by $S_m$.
The same definition and property holds for the notion of $\ICd$.
Now we proceed with the proof.

(1) For any $S$ in ENF, $J$ satisfies $\PUP{S_u}{S}$ iff it satisfies $S$ and all the properties that are not evaluated by $S$ satisfy $S_u$.
$$
\begin{array}{llllllll}
\multicolumn{3}{l}{
\PUP{S_u}{\qanyOf{ S_1,   \ldots, S_n}} 
}
\\
\mbox{\ \ } \ \ \ =  \JObjOpen \qany : \!\!&[ \!\!&\JObj{ \qall : [ S_1, \JObj{\key{pProps}(\ExEP(S_1)), \qaddProps : S_u }]}, & \qquad \\
& & \ldots\\
&&\JObj{ \qall : [ S_n, \JObj{\key{pProps}(\ExEP(S_n)), \qaddProps : S_u }]} & \qquad \\
&]  \JObjClose
\end{array}
$$
($\Rightarrow$). 
If $J$ satisfies $\PUP{S_u}{S}$, where $S=\qanyOf{ S_1,   \ldots, S_n}$,
then exists $l$ such that it satisfies $\qallOf{S_{l},\ldots}$,
hence it satisfies $S_{l}$, hence it satisfies $S$.
If a property $p$ is not evaluated by $S$ then it is not evaluated by the branch $S_{l}$,
hence, by Property \ref{pro:ex}, this property does not belong to $\ExEP(l)$, hence, since
$J$ satisfies $\JObj{\key{pProps}(\ExEP(S_{l})), \qaddProps : S_u }$
we conclude that $p$ satisfies $S_u$.

($\Leftarrow$). 
Let us assume that $J$ satisfies $S$ and that all the properties of $J$ that are not evaluated by $S$ satisfy $S_u$. 
Let $S=\qanyOf{ S_1,   \ldots, S_n}$ and let $\Cal{S}$ be the set of all schemas $S_i$ such that
$J$ satisfies $S_{i}$.
Since $S$ is in ENF, there exists $S_l\in\Set{S_1,   \ldots, S_n}$ such that $S_l\,\Cs\,\Cal{S}$.
Since $J$ satisfies all schemas in $\Cal{S}$ and $S_l\,\Cs\,\Cal{S}$, we deduce that $J$ satisfies $S_l$  (that is, $S_l\in\Cal{S}$).
Each property that is evaluated by $S$ is evaluated by some $S_i\in\Cal{S}$, hence, since $S_l\,\Cs\,\Cal{S}$, it is also evaluated
by $S_l$, hence, by Property \ref{pro:ex}, each evaluated property matches $\ExEP(S_l)$.
Hence, every property that does not match $\ExEP(S_l)$ is not evaluated by $S$ hence, by hypothesis, satisfies $S_u$.
Hence, $J$ satisfies $ \JObj{\key{pProps}(\ExEP(S_l)), \qaddProps : S_u }$. Since it also satisfies $S_l$, then $J$ satisfies the branch
 $\qallOf{S_{l},\JObj{\ldots}}$, of $\PUP{S_u}{S}$, hence it satisfies $\PUP{S_u}{S}$.
 
(2) For any $S$ in ENF, $J$ satisfies $\PUI{S_u}{S}$ iff it satisfies $S$ and all the items that are not evaluated by $S$ satisfy $S_u$.
$$
\begin{array}{llllllll}
\multicolumn{3}{l}{
\PUI{S_u}{\qanyOf{ S_1,   \ldots, S_n}}
}
\\
=  \JObjOpen\qany : \!\!&\!\!\![ &\!\JObj{ \qall : [ S_1, \JObj{\key{prefIts}(\ExEI_1(S_1)), \qits : \qanyOf{\ExEI_2(S_1),S_u} }]}, & \qquad \\
& & \!\ldots\\
&& \!\JObj{ \qall : [ S_n, \JObj{\key{prefIts}(\ExEI_1(S_n)), \qits : \qanyOf{\ExEI_2(S_n),S_u} }]} & \qquad \\
&
\multicolumn{2}{l}{ \!\!]\JObjClose }
\end{array}
$$
($\Rightarrow$). 
If $J$ satisfies $\PUI{S_u}{S}$, where $S=\qanyOf{ S_1,   \ldots, S_n}$,
then exists $l$ such that $J$ satisfies $\qallOf{S_{l},\ldots}$,
hence $J$ satisfies $S_{l}$, hence $J$ satisfies $S$.
If a item $J_i$ of $J$ is not evaluated by $S$ then it is not evaluated by the branch $S_{l}$,
hence, by Property \ref{pro:ex}, this item does not satisfy neither  $\ExEI_1(l)$  nor $\ExEI_2(l)$, 
that is, its position is $>\ExEI_1(l)$ and it does not satisfy $\ExEI_2(l)$, hence, since
$J$ satisfies $\JObj{\key{prefIts}(\ExEI_1(S_1)), \qits : \qanyOf{\ExEI_2(S_1),S_u} }$,
we conclude that $J_i$ satisfies $S_u$.

($\Leftarrow$). 
Let us assume that $J$ satisfies $S$ and that all the items of $J$ that are not evaluated by $S$ satisfy $S_u$. 
Let $S=\qanyOf{ S_1,   \ldots, S_n}$ and let $\Cal{S}$ be the set of all schemas $S_i$ such that
$J$ satisfies $S_{i}$.
Since $S$ is in ENF, there exists $S_l\in\Set{S_1,   \ldots, S_n}$ such that $S_l\,\Cs\,\Cal{S}$.
Since $J$ satisfies all schemas in $\Cal{S}$ and $S_l\,\Cs\,\Cal{S}$, we deduce that $J$ satisfies $S_l$.
Each item that is evaluated by $S$ is evaluated by some $S_i\in\Cal{S}$, hence, since $S_l\,\Cs\,\Cal{S}$, it is also evaluated
by $S_l$, hence, by Property \ref{pro:ex}, 
this item satisfies $\ExEI(S_l)$. 
Hence, every item that does not satisfy $\ExEI(S_l)$ is not evaluated by $S$ hence, by hypothesis, this item satisfies $S_u$;
we use (*) to refer to this fact.
We show that $J$ satisfies $ \JObj{\key{prefIts}(\ExEI_1(S_l)), \qits : \qanyOf{\ExEI_2(S_l),S_u} }$:
if an item satisfies $\ExEI_1(S_l)$, then it is not examined by $\qits$; it an item satisfies $\ExEI_2(S_l)$ then it satisfies
$\qanyOf{\ExEI_2(S_l),\ldots}$, hence it passes the $\qits$ test,
and if an item does not satisfy neither $\ExEI_1(S_l)$ nor $\ExEI_2(S_l)$, then it does not satisfy $\ExEI(S_l)$,
hence it satisfies $S_u$ (by *), else it satisfies
$\qanyOf{\ldots,S_u}$, hence also this item passes the $\qits$ test, hence $J$ satisfies
$ \JObj{\key{prefIts}(\ExEI_1(S_1)), \qits : \qanyOf{\ExEI_2(S_1),S_u} }$.
Since $J$ also satisfies $S_l$, then $J$ satisfies the branch
 $\qallOf{S_{l},\JObj{\ldots}}$ of $\PUI{S_u}{S}$, hence it satisfies $\PUI{S_u}{S}$.
 
(3) For any $J$ and $S$, $J$ satisfies $\UElim(S)$ iff $J$ satisfies $S$.

 $\UElim(S)$ is defined by the first rule that can be applied to $S$ among the four rules that follows.
$$
\begin{array}{lllll}
\multicolumn{3}{l}{ \UElim(\JObj{K_1,\ldots,K_n, \qunProps: S_p,K_n, \qunIts: S_i }) }  \\[\NL]
\multicolumn{3}{l}{ = \qallOf{ \PUP{S_p}{\ENF (\JObj{K_1,\ldots,K_n})} ,\PUI{S_i}{\ENF (\JObj{K_1,\ldots,K_n})} } } \\[\NL]
\multicolumn{3}{l}{ \UElim(\JObj{K_1,\ldots,K_n, \qunProps: S_p})  =   \PUP{S_p}{\ENF (\JObj{K_1,\ldots,K_n})}} \\[\NL]
\multicolumn{3}{l}{ \UElim(\JObj{K_1,\ldots,K_n, \qunIts: S_i})  =   \PUI{S_i}{\ENF (\JObj{K_1,\ldots,K_n})}} \\[\NL]
\UElim(S)  &=& S\\[3\NL]
\end{array}
$$
The first three cases are immediate consequences of cases (1)+Property \ref{pro:ENFcorrect} and (2)+Property \ref{pro:ENFcorrect}. The last case is trivial.

\end{appendixproof}

\begin{propxrep}\label{pro:elim}
A closed schema $S$  that has been unnested and where every {\ddefined} schema $S$ has been substituted
with $\UElim(S)$, does not contain any residual instance of $\qunStar$.
\end{propxrep}

\begin{appendixproof}
 $\UElim(S)$ is defined by the first rule that can be applied to $S$ among the four rules that follows.
$$
\begin{array}{lllll}
\multicolumn{3}{l}{ \UElim(\JObj{K_1,\ldots,K_n, \qunProps: S_p,K_n, \qunIts: S_i }) }  \\[\NL]
\multicolumn{3}{l}{ = \qallOf{ \PUP{S_p}{\ENF (\JObj{K_1,\ldots,K_n})} ,\PUI{S_i}{\ENF (\JObj{K_1,\ldots,K_n})} } } \\[\NL]
\multicolumn{3}{l}{ \UElim(\JObj{K_1,\ldots,K_n, \qunProps: S_p})  =   \PUP{S_p}{\ENF (\JObj{K_1,\ldots,K_n})}} \\[\NL]
\multicolumn{3}{l}{ \UElim(\JObj{K_1,\ldots,K_n, \qunIts: S_i})  =   \PUI{S_i}{\ENF (\JObj{K_1,\ldots,K_n})}} \\[\NL]
\UElim(S)  &=& S\\[3\NL]
\end{array}
$$

After all schemas are unnested,  neither the keywords $K_i$ nor the schemas $S$ and $S_u$ contain any instance of $\qunStar$.
We then prove by induction in the in-place depth  that the application of $\ENF$ to a schema $\JObj{K_1,\ldots,K_n}$ that contains no instance of $\qunStar$
returns a schema with the same property, 
and using the fact that the list manipulation operators, $\LAnd{\Cal{L}}$, $\LAnd{\Cal{L}}$, and $L' \LC L''$, do not
insert in the result any keyword that was not present in the parameters, with the only exception of $\qall$.
Finally, we observe that, when $S_u$ and $S_{ENF}$ contain no instance of $\qunStar$, then the same property holds for
$\PUP{S_u}{S_{ENF}}$ and for $\PUI{S_u}{S_{ENF}}$.
\end{appendixproof}

\section{Experimental evaluation}\label{sec:experiments}
\subsection{Implementation and Execution Environment}

Our research prototype implements our {\xunStar} elimination algorithm and is written in Scala 2.13. The dispatcher and evaluation scripts are written in Bash and Python 3.10. 
The experiments were executed in a Docker container running Ubuntu 22.0.4.
We configured a maximum heap space of 10 GB for the Java Virtual Machine (JVM).

Our execution platform is a server with two 20-core Intel Xeon Gold 6242R 3,1 GHz processors and 192~GB of RAM.

\subsection{Research Hypotheses}
We test the following hypotheses using our research prototype: 

\begin{compactitem}

\item H0 / Correctness of the implementation: Schemas before/after {\xunStar} elimination are equivalent.

\item H1 / Runtime: Given real-world schemas, the implementation has acceptable runtime, despite the exponential worst-case lower bound of the {\xunStar} elimination algorithm.

\item H2 / Size blow-up: Given real-world schemas, the blow-up in size is still reasonable, despite the exponential worst-case lower bound of the {\xunStar} elimination algorithm.
\end{compactitem}

H0 is not really a research hypothesis, 
it is rather an hypothesis about the quality of our code.
We describe its verification in this section for two reasons.
First of all, the reliability of the results regarding H1 and H2 depends on the reliability of the code;
therefore, we find it appropriate to describe here how the code has been tested.
Second, because, in order to test for H0, we developed a corpus of schemas and witnesses, and these artifacts may be
reused by other research groups. 

\subsection{Software Tools}
\label{sec:exp_tools}

In our experiments, we use several external tools: JSON Schema validators, a JSON Schema data generator, and a tool for equivalence checking of JSON Schema. We provide a brief overview of these tools here. In Section~\ref{sec:exp_check_correctness}, we explain their role
in our experiments.

\paragraph{Validators}
JSON Schema validators assess the validity of a JSON instance against a JSON Schema.  A plethora of JSON Schema validators is available which
unfortunately do not agree on some specific tests. To solve this issue, we employ Bowtie~\cite{bowtie}, a framework that offers a uniform
interface to different validators, to systematically compare the results of 22 validators that are integrated into the tool and that support {\mJS}.
We discovered that some of them never failed on our test cases, while others would fail on different cases. Hence, we decided to run each test
with the 22 validators and adopt the majority result. In practice, the result is the same as what we would get by choosing one of the robust tools,
but the approach protects us against overconfidence on a single tool.

\paragraph{Data Generator}
JSON data generators synthesize instances from a structural description, such as JSON Schema. We use JSON Generator~\cite{bl} (version 0.4.7), a tool which employs a heuristic approach to randomly create instances that conform to the provided JSON Schema. 

These instances are validated, and the program terminates once a valid instance is found.
While JSON Generator aims to generate valid instances, the integrated validator may occasionally produce false positives, leading to the creation of invalid instances. We leverage this behavior to generate both valid and invalid instances, assigning a ground truth (valid/invalid with respect to the schema) to each using the Bowtie framework, as described above.

\paragraph{Equivalence Checking}
To check the equivalence between manually eliminated schemas and the eliminated schemas produced by our implementation, as we will describe in Section~\ref{sec:exp_check_correctness}, we rely on the {\JS} witness generation tool described in~\cite{DBLP:journals/pvldb/AttoucheBCGSS22}.
While it can be quite slow, this tool was shown to be very accurate, and thus serves as a
reliable indicator of the correctness of our implementation. 

\subsection{Schema Collections and Instances}
\label{sec:exp_datasets}

\begin{table}[ht]
	\caption{Schema collections with number and size of schemas, and with the number of valid and invalid instances which were either generated using the data generator (\emph{Random}), or which were written by hand (\emph{Manual}).}
	\label{tab:exp_datasets}
	\centering
	\begin{tabular}{lrrrrrrc}
		\toprule
		\textbf{Collection} & \textbf{\#Schemas} & \multicolumn{1}{c}{\textbf{Average}}  
		& \multicolumn{1}{c}{\textbf{Maximum}}
		  & \multicolumn{1}{c}{\textbf{\#Total}} 
		  & \multicolumn{1}{c}{\textbf{\#Valid}}  & \multicolumn{1}{c}{\textbf{\#Invalid}}
		  & \multicolumn{1}{c}{\textbf{Generation}}  \\
		\textbf{} & \textbf{} 
		& \multicolumn{1}{c}{\textbf{Size (KB)}}  
		& \multicolumn{1}{c}{\textbf{Size (KB)}}
		& \multicolumn{1}{c}{\textbf{Instances}}  
		& \multicolumn{1}{c}{\textbf{Instances}}  & \multicolumn{1}{c}{\textbf{Instances}}  
		& \multicolumn{1}{c}{\textbf{Method}} \\
		\bottomrule
		GitHub & 305 &  63.21 &  429.90 & 1,347 & 515 &  832  &  \textit{Random} \\ 
		\midrule
		Test Suite & 65 &  0.34 & 1.58  & 182 & 99 &  83  &  \textit{Manual}\\ 
		\midrule
		Handwritten & 60 &  2.27 &  8.10 & 387 & 138 &  249 &  \textit{Manual}\\
		\bottomrule
	\end{tabular}

\end{table}

We analyze our approach using a diverse set of schemas and instances, consisting of real-world examples
and of  carefully hand-crafted tests. 
Table~\ref{tab:exp_datasets} describes these collections.

\paragraph{GitHub}
We systematically crawled schemas from GitHub using the GitHub code search API. Specifically, we downloaded all JSON files that contain the properties $\qunProps$ or $\qunIts$.

We removed duplicates as well as ill-formed JSON Schemas. 
We further removed 34 schemas that use features that are not yet supported by our implementation: 24 schemas referencing nested definitions, 6 schemas using {\qdepS}, 3 schema using {\qdid} and {\qda} keywords, 1 schema expressing dynamic references through {\qddRef} and {\qdda}. Notably, these are not inherent limitations of our approach, but simply deliberate restrictions of our research prototype.

Manual inspection of the 305 remaining schemas revealed that around 20\% of these schemas are toy examples or were designed to test validators (such as the JSON Schema Test Suite~\cite{testsuite}), while the remaining schemas seem to be used in practical, real-world applications.

We generated valid and invalid instances using the \emph{JSON Generator}~\cite{bl} data generator, as explained in Section~\ref{sec:exp_tools}. 
Therefore, we consider most of the schemas to constitute a realistic snapshot of real-world schemas that developers use, while all instances are artificial\footnote{Unfortunately, licensing restrictions prevent us from making this collection publicly available.}.

\paragraph{Test Suite}
The JSON Schema Test Suite~\cite{testsuite} is a community-curated collection of schemas with valid and invalid instances designed to benchmark validators. The schemas and instances are handcrafted to broadly cover the JSON Schema language and are very small, following a unit-test style.
The collection contains 70 schemas with $\qunStar$ keywords.
As with the GitHub collection, we removed 5 schemas because of features that are
not supported by our prototype: 3 schemas expressing dynamic references through {\qddRef} and {\qdda}, and 2~schemas using {\qdepS}.

\paragraph{Handwritten}
We carefully hand-crafted 60 schemas that complement the GitHub and test suite collections, as will be explained in Section~\ref{sec:exp_check_correctness}. 
For each handwritten schema, we created several valid and invalid instances, along with an equivalent schema in which we manually eliminated $\qunProps$ and $\qunIts$, as a ground truth for equivalence checking. 

We made the Handwritten collection available on GitHub~\cite{unevalcoll}.

\subsection{How we Test Correctness of the Implementation}
\label{sec:exp_check_correctness}


We developed three different approaches to spot correctness problems: (1)~random witness testing, (2)~manual witness testing, and (3)~handwritten translation testing.

(1) In \emph{random witness testing}, given an input {\mJS} schema $S_{i}$ from the GitHub collection, 
we use an external tool to generate positive witnesses, 
which are instances $J_y$
such that $S_i$ validates $J_y$, and also negative witnesses, which are instances $J_n$ such that $S_i$ does not validate $J_n$.
If $T(S_i)$ is the {\cJS} result of our {\xunStar} elimination tool, we use external validators to check that $T(S_i)$ validates every
positive witness $J_y$ and also does not validate any $J_n$ negative witness; if this does not hold, our tool has an error.

Of course, this ``pointwise'' approach is not exhaustive: the fact that $T(S_i)$ is equivalent to~$S_i$ on a limited set of instances does not
ensure that the two are equivalent, but every approach based on a finite set of tests has this problem.
A more serious limitation is that many, or even most, of the generated ``random'' witnesses may not really depend on how $S_i$ deals with
unevaluated properties; for example, when the
positive witness $J_y$ is just the empty record ``$\JObj{}$'', 
then its validation does not depend on the
presence of an $\qunProps:O$ keyword in $S_i$.

(2) In \emph{manual witness testing}, the schema and the witness have been designed purposely to test correctness.
To this aim, we exploited the standard JSON Schema Test Suite, which contains some examples of schemas that contain $\qunStar$ keywords
with positive and negative witnesses, and, more importantly, we designed and wrote our own Handwritten collection.
This is a collection of {\mJS} schemas that use the
$\qunStar$ keywords in a variety of
ways, in combination with a variety of other keywords. For each schema, we prepared a set of positive and negative examples whose validation,
or non-validation, depends on the $\qunStar$ keyword and on the way that annotations are passed by the other operators; this is similar
in spirit to the JSON Schema Test Suite, but is much more focused on this specific test case, and hence is much more thorough.

For each such schema $S_h$ and for each positive and negative witness, we then used an external validator to check that the witness is
actually positive (or negative), and then to check whether the {\xunStar}-eliminated schema $T(S_h)$ is equivalent to $S_h$ on these specific test cases.

This approach is much more likely to identify translation problems than the previous approach, as witnesses are designed to depend on the $\qunStar$ keyword and its precise placement. However, design and production of this kind of test case requires significant effort.

(3) In \emph{handwritten translation testing}, we prepared a set of {\mJS} schemas that use the $\qunStar$ keywords in a variety of ways and, for each schema $S_h$, we prepared a manual {\cJS} translation $S_m$ that is supposedly equivalent to $S_h$.
Then, for each schema $S_h$, we used an external {\cJS} equivalence checker to verify that the algorithmic translation
$T(S_h)$ was equivalent to the handwritten translation $S_m$.

Before carrying out the actual experiment, we verified the correctness of our translation using approach~(2), which is quite reliable. However, we cannot guarantee that the manual translation is correct, which could lead to false positives or false negatives. False positives are not a concern, as they are inherent to finite testing. False negatives have been removed, as follows:
when the external tool discovers that $T(S_h)$ is not equivalent to $S_m$, it also returns a non-equivalence witness $J$, that is,
an instance that is validated by $T(S_h)$ but not by $S_m$, or vice versa;
e.g., we may have $\SJudgCJ{T(S_h)}$ and not $\SJudgCJ{S_m}$.
At this point, we use an external validator to check the validity of $J$
for $S_h$. If $S_h$ and $T(S_h)$ are equivalent on $J$, e.g.\ $\SJudgCJ{T(S_h)}$ and $\SJudgCJ{S_h}$,
this means that $S_m$ was not a correct translation of $S_h$, and we can correct it;
If $S_h$ and $T(S_h)$ are not equivalent on $J$, in our example $\SJudgCJ{T(S_h)}$ and not $\SJudgCJ{S_h}$,
then we have a real, verified logical error of the translation tool.

This third approach is much more effective than the previous two, as it is not based on ``pointwise'' tests on some specific witnesses,
but on a full equivalence analysis between $T(S_h)$ and~$S_m$. However, it has its own drawbacks. 
First, its preparation is even more costly than that of the other approaches, as the manual translation of a complex schema requires considerable effort.
Secondly, equivalence checking for {\cJS} schemas containing complex combinations of boolean operators is computationally expensive. As we will see later, the only external tool available that supports the required set of {\JS} features often failed to complete the equivalence test within the allotted time.

To reduce manual effort, we use the same handwritten schemas $S_h$ with their positive and negative witnesses in two experiments.
 In \emph{handwritten translation testing}~(3), we use the witnesses in the preliminary phase where we verify the equivalence between $S_h$ and $S_m$, while they play no role in the $S_m \sim T(S_h)$ equivalence test.
 In  \emph{manual witness testing}~(2), we use the witnesses for the ``pointwise" comparison between $S_h$ and $T(S_h)$,
 while the manually translated schemas $S_m$ play no role.

An external equivalence tool that can directly check the equivalence between $S_i$ and $T(S_i)$ would give us a much more robust and easier way to check
the correctness of our implementation. However, as we specified in the Introduction, in this moment, no algorithm has yet been described to check equivalence
between two schemas that use {\qunStar} keywords, apart from the one we provide in this paper: we use our algorithm to eliminate
{\qunStar} from the schemas, and then the algorithm described in \cite{DBLP:journals/pvldb/AttoucheBCGSS22} to check the equivalence of the results.

\subsection{Experimental Results}

\begin{table}[ht]
	\caption{Correctness results of random/manual witness testing for each collection.}
	\label{tab:exp_correctness}
	\centering

	\begin{tabular}{lrrlrr}
		\toprule
		\textbf{Collection} & \textbf{\#Schemas} & \textbf{\#Instances}& \textbf{Method} & \textbf{Success} & \textbf{Errors}  \\
		\bottomrule
		GitHub & 305 & 1,347 & \textit{Random} & 100\% & 0\%     \\
		\midrule
		Test Suite & 65 & 182 & \textit{Manual} & 100\%  & 0\%    \\
		\midrule
		Handwritten & 60 & 387 &\textit{Manual} &  100\% & 0\%    \\
		\bottomrule
	\end{tabular}

\end{table}

\begin{table}[ht]
	\caption{Correctness results of handwritten translation testing.}
	\label{tab:exp_equivalence}
    
	\centering

	\begin{tabular}{lrrrrr}
		\toprule
		\textbf{Collection} & \textbf{\#Schemas} & \textbf{\#Success} & \textbf{\#Logical errors} & \textbf{\#Timeout} & \textbf{\#Unsupported}\\
		\toprule
		Handwritten & 60 & 30 & 0 &  29 & 1 \\
		\bottomrule
	\end{tabular}

\end{table}
\paragraph{Correctness of the implementation}
To verify hypothesis H0, we used three collections. The JSON Schema Test Suite offers complete coverage of the language, the GitHub
collection covers realistic use cases, and the Handwritten collection covers complex interactions between different operators.

Table \ref{tab:exp_correctness} reports the results of random and manual \emph{witness testing} (Section \ref{sec:exp_check_correctness}),
and it shows that almost 2,000 validity tests, on three different collections built in three very different ways, did not spot any problems
with our implementation of the algorithm described here.


Table~\ref{tab:exp_equivalence} shows the result of \emph{handwritten translation testing}
(Section \ref{sec:exp_check_correctness}). In this case, for each file $S_h$ in the 
Handwritten test, we manually prepared an equivalent file $S_m$ written in {\cJS}.
It is worth specifying that our manual translation was not just an application of our algorithm; for example, for all schemas that feature
the $\qone$ operator, we used a completely different technique, not described in the paper for space reasons,
that is much easier to compute by hand. As described in Section \ref{sec:exp_check_correctness},
we tested the actual equivalence between $S_m$ and  $S_h$
by the same pointwise testing that we documented in Table \ref{tab:exp_correctness}.
Further, we used the tool described in Section \ref{sec:exp_tools}
to verify the equivalence between the result of our algorithm on $S_h$ and the manual translation $S_m$.

Table~\ref{tab:exp_equivalence} shows the results. \emph{Success} states the number of eliminated schemas that are
proved equivalent to the manually translated schema, while \emph{Logical Error} states the number of eliminated schemas that are not equivalent.
Equivalence checking was performed with a timeout of three hours per pair of schemas; 
pairs exceeding this timeout are counted in \emph{Timeout}, while \emph{Unsupported} indicates runtime errors in the equivalence testing tool.
Unfortunately, equivalence testing is an exponential problem
and some of the Handwritten schemas exhibit complex combinations of boolean and structural operators that are quite challenging.
As a result, the external tool could complete the equivalence analysis for only half of our schemas;
all completed analyzes confirmed the expected perfect equivalence between the algorithm result
and manual elimination.

\begin{table}[ht]
	\centering
    	\caption{Runtime results (average of 100 runs), showing median, 95th percentile, and average runtime of each collection.}
	\label{tab:exp_runtime}
	\begin{tabular}{lrrrrr}
	\toprule
	\textbf{Collection} & \textbf{\#Schemas} & \textbf{Avg.\ Size (KB)} & \textbf{Median Runtime} & \textbf{95\%-tile Runtime} & \textbf{Avg.\ Runtime} \\
	\bottomrule
	GitHub & 305 & 63.21 & 0.13 ms & 0.68 ms & 0.26 ms \\
	\midrule
	Test Suite &  65 & 0.34 & 0.14 ms & 0.33 ms & 0.17 ms \\
	\midrule
	Handwritten & 60 & 2.27 & 0.14 ms & 0.39 ms & 0.18 ms \\
		\bottomrule
	\end{tabular}

\end{table}

\paragraph{Runtime}
\begin{figure}[ht!]
	\centering
	\begin{subcaptionblock}{0.32\textwidth}
		\centering
		\includegraphics[width=\textwidth]{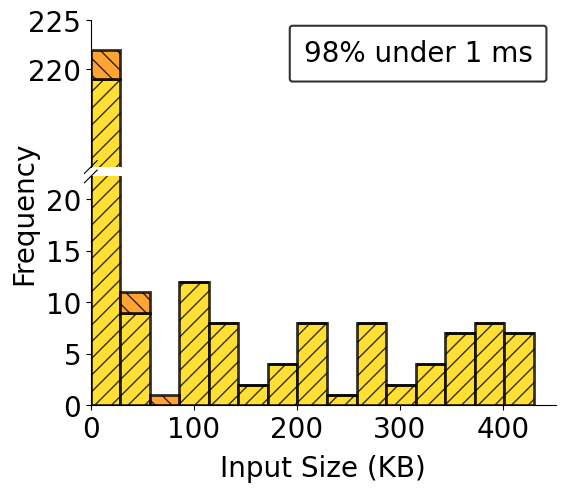}
		\caption{GitHub}
		\label{fig:runtime_github}
	\end{subcaptionblock}
	\hfill
	\begin{subcaptionblock}{0.32\textwidth}
		\centering
		\includegraphics[width=\textwidth]{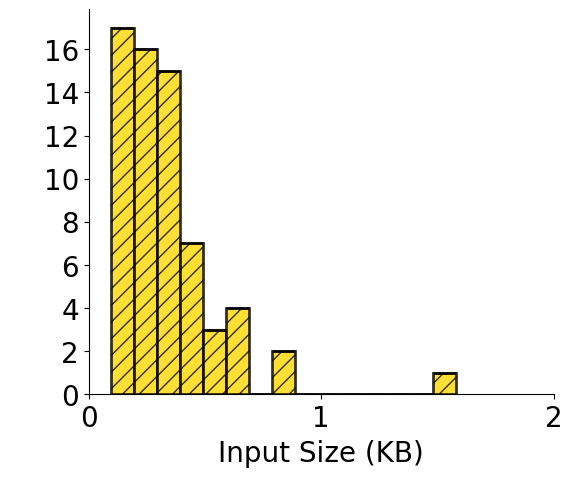}
		\caption{Test Suite}
		\label{fig:runtime_testsuite}
	\end{subcaptionblock}
	\hfill
	\begin{subcaptionblock}{0.32\textwidth}
		\centering
		\includegraphics[width=\textwidth]{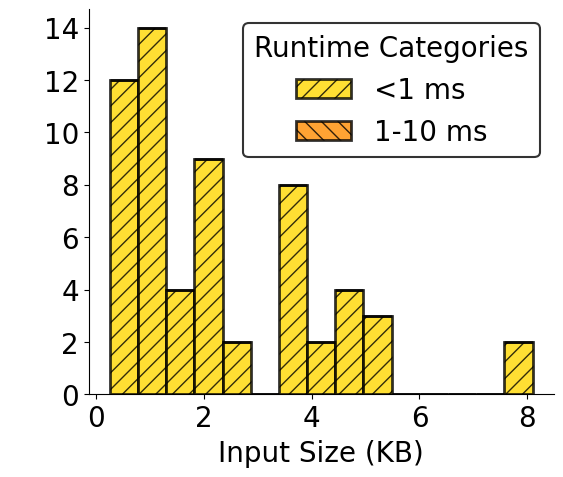}
		\caption{Handwritten}
		\label{fig:runtime_handwritten}
	\end{subcaptionblock}
	\caption{Histogram over the schema sizes for the schema collections. Yellow bars for {\xunStar} elimination within less than 1~ms, orange bars for runtimes of 1 to 10 ms. Runtimes averaged over 100 runs.}
    
	\label{fig:runtime}
\end{figure}

We next verify hypothesis H1 by measuring the runtime of {\xunStar} elimination on our collections. Table~\ref{tab:exp_runtime} shows the runtimes averaged over 100 runs for each collection, specifically the median, 95th percentile, and average runtime. 
The measurements exclude the initial schema parsing.
As runtime measurements under 1~millisecond are subject to measurement errors, these statistics should only be viewed as rough indicators.

Figure~\ref{fig:runtime} shows the distribution of schemas with respect to input size. Runtimes are visualized as stacked bars, distinguishing schemas processed in under 1 ms (yellow) and those taking between 1 and 10 ms (orange). In the GitHub collection, 98\% of schemas finish in under 1 millisecond and all remaining schemas take between 1 and 10 milliseconds. In the other collections, elimination for each schema is always performed in less than 1 millisecond. Interestingly, there is little correlation between input size and runtime, which may be explained by the fact that we do not include the parsing time in these runtime measurements. If we also consider parsing time, runtimes for all schemas larger than 50~kilobytes in the GitHub collection exceed 1 millisecond, yet no schema exceeds a runtime of 10~milliseconds, and 70\% remain below 1 millisecond.

These results confirm hypothesis H1: Our implementation demonstrates perfectly acceptable runtime performance on real-world schemas, processing over 98\% of schemas in under 1 millisecond each and never exceeding 10 milliseconds.

\begin{figure}[h!]
	\centering
	\begin{subcaptionblock}{0.32\textwidth}
		\centering
		\includegraphics[width=\textwidth]{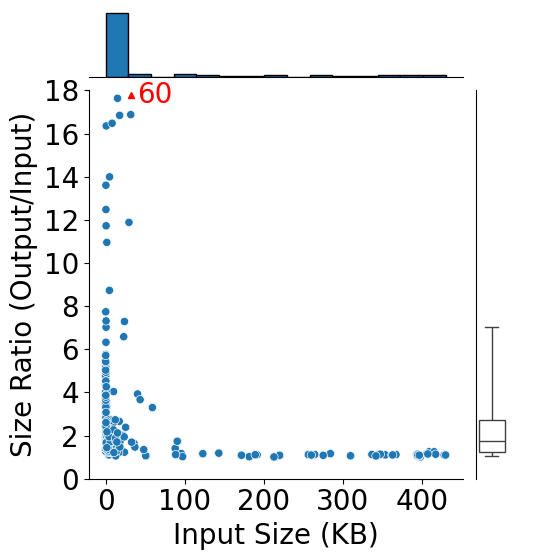}
		\caption{GitHub}
		\label{fig:github_size_ratio}
	\end{subcaptionblock}
	\hfill
	\begin{subcaptionblock}{0.32\textwidth}
		\centering
		\includegraphics[width=\textwidth]{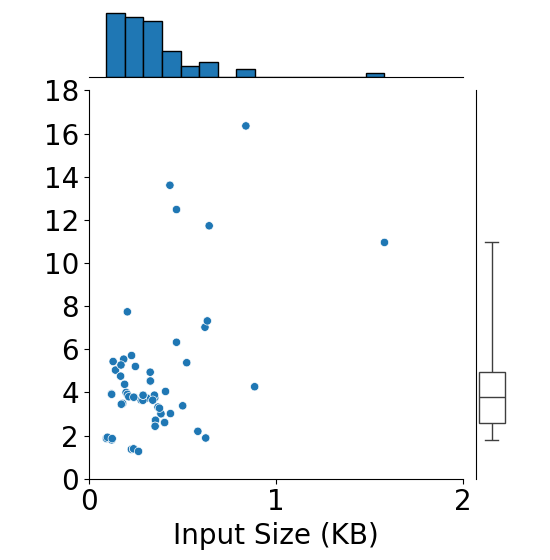}
		\caption{Test Suite}
		\label{fig:testsuite_size_ratio}
	\end{subcaptionblock}
	\hfill
	\begin{subcaptionblock}{0.32\textwidth}
		\centering
		\includegraphics[width=\textwidth]{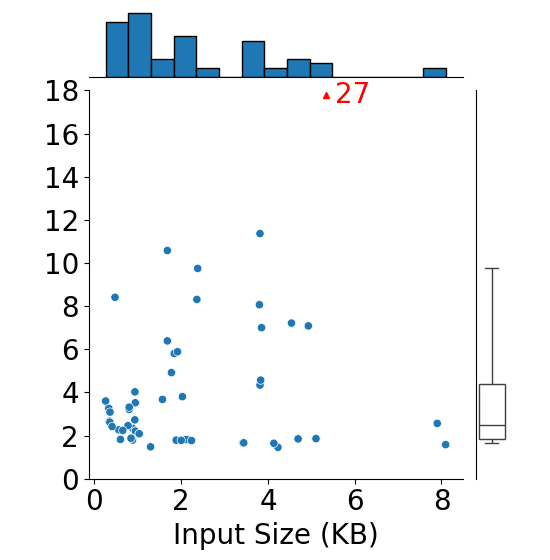}
		\caption{Handwritten}
		\label{fig:handwritten_size_ratio}
	\end{subcaptionblock}
	\caption{Investigating size blow-up: Schema input size vs.\ size ratio (output size/input size) of schemas for the schema collections. Outliers and their size ratio (rounded) are shown in red. Histograms along the horizontal axis show the distribution of schemas by input size. Boxplots along the  vertical axis show distribution of size ratios, with whiskers reaching to the 5th/95th percentile.}
	\label{fig:size_ratio}
\end{figure}

\paragraph{Size blow-up}
We test hypothesis H2 by comparing the size of the eliminated schemas to the size of the original schemas. Figure~\ref{fig:size_ratio} shows the size ratio (output size divided by input size) of schemas processed by our implementation compared to the input size of the schemas. 87\% of schemas across all collections have a size ratio below 5, while less than 5\% of schemas exceed a size ratio of 10.

While the Test Suite and Handwritten collections contain only small schemas of up to 8 kilobytes, the GitHub collection also contains a number of larger schemas, up to 430 kilobytes, with 75\% of schemas in the GitHub collection having an input size below 50 kilobytes. Noticeably, almost all larger schemas (more than 50 kilobytes) in the GitHub collection have a size ratio close to~1. This can be explained by two factors: First, we measure the blow-up relative to the size of the input schema, and thus the same absolute size blow-up caused by {\xunStar} elimination is more noticeable in small schemas than in large ones.
Second, all 59 schemas in the GitHub collection that are larger than 120 kilobytes originate from the same GitHub repository, exhibiting considerable similarities in their use of $\qunStar$ keywords, which may explain why the size ratio shows very little variation for these schemas.


There are two noticeable outliers, one in the GitHub collection, with a size ratio of 60 (rounded) and one in the Handwritten collection, with a size ratio of 27 (rounded). The high size ratio of these outliers is caused by an unusually extensive use of logical operators.

These results support hypothesis~H2, with the vast majority of analyzed real-world schemas exhibiting a size ratio below 5. Although some schemas exceed a size ratio of 10 and one outlier even reaches a size ratio of 60, we observe these higher values only for schemas with an input size below 50 KB, meaning that the absolute output size is still reasonable: on average, schemas with a size ratio above 10 have an output size of around 200~KB.

\subsection{Discussion}
We evaluated the correctness of our research prototype on more than 400 real-world and artificial schemas. Using almost 2,000 instances for random and manual witness testing, along with a set of manually translated schemas for handwritten translation testing, we found no correctness issues with our implementation. Runtime analysis showed that our algorithm processes most schemas in under 1~millisecond and never exceeds 10~milliseconds, suggesting that it is well suited for practical use. Despite the exponential worst-case lower bound, the observed size blow-up remained manageable for most real-world schemas, with only one noticeable outlier.

These results are coherent with our research hypotheses and show the applicability of our algorithm for real-world schemas.

\section{Conclusions}

The $\qunStar$ keywords of {\mJS} are very useful, but they are not compatible with the algorithms that
have been defined to decide inclusion, equivalence and satisfiability of {\JS}. The automatic elimination of these
keywords is the most natural solution.

In this paper, we have shown that no elimination technique can avoid an exponential explosion of the schema size,
for some specific families of schemas.
We have shown that this property holds for both the $\qunProps$ and $\qunIts$ keywords, despite the important technical differences between the two.

We have then defined an elimination algorithm, based on the static characterization 
 of the evaluated properties and
items, on the notion of cover closure, and on the Evaluation Normal Form (ENF).

The ENF approach is designed in order to keep the size of the output, and hence the computation time, under strict control
on real-world schemas.
We have designed and executed a set of experiments to validate the algorithm and to analyze its performance on real-world schemas, and these
experiments confirm our hypothesis.

{\mJS} adds both the $\qunStar$ keywords and dynamic references to {\cJS}; the elimination of dynamic references has been proved
to require an exponential explosion in \cite{DBLP:journals/pacmpl/AttoucheBCGSS24}, and here we prove that the same holds for the $\qunStar$ keywords.
It is difficult to guess whether it would be possible to eliminate both classes of operators with a single exponential blow-up or whether
there exist schemas where a double exponential blow-up is unavoidable --- we leave this as an open problem.

We are in contact with the community in charge of maintaining the standard, have communicated our results, and have received very valuable feedback; we thank them here for their support, questions, and suggestions.

\bibliographystyle{elsarticle-num-names}
\bibliography{references}

\end{document}